\documentclass[numberedappendix,onecolumn,apj]{openjournal}
\usepackage{array}[=2016-10-06]
\usepackage[english]{babel}
\usepackage{xcolor}

\newcommand{\grayrowstart}[1]{%
  \begingroup
  \setlength{\fboxsep}{0pt}%
  \leavevmode
  \rlap{%
    \hspace*{-3pt}%
    \colorbox[gray]{0.9}{%
      \makebox[\dimexpr0.90\textwidth\relax][l]{\strut}%
    }%
  }%
  #1%
  \endgroup
}

\usepackage{amsmath}
\usepackage{bbm}
\usepackage{graphicx}
\usepackage{hyperref}
\hypersetup{colorlinks=true, allcolors=blue}

\usepackage{enumitem}
\usepackage{acronym}
\newcolumntype{L}[1]{>{\hspace{3pt}\arraybackslash}p{#1}<{\hspace{3pt}}}
\newcolumntype{C}[1]{>{\hspace{3pt}\centering\arraybackslash}p{#1}<{\hspace{3pt}}}

\usepackage[caption=false]{subfig}

\makeatletter
\providecommand{\@eapj@font}{\footnotesize}
\makeatother

\newacro{ICM}{Intra-Cluster Medium}
\newacro{SED}{Spectral Energy Density}
\newacro{CMB}{Cosmic Microwave Background}
\newacro{CIB}{Cosmic Infrared Background}
\newacro{ACT}{Atacama Cosmology Telescope}
\newacro{tSZ}{thermal Sunyaev-Zel'dovich}
\newacro{pa}{polarized array}
\newacro{ISW}{Integrated Sachs-Wolfe}
\newacro{kSZ}{kinetic Sunyaev-Zel'dovich}
\newacro{BIC}{Bayesian Information Criterion}
\newacro{MCMC}{Markov-Chain Monte Carlo}
\newacro{HOD}{Halo Occupation Distribution}
\newacro{NFW}{Navarro–Frenk–White}

\begin{document}


\title{Thermal Sunyaev-Zel'dovich cross-correlations with unWISE galaxies: disentangling radio contamination, dust properties, and electron pressure}

\author{Guandi Zhao}
\affiliation{Department of Physics, ETH Z\"urich, R\"amistrasse 101, 8092 Z\"urich, Switzerland}

\author{Alex Krolewski}
\affiliation{Waterloo Centre for Astrophysics, University of Waterloo, 200 University Avenue West, Waterloo, ON N2L 3G1, Canada}
\affiliation{Department of Physics and Astronomy, University of Waterloo, 200 University Avenue West, Waterloo, ON N2L 3G1, Canada}
\affiliation{California Institute of Technology, 1200~East California Boulevard, Pasadena, CA 91125, USA}
\affiliation{CITA National Fellow}

\author{Niayesh Afshordi}
\affiliation{Waterloo Centre for Astrophysics,  University of Waterloo, 200 University Avenue West, Waterloo, ON N2L 3G1, Canada}
\affiliation{Department of Physics and Astronomy, University of Waterloo, 200 University Avenue West, Waterloo, ON N2L 3G1, Canada}
\affiliation{Perimeter Institute for Theoretical Physics, Waterloo, ON N2L 2Y5, Canada}


\begin{abstract}
Cross-correlations between the thermal Sunyaev-Zel'dovich (tSZ) effect and galaxy surveys provide a sensitive probe of hot gas in low-mass halos, but on the angular scales of greatest astrophysical interest they are also highly vulnerable to residual foreground contamination. We analyze the cross-correlation of two low-redshift unWISE galaxy samples with \textit{Planck} PR4 and ACT DR6 microwave temperature sky maps directly in harmonic space, fitting simultaneously for correlated tSZ, cosmic infrared background (CIB), and radio emission. Using the nine Planck\ bands to perform model selection, we find that a three-component model consisting of tSZ, radio emission, and a CIB amplitude term is strongly preferred over a model that omits radio contamination, with a significance of $9.5\sigma$ for the unWISE Low-z sample and $11\sigma$ for the unWISE Mid-z sample  respectively. For the unWISE Low-z sample  ($\langle z \rangle_L = 0.14$) the preferred \textit{Planck} fit gives an effective CIB emissivity index $\beta_0 = 1.79 \pm 0.29$, an effective CIB dust temperature $T_0 = 22.0 \pm 5.1~{\rm K}$, and a radio spectral index $\beta_r = -2.18 \pm 0.24$; for the unWISE Mid-z sample ($\langle z \rangle_M = 0.23$)  we find $\beta_0 = 1.41 \pm 0.13$, $T_0 = 28.0 \pm 4.2~{\rm K}$, and $\beta_r = -2.62 \pm 0.26$. We apply this radio-inclusive recipe to the ACT+\textit{Planck} 90, 150, and 220 GHz maps to obtain the best small-scale measurements. This removes the apparent negative galaxy-tSZ cross-correlation seen in the fiducial ACT DR6 NILC reconstruction and in no-radio fits. The cleaned tSZ$\,\times\,$unWISE spectra remain positive to $\ell \simeq 6000$ and are broadly consistent with the Planck-only reconstruction on overlapping scales. We show that these radio-cleaned spectra are well described by a conventional halo model calibrated to the galaxy population and a two-parameter generalized NFW electron pressure profile. Our main conclusion is that radio contamination separation is crucial to interpreting small-scale tSZ$\,\times\,$galaxy measurements.
\end{abstract}

\maketitle

\section{Introduction}


The \ac{tSZ} effect is one of the most important probes of hot ionized gas in the late-time universe. Inverse-Compton scattering of \ac{CMB} photons by energetic electrons imprints a characteristic frequency-dependent distortion that is proportional to the line-of-sight integral of the electron pressure \citep{szeffectsz,szbirkshaw,szcosmology2002}. In massive clusters, this effect has long been used together with X-ray observations to study ICM thermodynamics, pressure profiles, gas fractions, and the impact of baryonic feedback \citep{XrayICM1986,tracingicm,clusterformationrev,ArnaudGNFW,Battaglia2012_YM,Battaglia2012_tSZpower,Battaglia2013_fgas,Eckert2017_xcop,Walker2019_outskirts}. These same baryonic processes are now understood to have important consequences for cosmological observables on non-linear scales, motivating feedback-calibrated modeling frameworks \citep{vanDaalen2011_baryonpk,McCarthy2017_BAHAMAS,Henden2019_FABLE}. The more difficult and arguably more interesting regime is the lower-mass halo population, where the \ac{tSZ} signal is weaker but carries direct information about circumgalactic gas, baryon redistribution, and the missing-baryon problem \citep{wherearethebaryons,missingbaryonniayesh,missingbaryonstacking,szstackingtanumura,Chadayammuri2022_CGM,Zhang2024_eROSITA_profile,Zhang2024_eROSITA_scaling}.

Galaxy--CMB cross-correlations are a natural way to reach this regime because they amplify the signal from halos that are too faint to detect individually. The unWISE catalog is especially useful in this context: it provides all-sky infrared-selected tracers with broad redshift coverage and high surface density \citep{unWISE_survey,unWISE_catalog}. 
In this work, we are particularly
interested in probing the tSZ effect at low halo masses, and so rather than using the unWISE Blue, Green and Red samples used in previous work \citep{unWISE_catalog,krolewski_unwise_2020}, we use galaxy samples drawn from a different part of the \textit{WISE} color space,  with a selection function
chosen to pick out low-redshift, low-mass galaxies. We use two different color selections, referred to as the unWISE Low-z ($\bar{z} = 0.14$) and unWISE Mid-z ($\bar{z} = 0.23$) samples.

On the microwave side, \textit{Planck} provides broad frequency leverage from 30 to 857 GHz, while \ac{ACT}\ DR6 provides arcminute-scale resolution in the 90, 150, and 220 GHz bands \citep{ACTDR6maps_description,Coultontszcomponent}. Together, they offer a powerful combination: \textit{Planck} can identify the relevant spectral components, and \ac{ACT} can extend the measurement to the scales where halo-pressure modeling becomes most informative.

The main obstacle is that component-separated Compton-$y$ maps are not guaranteed to isolate the true low-redshift \ac{tSZ} signal on galaxy scales. Other temperature fluctuations from diffuse components such as \ac{CIB} can bury the true \ac{tSZ} signal, especially at angular scales $\ell>1000$ \citep{SPT_sz_2012}. The \ac{CIB} is mainly attributed to thermal dust emission, and, due to uncertainty in the CIB's \ac{SED}, it is difficult to remove via component separation. Both ACT and \textit{Planck} collaboration made map-level component separation for \ac{tSZ} to suppress the contamination using Needlet Internal Linear Combination (NILC) method, including additional marginalization over the uncertainty in the CIB spectral index $\beta$  \citep{mccarthy_component-separated_2023,improvedcomptonyplanck,Coultontszcomponent}. However, as discussed by \citet{Chandran_enhancednilc}, the NILC map products still show residual correlation with \ac{CIB} and need to be handled with caution in cross-correlation analysis. Recent work by \citet{LiustackACTtsz} also emphasized that small-scale galaxy--$y$ measurements can remain strongly dependent on the CIB model even after map-level \ac{CIB} deprojection. For low-redshift, infrared-selected galaxies such as unWISE, there is an additional concern: radio emission correlated with the galaxy sample can mimic or bias the frequency dependence of the \ac{tSZ} decrement at 90--150 GHz. If that contaminant is not removed, a physically sensible positive-pressure signal can be mistaken for a suppressed or even negative galaxy--$y$ correlation.

In this work, we disentangle the contamination by modeling the galaxy-temperature cross-spectra directly in harmonic space. We first use the nine \textit{Planck} bands to compare several spectral recipes and determine which components are required by the data. We then apply the preferred recipe to the ACT+\textit{Planck} frequency maps, where the highest multipole extends to $\ell \simeq 6000$. Then, we test whether the cleaned \ac{tSZ}$\,\times\,$unWISE spectra allow a conventional halo-model description using a Halo Occupation Distribution (HOD) fit to the galaxy clustering and a generalized Navarro-Frenk-White (GNFW) pressure profile for the gas. Finally, we show the inferred SZ flux--halo mass ($Y_{\rm SZ}-M$) relation, and compare to different SZ group and cluster observations.



\section{Method}\label{sec:method}

We model the galaxy--temperature cross-correlations directly in harmonic space. For each frequency band $f$, we measure the angular cross-spectrum between the galaxy overdensity field $g$ and the CMB temperature fluctuation map $T_f$, and describe it as a linear combination of components with distinct spectral energy distributions (SEDs). In this framework, the scale dependence is carried by the component cross-spectra, while the frequency dependence is encoded in band-averaged response coefficients. This allows us to perform component separation in each multipole bin, and therefore to account for scale-dependent contamination from radio and CIB-related signals.

The angular cross-correlation of two fields $u$ and $v$ on the sky is written as
\begin{align}
    w^{uv}(\hat n\cdot\hat n') &= \langle u(\hat n) v(\hat n') \rangle
    = \sum_\ell \frac{2\ell+1}{4\pi} C_\ell^{uv} P_\ell(\hat n\cdot\hat n'),\\
    C_\ell^{uv} &= \frac{1}{2\ell+1}\sum_m u_{\ell m} v_{\ell m}^\ast.
\end{align}
In this work, the relevant observables are the galaxy overdensity field $g$, the frequency-dependent CMB temperature fluctuation maps $T_f$, and the component amplitudes associated with tSZ, CIB, and radio emission described later in Sec.~\ref{sec:signal_responses}.

\subsection{Signal model}\label{sec:signalmodel}

For multipoles $\ell\gtrsim100$, the correlation between galaxy overdensity and the primary CMB is negligible apart from a small Integrated Sachs-Wolfe contribution on large scales. Therefore we apply a scale cut of $\ell_{\textrm{min}}=200$, and model the observed galaxy--temperature cross-spectrum in each frequency band as a linear superposition of several astrophysical components with different spectral responses. For a frequency channel $f$, the measured cross-spectrum can be written as
\begin{equation}
    \hat C_\ell^{gT_f}
    =
    B_\ell^f W_\ell
    \sum_{i=1}^{N_s} g_i^f\, C_\ell^{gS_i}
    +
    n_\ell^f,
\end{equation}
where $C_\ell^{gS_i}$ is the cross-spectrum between the galaxy field and component $S_i$, $N_s$ is the total number of components, and $g_i^f$ is the band-averaged spectral response of that component in frequency channel $f$, and $B_\ell^f$ is the beam transfer function, $W_\ell$ is the pixel window function, and $n_\ell^f$ denotes the measurement noise. In this description, the frequency dependence is encoded in the response coefficients $g_i^f$, while the scale dependence is carried by the component cross-spectra $C_\ell^{gS_i}$.

The goodness of fit to the observed cross-correlation in each frequency channel $C_\ell^{gT_f}$ is given by $\chi^2$:
\begin{equation}
    \chi^2 = \left[\hat{C}^{gT_f}_\ell - B_\ell W_\ell \sum_i g_i^f C_\ell^{gS_i}\right]^T \mathbf{Cov}^{-1}\left[\hat{C}^{gT_f}_\ell - B_\ell W_\ell \sum_i g_i^f C_\ell^{gS_i}\right].\label{eq:chisq_clean}
\end{equation}

Stacking all frequency channels and multipole bins into a single data vector, the problem can be written in matrix form as
\begin{equation}
    \mathbf{D} = \mathbf{G}\mathbf{v} + \mathbf{n},
\end{equation}
where $\mathbf{D}$ contains the measured galaxy--temperature cross-spectra, $\mathbf{v}$ contains the individual foreground component amplitudes in each multipole bin, and $\mathbf{G}$ is the response matrix constructed from the beam, pixel window, and band-averaged SED coefficients. Explicitly, for $N_f$ frequency channels, $N_s$ foreground components, and $N_\ell$ multipole bins, we write
\begin{equation}
\mathbf{D}=
\begin{pmatrix}
\hat C_{\ell_1}^{gT_1}\\
\hat C_{\ell_2}^{gT_1}\\
\vdots\\
\hat C_{\ell_{N_\ell-1}}^{gT_{N_f}}\\
\hat C_{\ell_{N_\ell}}^{gT_{N_f}}
\end{pmatrix},
\qquad
\mathbf{v}=
\begin{pmatrix}
C_{\ell_1}^{gS_1}\\
C_{\ell_2}^{gS_1}\\
\vdots\\
C_{\ell_{N_\ell-1}}^{gS_{N_s}}\\
C_{\ell_{N_\ell}}^{gS_{N_s}}
\end{pmatrix},
\end{equation}
with response matrix (of dimension $(N_f N_\ell)\times(N_s N_\ell)$)
\begin{equation}
\mathbf{G}=
\begin{pmatrix}
B_{\ell_1}^{1}W_{\ell_1}^1g_1^1 & 0 & \cdots & B_{\ell_1}^{1}W_{\ell_1}^1g_{n}^1 & 0 & \cdots & 0\\
0 & B_{\ell_2}^{1}W_{\ell_2}^1g_1^1 & \cdots & 0 & B_{\ell_2}^1W_{\ell_2}^1g_n^1 & \cdots & 0\\
\vdots & \vdots & \ddots & \vdots & \vdots & & \vdots\\
B_{\ell_1}^{k}W_{\ell_1}^{k}g_1^{k} & 0 & \cdots & B_{\ell_1}^{k}W_{\ell_1}^kg_{n}^{k} & 0 & \cdots & 0\\
0 & B_{\ell_2}^{k}W_{\ell_2}^{k}g_1^{k} & \cdots & 0 & B_{\ell_2}^{k}W_{\ell_2}^kg_n^k & \cdots & 0\\
\vdots & \vdots &  & \vdots & \vdots & \ddots & \vdots\\
0 & 0 & \cdots & 0 & 0 & \cdots & B_{\ell_{N_\ell}}^{N_f}W_{\ell_{N_\ell}}^{N_f}g_{N_s}^{N_f}
\end{pmatrix}.
\end{equation}
The matrix $\mathbf{G}$ is {\it assumed to be} block-diagonal over multipoles in each signal-frequency block, justifying that component separation can be performed independently in each $\ell$ bin. Then the $\chi^2$ is written as:
\begin{equation}
    \chi^2 = [\mathbf{D}-\mathbf{Gv}]^{\rm T}\mathbf{Cov}^{-1}[\mathbf{D}-\mathbf{Gv}].
\end{equation}

For fixed external \ac{SED} parameters, the foreground component amplitudes are obtained with a generalized least-squares estimator,
\begin{equation}
    \hat{\mathbf{v}}
    =
    \left(\mathbf{G}^{\mathrm T}\mathbf{Cov}^{-1}\mathbf{G}\right)^{-1}
    \mathbf{G}^{\mathrm T}\mathbf{Cov}^{-1}\mathbf{D},\label{eq:estimatorsol}
\end{equation}
with covariance
\begin{equation}
    \mathbf{\mathcal C}
    =
    \left(\mathbf{G}^{\mathrm T}\mathbf{Cov}^{-1}\mathbf{G}\right)^{-1}.
\end{equation}

A necessary condition for this component-separation problem to be well-posed is that the number of frequency channels be at least as large as the number of components, $N_f \geq N_s$, so that each multipole block is not underdetermined. This requirement is important to our analysis strategy. Because \textit{Planck} provides nine frequency channels, it allows us to test different multi-component SED recipes and assess which one is favored by the data. By contrast, ACT provides only three frequency channels, so ACT-only component separation is restricted to at most three components and cannot by itself determine which three-component recipe is preferred. For this reason, we use the broader frequency coverage of \textit{Planck} for SED model selection, and then apply the preferred recipe to the ACT+\textit{Planck} cross-spectra to recover the small-scale galaxy--tSZ signal.

\subsection{Signal responses}\label{sec:signal_responses}

We now specify the frequency response coefficients $g_i^f$ that enter the linear signal model in Sec.~\ref{sec:signalmodel}. These coefficients encode the spectral dependence of each component in frequency channel $f$, after averaging over the instrumental passband. In our analysis, the scale dependence is carried by the component cross-spectra $C_\ell^{gS_i}$, while the frequency dependence is encoded in the passband-averaged responses $g_i^f$.

For each observing channel, the response coefficient is computed by averaging the corresponding spectral signal over the measured passband of that channel. We follow the same passband convention as ACT DR6 throughout. The components considered in this work are the tSZ signal, the CIB amplitude term, the CIB spectral-variation term (as developed in \cite{mccarthy_component}), and the radio component.

\paragraph{tSZ response.}
The tSZ response is determined by its standard frequency dependence in thermodynamic CMB temperature units. Since this spectral shape is fixed, the corresponding passband-averaged response coefficient is fully specified once the channel passband is given. This defines the tSZ response used in the component-separation model.

\paragraph{CIB amplitude response.}
The clustered CIB contribution is modeled through a modified blackbody form, parameterized by the fiducial dust spectral index, $\beta_0$, and temperature, $T_0$. After conversion to thermodynamic CMB temperature units and averaging over the instrumental passband, this gives the response coefficient for the CIB amplitude component. This term captures the leading CIB contribution to the galaxy--temperature cross-correlation.

\paragraph{CIB spectral-variation response.}
To allow for deviations from a single fixed CIB spectral shape, we also include a first-order variation around the fiducial CIB spectral index \citep{rethinkingcomponentChluba,mccarthy_component}. This defines an additional CIB-like response component, which we denote as the CIB-$\delta\beta$ term. Its role is to capture leading-order spectral variation in the CIB contribution when testing more flexible SED recipes.

\paragraph{Radio response.}
The radio-correlated contribution is modeled phenomenologically with an effective spectral index, $\beta_r$. After conversion to thermodynamic CMB temperature units and passband averaging, this yields the radio response coefficient used in the signal model. In this work, the fitted radio spectral index should be interpreted as an effective parameter describing the net radio-correlated contribution to the cross-spectra.


In summary, the tSZ response is fixed by its known spectral form, while the CIB and radio responses depend on the external SED parameters introduced above. For any chosen set of these parameters, the passband-averaged coefficients $g_i^f$ are fully specified, and the component amplitudes $C_\ell^{gS_i}$ can then be solved linearly using the estimator in Sec.~\ref{sec:signalmodel}.

We now specify the frequency response coefficients $g_i^f$ that enter the linear signal model in Sec.~\ref{sec:signalmodel}. These coefficients encode the spectral dependence of each component in frequency channel $f$, after averaging over the instrumental passband. Here we adopt the ACT convention for passbands, and write the $\tau(\nu)$ passband as the response to a Rayleigh-Jeans ($\propto\nu^2$) source \citep{Coultontszcomponent}. The weighted response for each nominal band $f$ is:
\begin{equation}
    g_{i}^f=\frac{\int d\nu\ G_i(\nu)\left.\frac{dB_\nu(\nu,T)}{dT}\right|_{T=T_{\rm CMB}}\nu^{-2}\tau^f(\nu)}{\int d\nu\ \left.\frac{dB_\nu(\nu,T)}{dT}\right|_{T=T_{\rm CMB}}\nu^{-2}\tau^f(\nu)}.
    \label{eq:bandaverage}
\end{equation}

For each observing channel, the response coefficient is computed by averaging the corresponding spectral signal $G_i(\nu)$ over the passband of that channel. The components considered in this work are the tSZ signal, the CIB amplitude term, the CIB spectral-variation term, and a radio component.

\subsubsection{tSZ response}
The tSZ response is determined by its standard frequency dependence in thermodynamic CMB temperature units. In non-relativistic limit\citep{szeffectsz,szbirkshaw}: 
\begin{align}
    \delta T_{\rm SZ}(\nu,\vec{n})&=G_{\rm tSZ}(\nu)y(\vec{n}),\\
    G_{\rm tSZ}(\nu)&=T_{\rm CMB}\left[x\coth\left(\frac{x}{2}\right)-4\right], \label{eq:tszresponse}
\end{align}
where $y$ is the dimensionless Compton-y parameter and $x = h\nu/k_BT_{\rm CMB}$. At CMB temperature $T_{\rm CMB}=2.7255$ K \citep{Fixsencmbtemp}, $G_{\rm tSZ}(\nu)$ transitions from negative to positive near the null frequency $\nu_t\approx227\rm GHz$. Since this spectral shape is fixed, the corresponding passband-averaged response coefficient is fully specified without introducing external \ac{SED} parameters.

\subsubsection{CIB amplitude response}

The clustered CIB contribution is modeled through a modified blackbody form, parameterized by the fiducial dust spectral index and temperature, with the same parametrization as \citet{mccarthy_component-separated_2023,Coultontszcomponent}. The flux induced by \ac{CIB} is:
\begin{equation}
    \Delta I(\nu)\propto\left(\frac{\nu}{\nu_0}\right)^\beta_0 B_\nu(\nu,T),
\end{equation}
where $\nu_0$ is an arbitrary base frequency, and we define $\nu_0=220~\rm GHz$ in this work. Converting the SED into CMB thermodynamic temperature:
\begin{align}
    \Delta T_{\rm CIB}(\nu,\vec{n})&=G_{\rm CIB}(\nu)B(\vec{n})\\
    G_{\rm CIB}(\nu)&=\left(\frac{\nu}{\nu_0}\right)^{\beta_0}B_\nu(\nu,T_{0})\left(\left.\frac{dB_\nu(\nu,T)}{dT}\right|_{T=T_{\rm CMB}}\right)^{-1},
\end{align}
where
\begin{equation}
    B_\nu(\nu,T) = \frac{2h\nu^3}{c^3}\frac{1}{e^x-1},
\end{equation}
is the blackbody spectral radiance and $x=h\nu/(k_BT)$. As noted above, $\beta_0,T_0$ are modified blackbody spectral index and temperature parameters for the CIB component and are regarded as external SED parameters. 

\subsubsection{CIB spectral-variation response}
The clustered CIB signal could have varying $\beta_0$ and $T_0$ parameters across different galaxy mass, feedback strength, and gas cooling properties \citep[e.g., see][]{Planck2013CIB,CIBmodelmccarthy,SPIRECIB}. To allow for deviations from a single fixed CIB spectral shape (which may not be appropriate for all galaxies, and whose parameters $\beta_0$ and $T_0$ are uncertain anyway), we may also include a first-order variation around the fiducial CIB spectral index \citep{mccarthy_component-separated_2023}. This defines an additional CIB-like response component, which we denote as the CIB-$\delta\beta$ term \citep{rethinkingcomponentChluba}:
\begin{equation}
    \Delta I(\nu)\propto \left(\frac{\nu}{\nu_0}\right)^{\beta_0} B_\nu(\nu,T_0)+\left(\frac{\nu}{\nu_0}\right)^{\beta_0}\ln\left(\frac{\nu}{\nu_0}\right) B_\nu(\nu,T_0)\delta\beta + {\cal O}(\delta\beta^2).
\end{equation}
This translates into another spatially varying linear component from \ac{CIB}:
\begin{align}
\Delta T_\beta(\nu, \vec{n})&=G_\beta(\nu)\delta\beta(\vec{n}),
\end{align}
where
\begin{align}
    G_\beta(\nu) &= \left(\frac{\nu}{\nu_0}\right)^{\beta_0}\ln\left(\frac{\nu}{\nu_0}\right)B_\nu(\nu,T_0)\left(\left.\frac{dB_\nu(\nu,T)}{dT}\right|_{T=T_{\rm CMB}}\right)^{-1}.
\end{align}

\subsubsection{Radio response}
The radio-correlated contribution is modeled phenomenologically with an effective spectral index. Following conventions from \citet{Planckdiffusemap2015}, we define the spectral parameter of radio component in Rayleigh-Jeans temperature and convert to CMB temperature:
\begin{align}
    \Delta T_{\rm r}&=G_{\rm r}(\nu)r(\vec{n}),\\
    G_{\rm r} &= \left(\frac{\nu}{\nu_0}\right)^{\beta_r} \frac{(e^x-1)^2}{x^2 e^x},
\end{align}
Two important radio contributions to the millimeter-wave band are free-free emission and synchrotron radiation. The free-free \ac{SED} scales with $\beta_{\rm ff}=-2.14\pm 0.01$ for clusters with electron temperature between $500\rm K-20000 K$, while synchrotron has a steeper spectral index $\beta_s=-3.11$ for Milky Way-like galaxies \citep{Planckdiffusemap2015}.

The spectral parameter $\beta_r$ in this work should be regarded as an effective spectral parameter as averaged over signals with different \ac{SED}, and describes the net radio-correlated contribution to the cross-spectra. 

\begin{figure}[h!]
    \centering
    \includegraphics[width=0.70\linewidth]{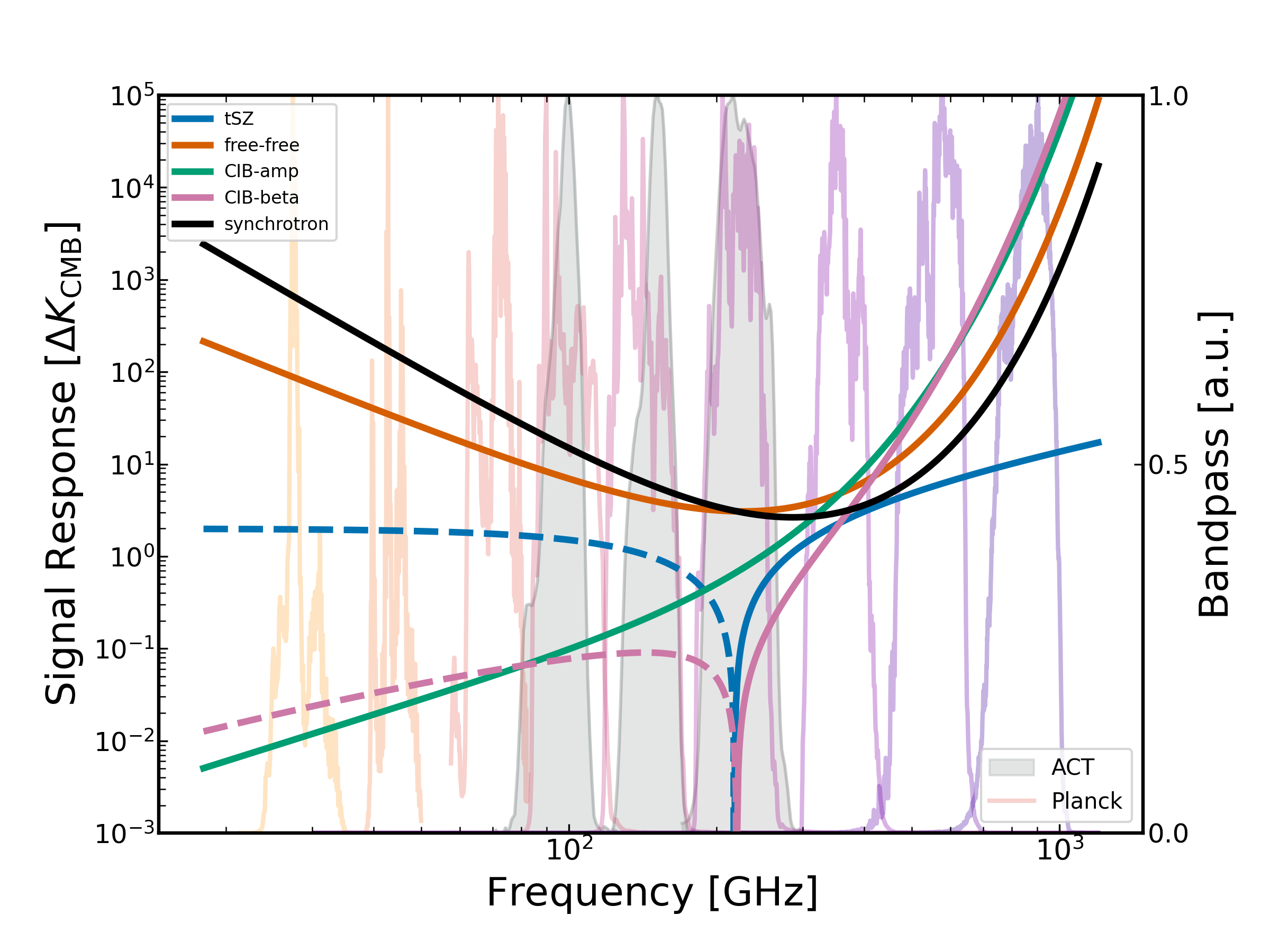}
    \caption{Signal responses of the four components considered in this work. Responses that are negative are plotted with dashed curves. For the radio sources, the two scenarios $\beta_{r}=\beta_{\rm ff}$ (free-free) and $\beta_r=\beta_{\rm syn}$ (synchrotron) are both shown. The \ac{SED} parameters for the CIB-amplitude and CIB-$\delta \beta$ response is taken as $\beta_0=1.7,T_0=10.7$ K, consistent with the fiducial values by \citet{Coultontszcomponent}. The responses are in the unit of CMB temperature $\Delta K_{\rm CMB}$. The passbands of \textit{Planck} and ACT instruments are also plotted for reference.}
    \label{fig:passbandsresponses}
\end{figure}

In summary, the tSZ response is fixed by its known spectral form, while the CIB and radio responses depend on the external SED parameters $\beta_0,T_0,\beta_r$ introduced above. For any chosen set of these parameters, the passband-averaged coefficients $g_i^f$ are fully specified, and the component amplitudes $C_\ell^{gS_i}$ can then be solved linearly using the estimator in Sec.~\ref{sec:signalmodel}. Figure \ref{fig:passbandsresponses} shows the response of \ac{tSZ}, CIB-amplitude, CIB-$\delta\beta$, and radio in the frequency range of $\rm 10~GHz - 2000 ~GHz$, where positive responses are in solid lines and negative responses are in dashed lines. The relevant ACT and \textit{Planck} passbands are also shown for reference.

\subsection{Recipe selection strategy}

ACT provides only three frequency bands, so ACT-only component separation is limited to at most three components. We therefore use the broader frequency coverage of \textit{Planck} to determine which three-component SED recipe is preferred by the data, and then apply that recipe to the ACT+\textit{Planck} cross-spectra to recover the small-scale $g\times y$ signal.

For our purposes we focus the discussion of the following recipes:
\begin{enumerate}
    \item tSZ ($C_\ell^{gy}$) + Radio ($C_\ell^{gr}$) + CIB-amplitude ($C_\ell^{gB}$) + CIB-$\delta\beta$ ($C_\ell^{g\beta}$), (which can only be studied using \textit{Planck} since there are 4 components).
    \item tSZ ($C_\ell^{gy}$) + Radio ($C_\ell^{gr}$) + CIB-amplitude ($C_\ell^{gB}$),
    \item tSZ ($C_\ell^{gy}$) + CIB-amplitude ($C_\ell^{gB}$) + CIB-$\delta \beta$ ($C_\ell^{g\beta}$).
\end{enumerate}

For each signal recipe, we first marginalize the residual $\chi^2$ of the signal model in equation \ref{eq:chisq_clean} on the $C_\ell^{gS_i}$ degrees of freedom using the solution \ref{eq:estimatorsol}, this yields
\begin{equation}
\tilde{\chi}^2=\min_{C_\ell^{gS_i}}\chi^2(\beta_0,T_0,\beta_r,C_\ell^{gS_i})=\tilde{\chi}^2(\beta_0,T_0,\beta_r)\label{eq:marginalized_chisq}
\end{equation}
We then marginalize $\chi^2$ on the $(\beta_0,T_0,\beta_r)$ parameter set for a best-fit $\chi^2_{\rm bf}=\tilde{\chi}^2(\beta_0^{\rm bf},T_0^{\rm bf},\beta_r^{\rm bf})$. We determine the evidence of the recipe using \ac{BIC} defined as:
\begin{align}
    \mathrm{BIC}&=\chi^2_{\rm bf}+(\mathbf{\#\ of\ parameters})\ln(\mathbf{\#\ of \ data})\nonumber\\
    &=\chi^2_{\rm bf}+(N_sN_\ell+N_{\rm ext})\ln(N_\ell N_f),
\end{align}
where $N_{\rm ext}$ denotes the number of external parameters (2 for $\beta_0,T_0$, 3 for $\beta_0,T_0,\beta_r$). The \ac{BIC} penalizes models with more foreground components and more external parameters; a larger BIC for the best-fitting model indicates either that the overall fit is poor or that the model is overfitting.

\section{Datasets}\label{sec:dataset}
\subsection{The ACT DR6 CMB temperature}\label{sec:ACT_temperature}
The ACT DR6 data release provides CMB temperature maps in three nominal frequency bands centered at 90, 150, and 220\,GHz. In this work, we use the ACT DR6 maps coadded with \textit{Planck} and with source subtraction applied, in which all point sources detected at $5\sigma$ in the full ACT coadd are removed, described in detail in Sec.~4 of \citet{ACTDR6maps_description}. In these coadds, the ACT DR6 and \textit{Planck} data are convolved to a common ACT beam. The corresponding beam specifications are listed in Tab.~\ref{tab:actbandsbeams}, and the harmonic beam profiles are shown in Fig.~\ref{fig:allbeams}.

A detailed characterization of the instrumental passbands is required for the component-separation analysis. Because the 90 and 150\,GHz maps are coadded from multiple ACT polarization arrays (PAs), we adopt the PA5 passbands for the 90 and 150\,GHz channels, and the PA4 passband for the 220\,GHz channel. The impact of this choice on the band-averaged responses is discussed in Appendix~\ref{sec:ACTpassbandcomp}.

\begin{figure}[h!]
  \centering

  \includegraphics[
    width=0.32\textwidth,
    trim=1.5cm 0.8cm 1.5cm 1cm,
    clip
  ]{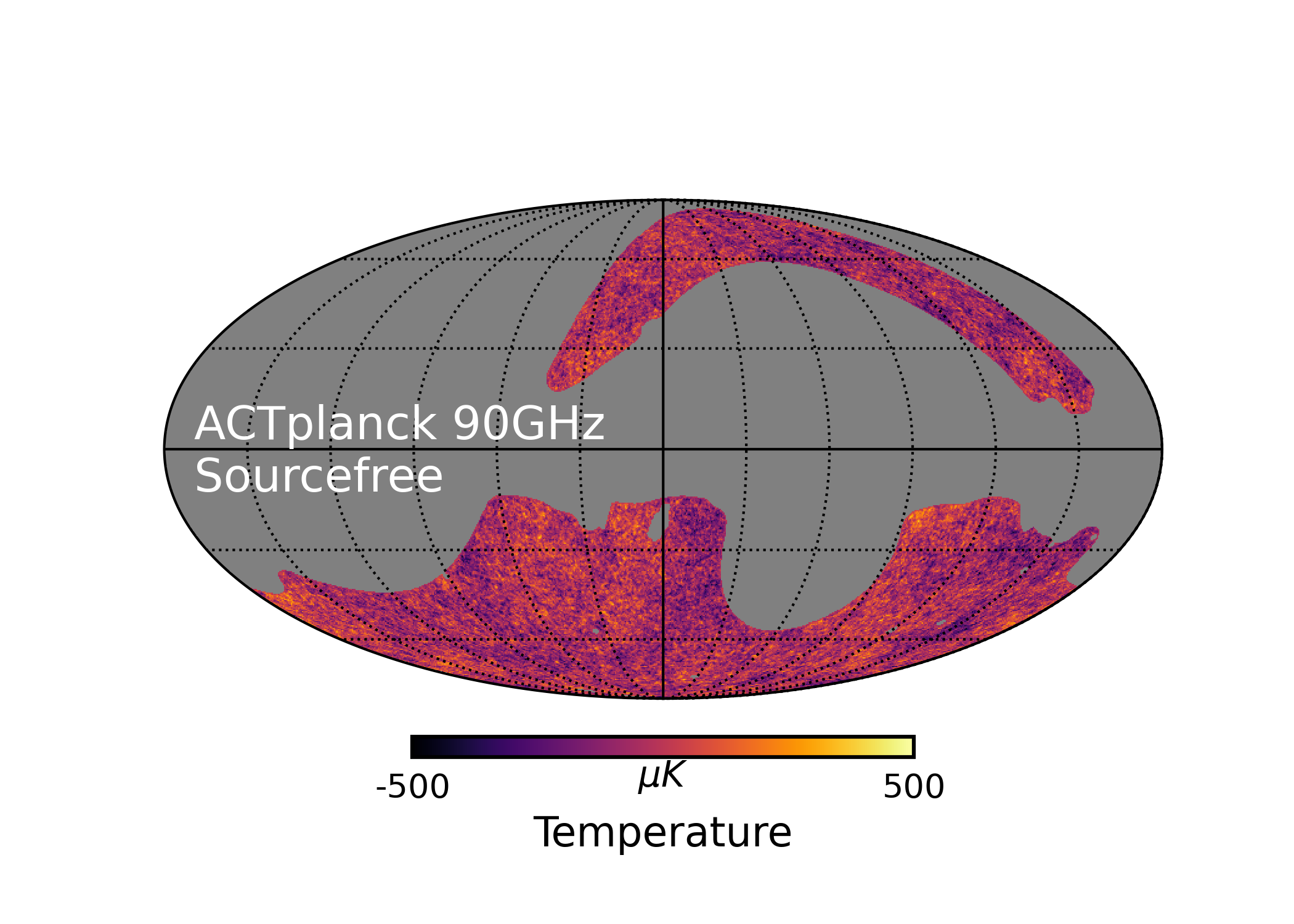}%
  \hfill
  \includegraphics[
    width=0.32\textwidth,
    trim=1.5cm 0.8cm 1.5cm 1cm,
    clip
  ]{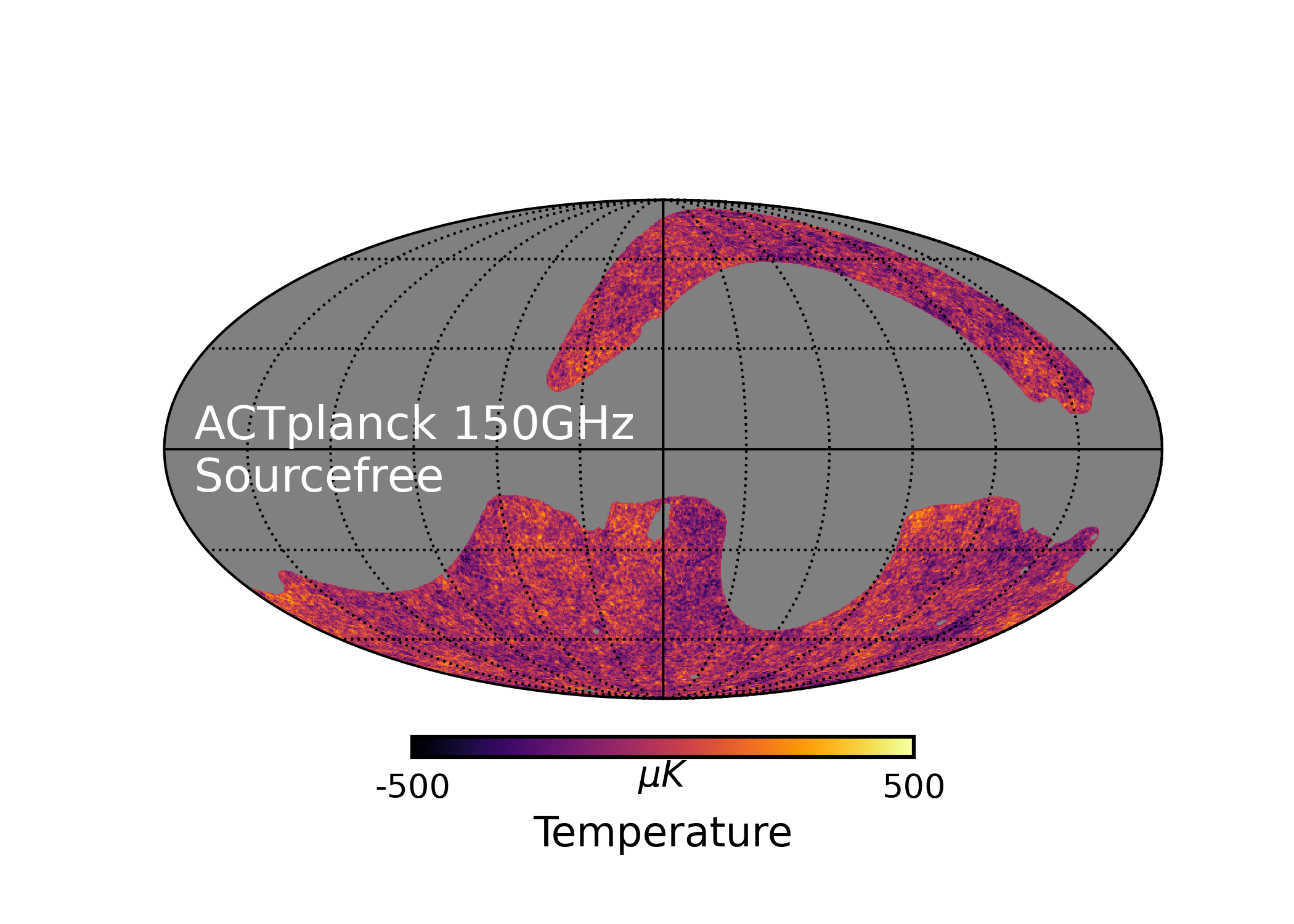}%
  \hfill
  \includegraphics[
    width=0.32\textwidth,
    trim=1.5cm 0.8cm 1.5cm 1cm,
    clip
  ]{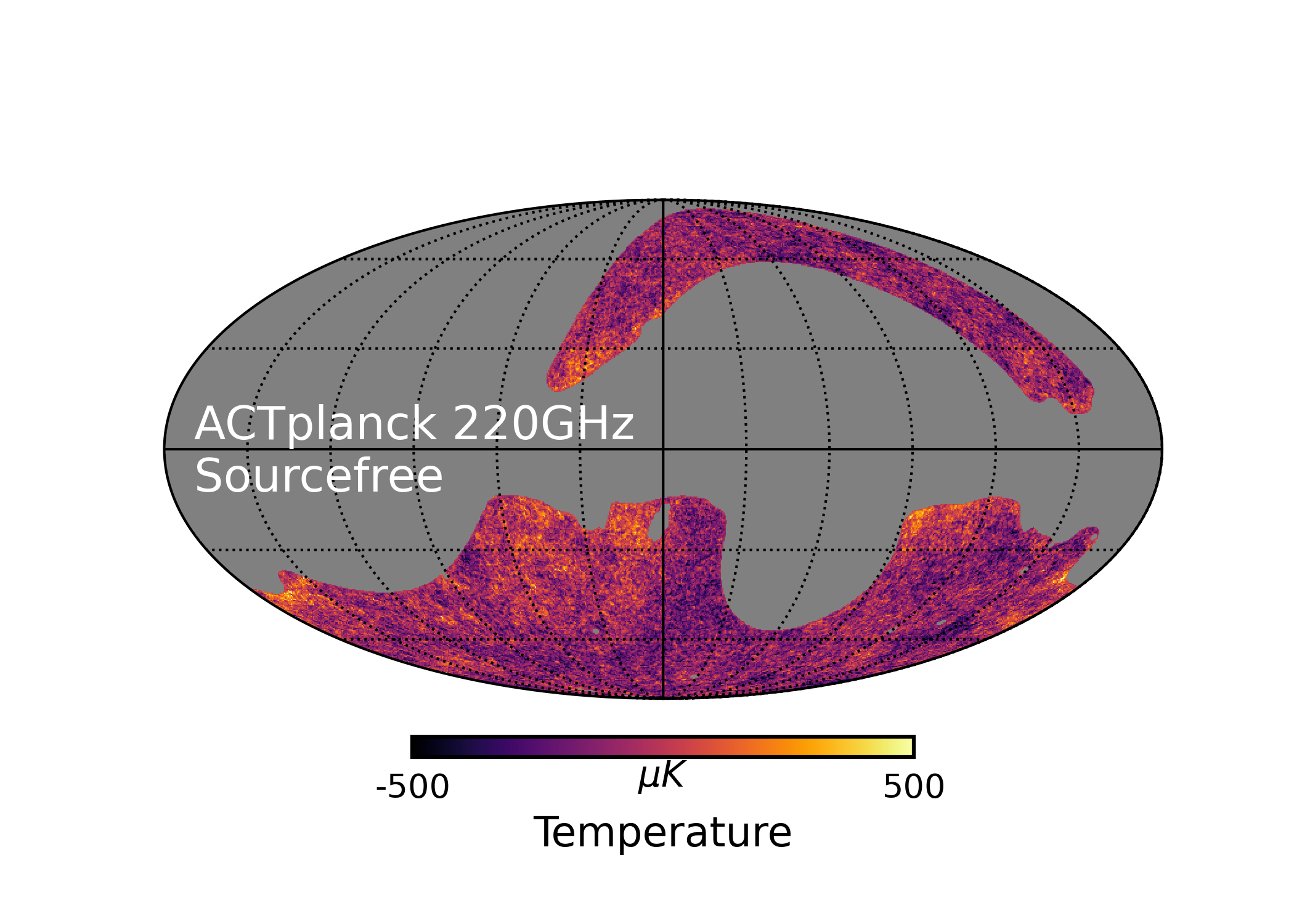}%

  \caption{
    ACT+\textit{Planck} source-subtracted coadded maps from ACT DR6,
    described in \protect\citep{ACTDR6maps_description}, and used in
    this work. The maps are shown in Galactic coordinates after
    applying the ACT footprint mask.
  }
  \label{fig:ACTsinglefreqcoadds}
\end{figure}

We apply the ACT footprint mask, described in \citet{ACTDR6lensingpwr, Coultontszcomponent}\footnote{The masks are available at \href{https://portal.nersc.gov/project/act/dr6_nilc/ymaps_20230220/masks/}{\texttt{https://portal.nersc.gov/project/act/dr6\_nilc/ymaps\_20230220/masks/}}. In our analysis we use the ACT footprint mask.}. The mask is apodized with a scale of $1.5^\prime$ and then reprojected from Plate Carr\'ee to \texttt{healpix} using spline interpolation in the \texttt{pixell} package. The resulting coadded maps in the three ACT frequency bands are shown in Fig.~\ref{fig:ACTsinglefreqcoadds}.

\begin{table}[h!]
    \centering
    \begin{tabular}{ccc}
         \hline
         Nominal Band & Beam file   & passbands used \\
         \hline
         90GHz        & coadd\_pa5\_f090\_night\_beam\_tform\_jitter\_cmb.txt & pa6, pa5\\
         150GHz       & coadd\_pa5\_f150\_night\_beam\_tform\_jitter\_cmb.txt & pa4, pa5, pa6     \\
         220GHz       & coadd\_pa4\_f220\_night\_beam\_tform\_jitter\_cmb.txt & pa4   \\
         \hline
    
    \end{tabular}
    \caption{Nominal bands of \ac{ACT} DR6 and the associated beam files on \href{https://lambda.gsfc.nasa.gov/product/act/act_dr6.02/act_dr6.02_maps_info.html}{https://lambda.gsfc.nasa.gov/product/act/act\_dr6.02/act\_dr6.02\_maps\_info.html}} used in this work.
    \label{tab:actbandsbeams}
\end{table}

\subsection{The \textit{Planck} temperature map}\label{sec:planckmaps}

\begin{figure}[h!]
  \centering

  \includegraphics[
    width=0.32\textwidth,
    trim=1.5cm 0.8cm 1.5cm 1cm,
    clip
  ]{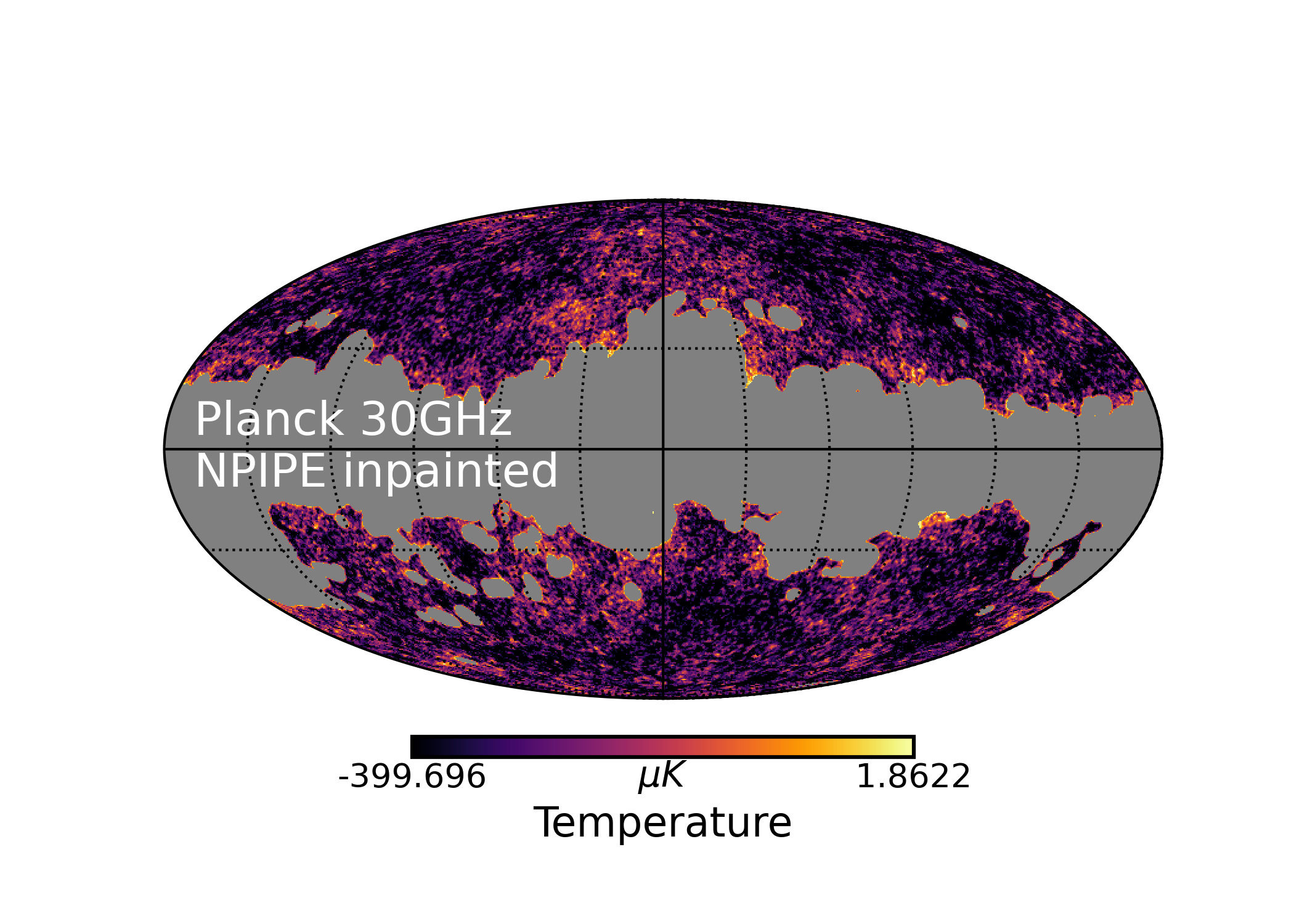}%
  \hfill
  \includegraphics[
    width=0.32\textwidth,
    trim=1.5cm 0.8cm 1.5cm 1cm,
    clip
  ]{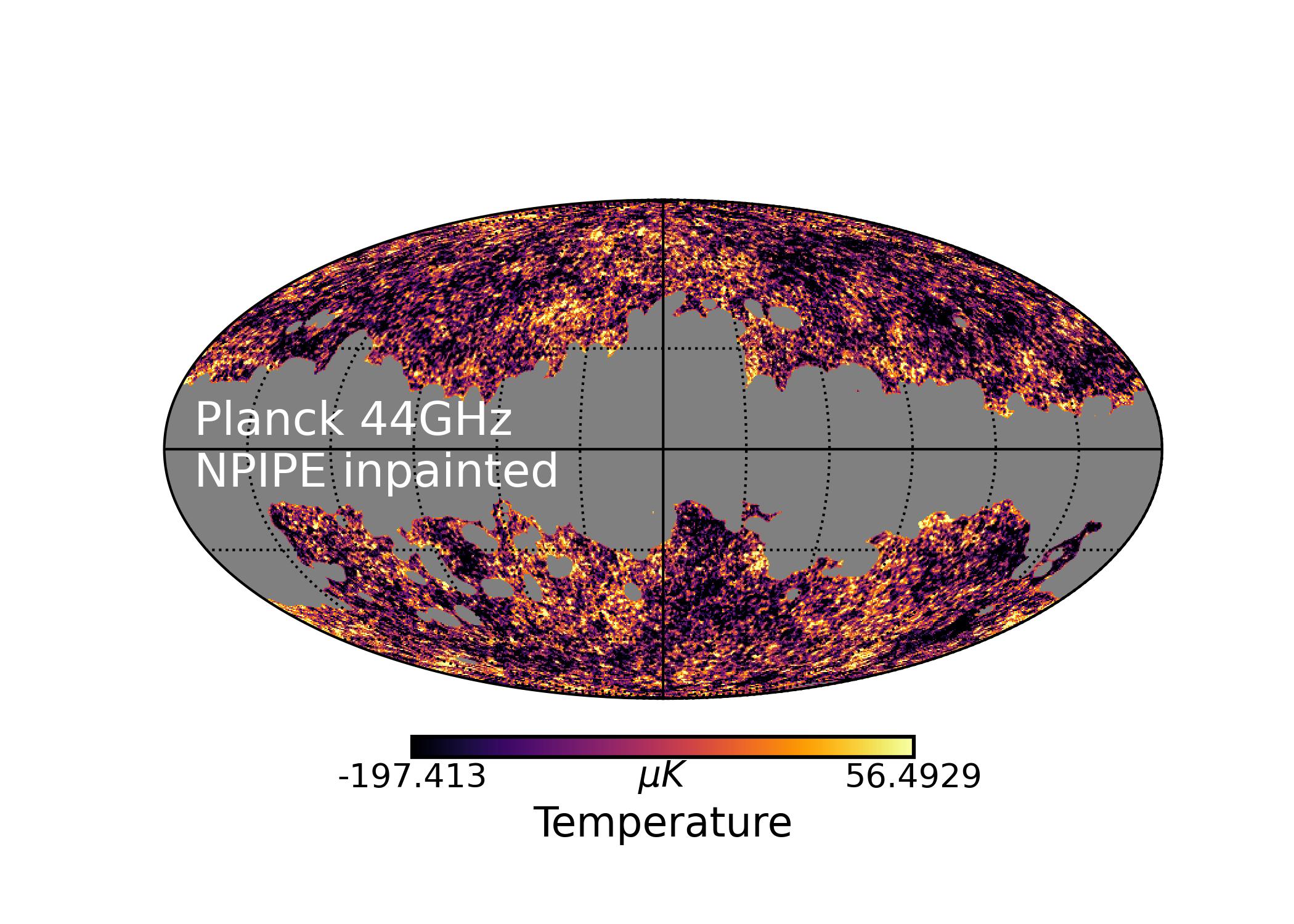}%
  \hfill
  \includegraphics[
    width=0.32\textwidth,
    trim=1.5cm 0.8cm 1.5cm 1cm,
    clip
  ]{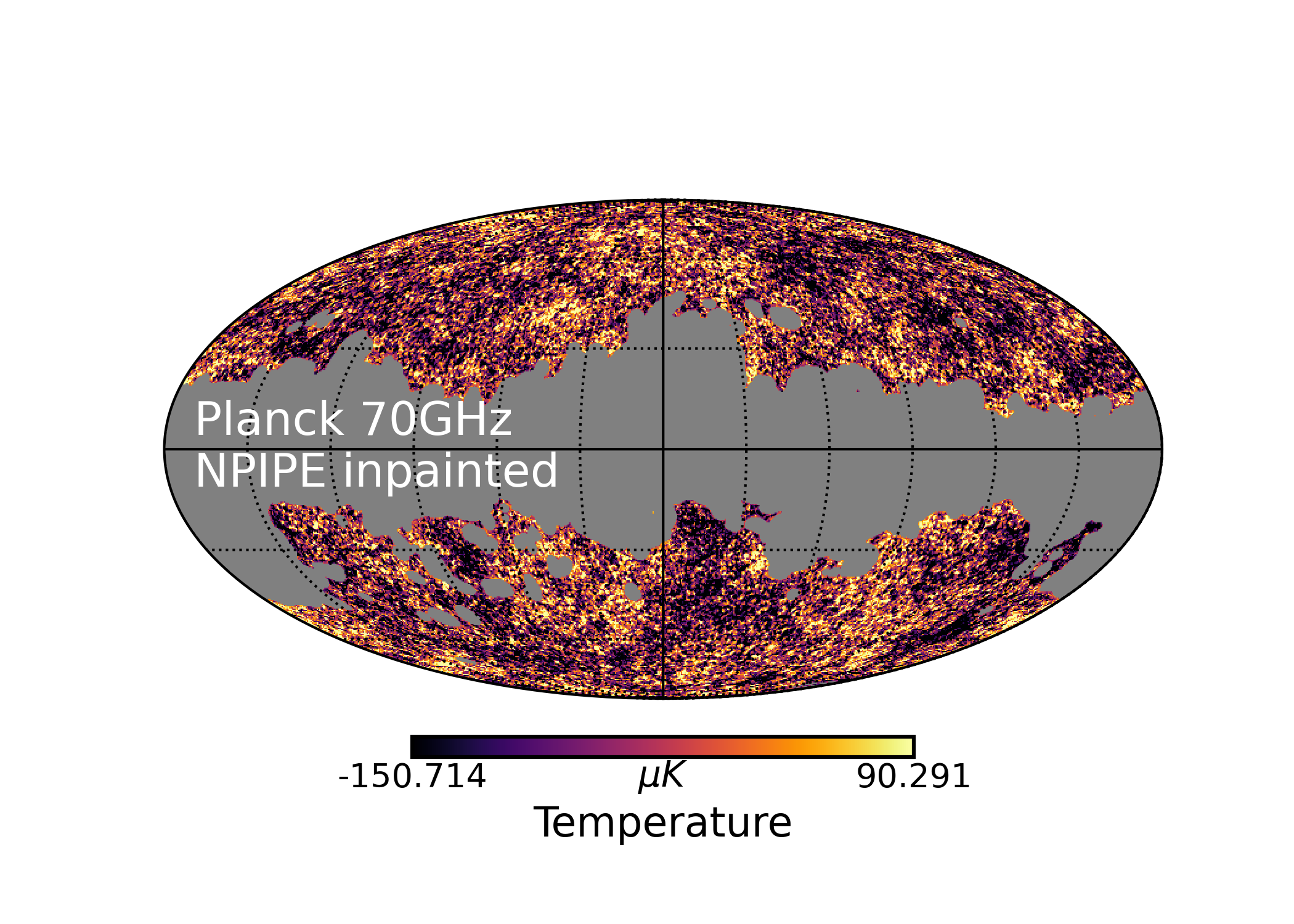}%

  \par\medskip

  \includegraphics[
    width=0.32\textwidth,
    trim=1.5cm 0.8cm 1.5cm 1cm,
    clip
  ]{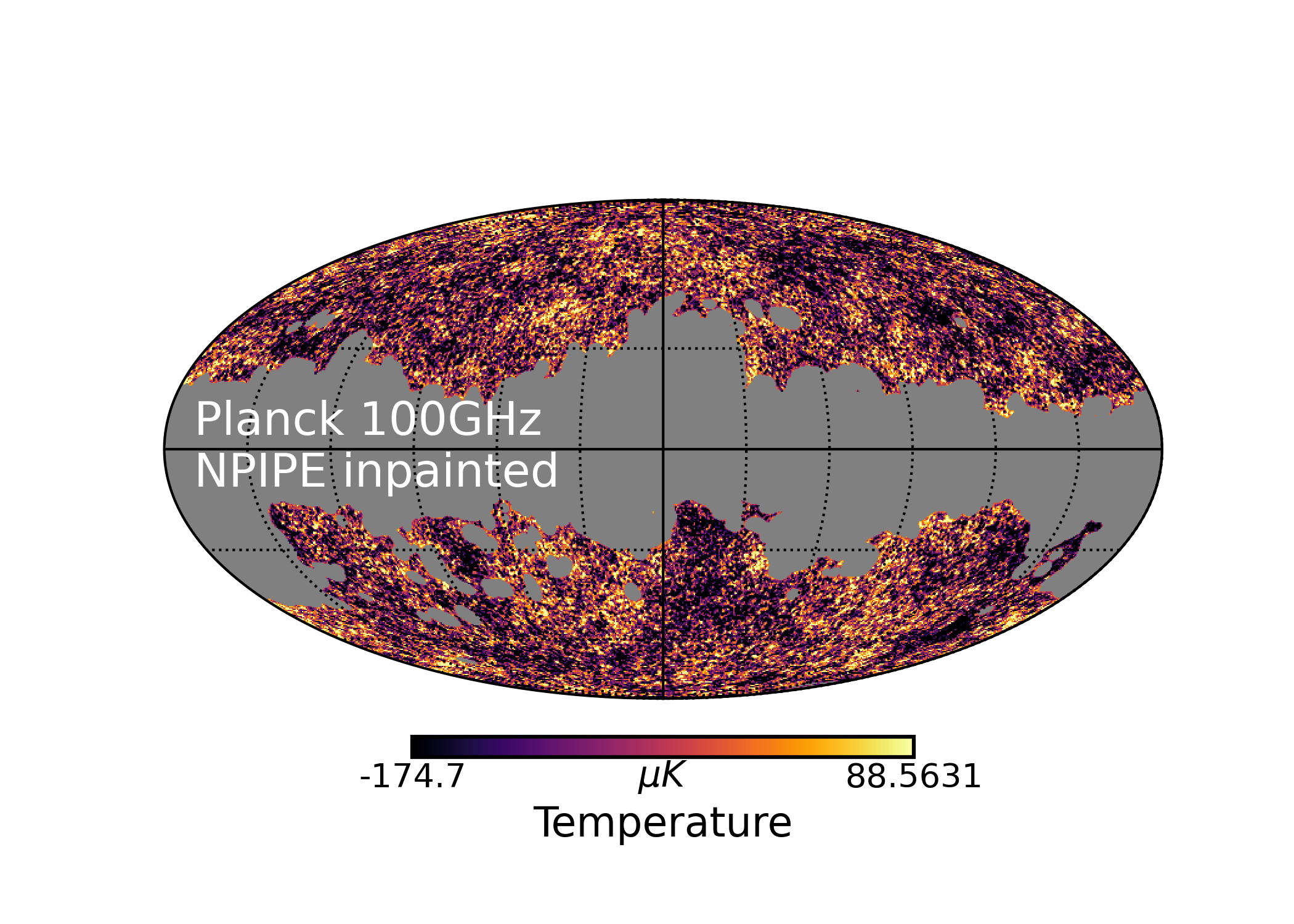}%
  \hfill
  \includegraphics[
    width=0.32\textwidth,
    trim=1.5cm 0.8cm 1.5cm 1cm,
    clip
  ]{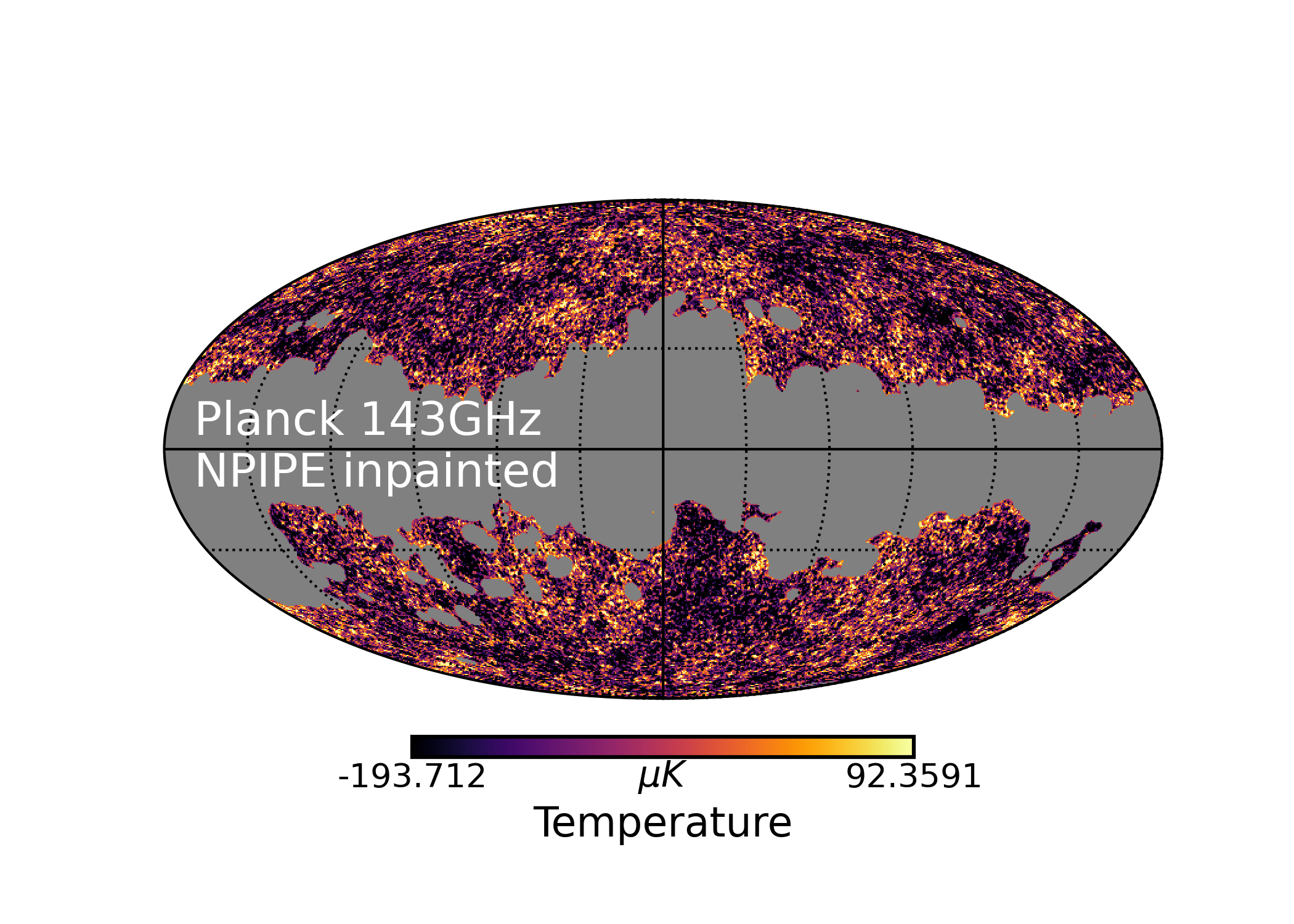}%
  \hfill
  \includegraphics[
    width=0.32\textwidth,
    trim=1.5cm 0.8cm 1.5cm 1cm,
    clip
  ]{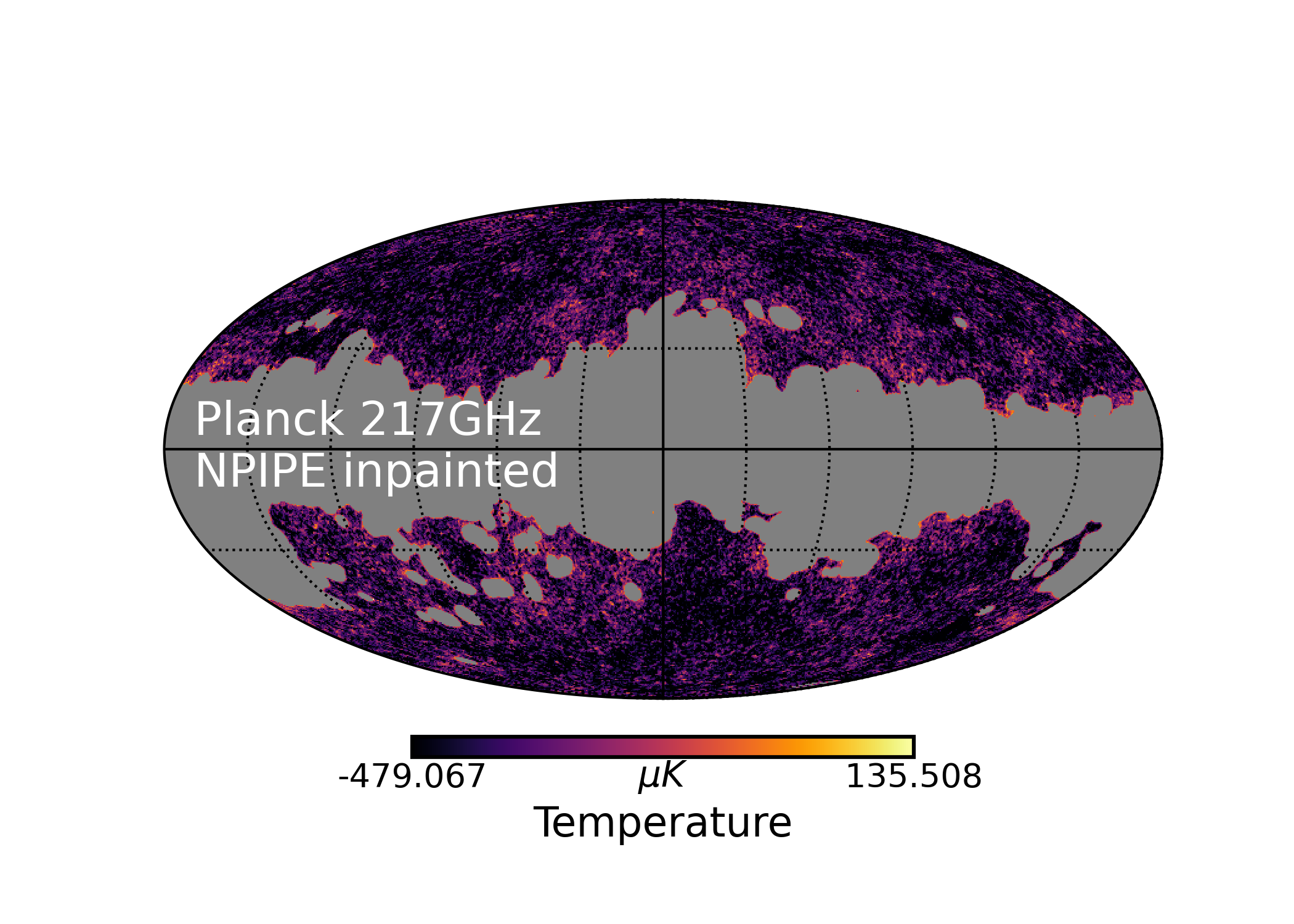}%

  \par\medskip

  \includegraphics[
    width=0.32\textwidth,
    trim=1.5cm 0.8cm 1.5cm 1cm,
    clip
  ]{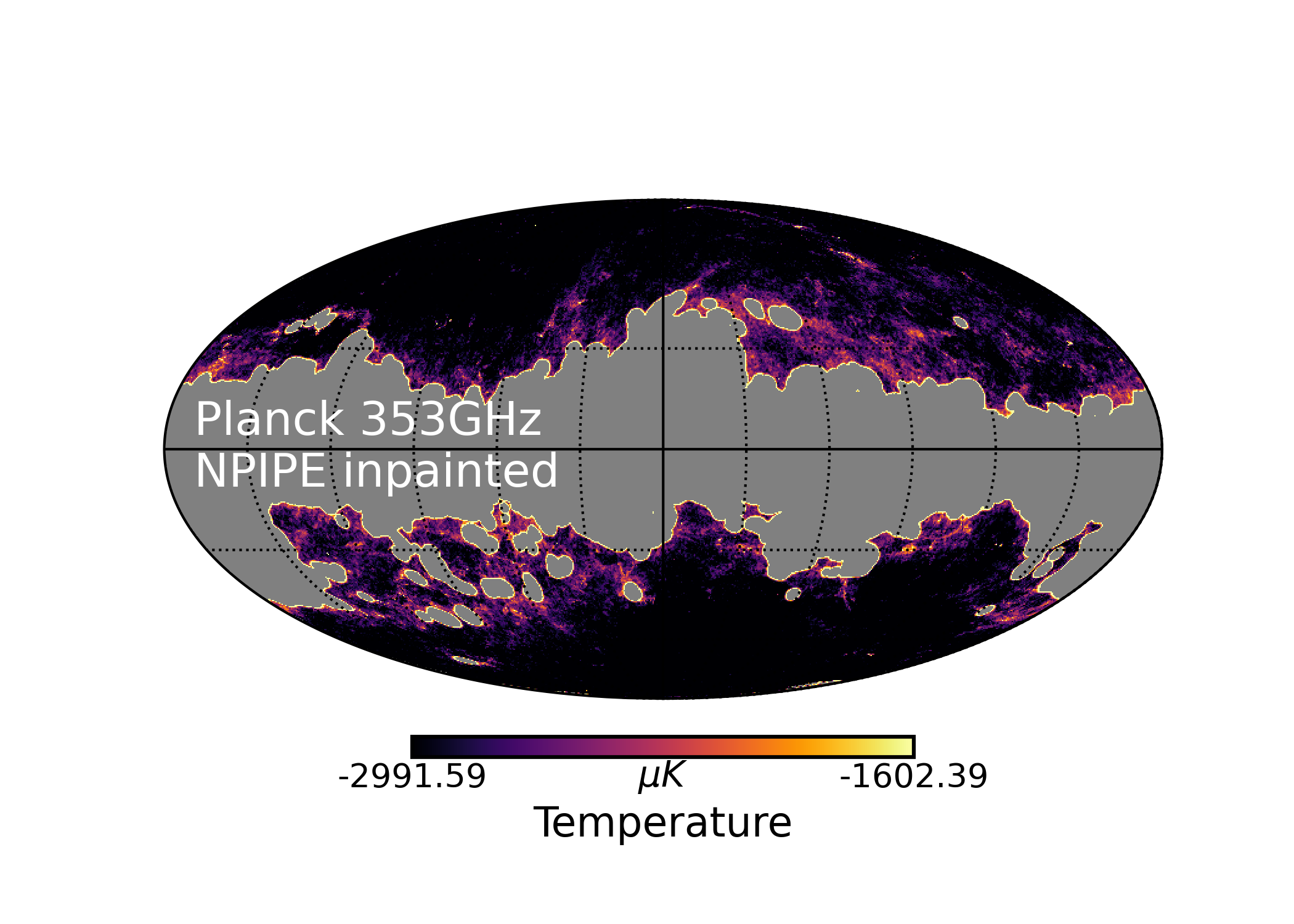}%
  \hfill
  \includegraphics[
    width=0.32\textwidth,
    trim=1.5cm 0.8cm 1.5cm 1cm,
    clip
  ]{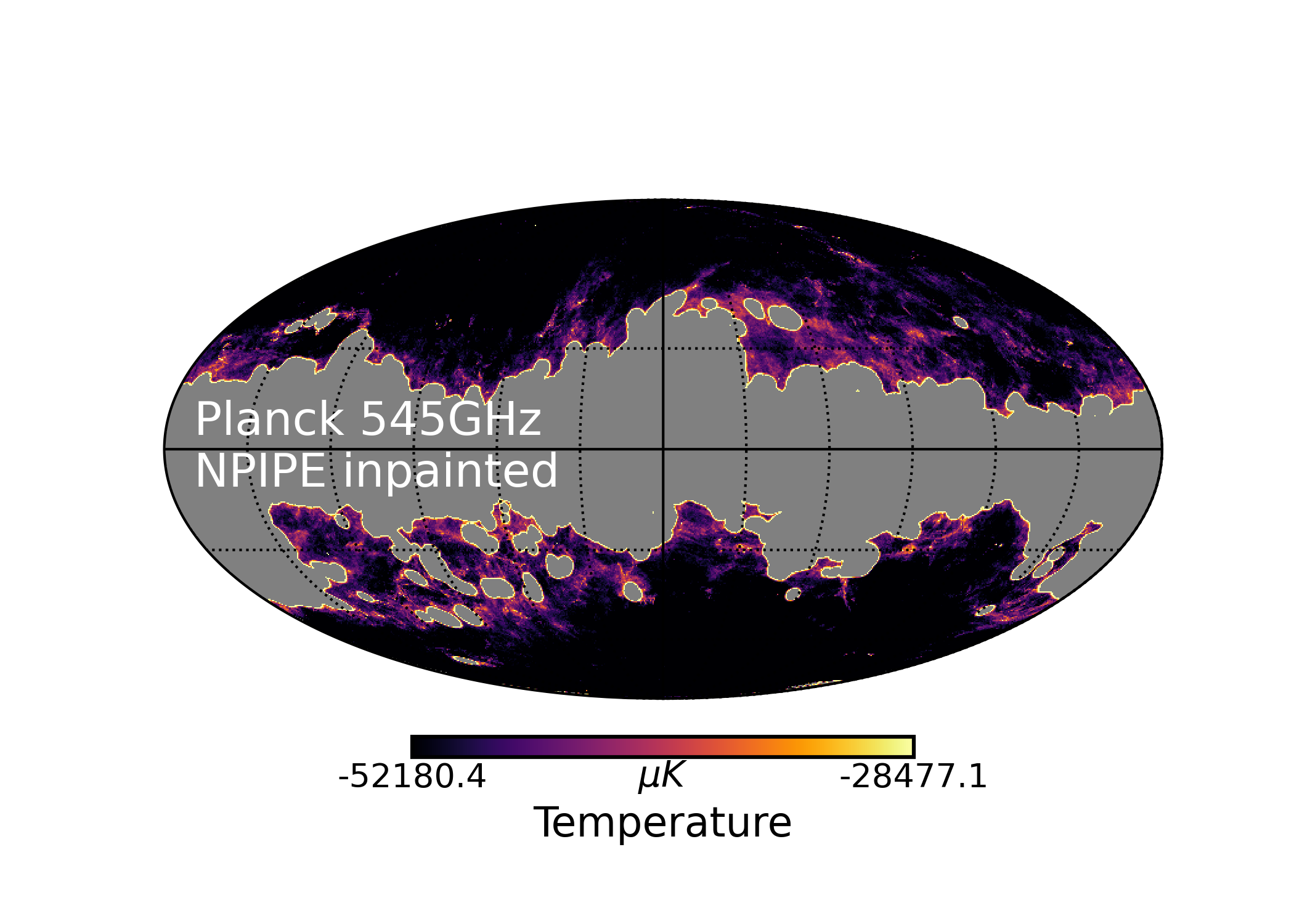}%
  \hfill
  \includegraphics[
    width=0.32\textwidth,
    trim=1.5cm 0.8cm 1.5cm 1cm,
    clip
  ]{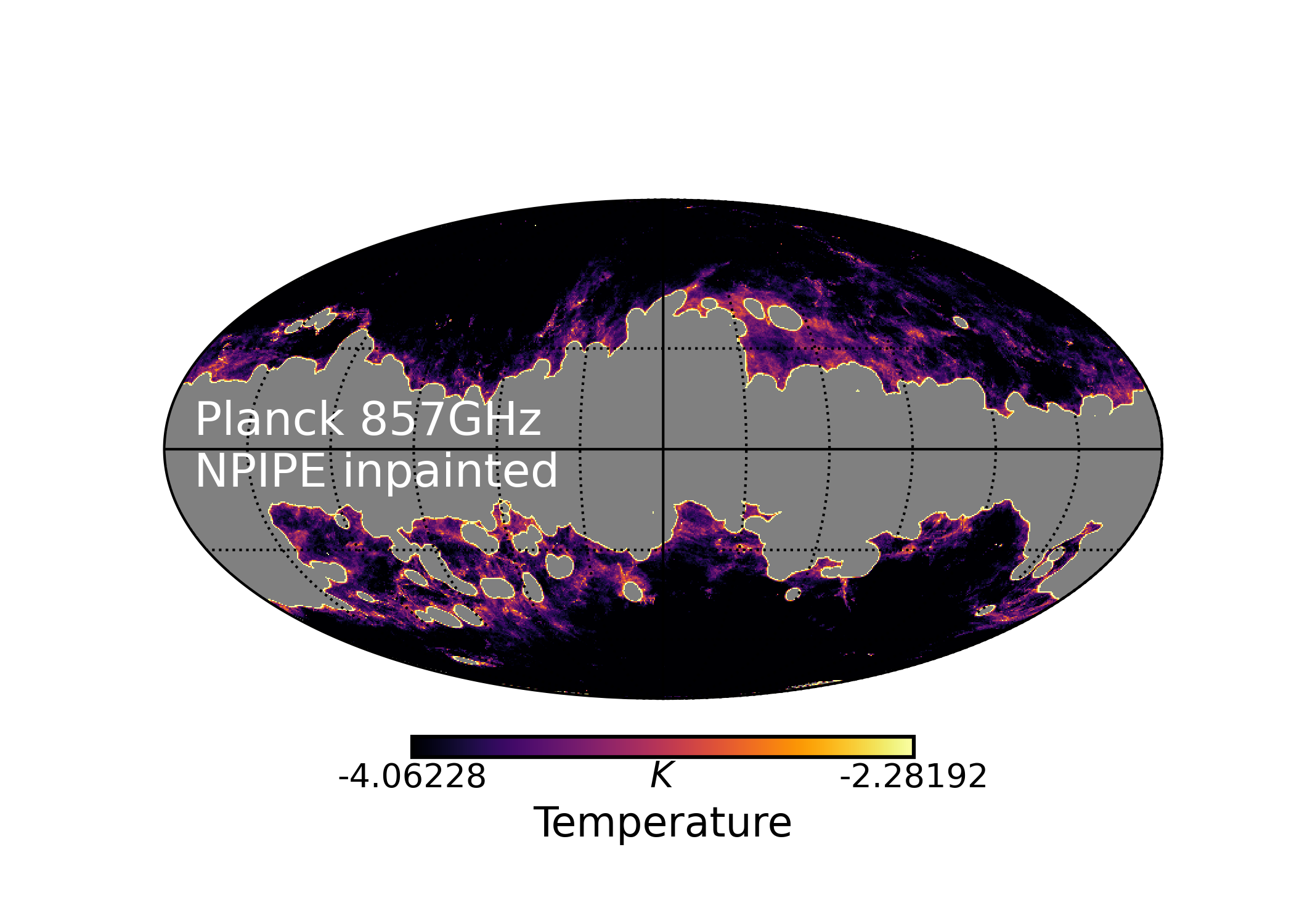}%

  \caption{
    \textit{Planck} PR4 NPIPE inpainted maps with the dipole and
    monopole subtracted, as described in detail by
    \protect\citet{mccarthy_component-separated_2023}, and used in
    this work. A footprint mask is applied to the maps shown here.
  }
  \label{fig:PlanckNPIPEinpainted}
\end{figure}

While ACT DR6 provides high angular resolution, it includes only three nominal frequency bands. To supplement the frequency information used for SED modeling, we also use the \textit{Planck} PR4 temperature maps. Specifically, we adopt the single-frequency CMB maps, together with the associated footprint and inpainted masks, provided by \citet{mccarthy_component-separated_2023}. These products are based on the NPIPE single-frequency maps \citep{npipe}, with point sources and the kinematic dipole removed, and were originally constructed for internal linear combination analyses. In our analysis, we additionally mask the inpainted regions and apodize the resulting mask with a scale of $1.5^\circ$. The maps in the nine \textit{Planck} frequency bands are shown in Fig.~\ref{fig:PlanckNPIPEinpainted}.

\begin{figure}
    \centering
    \includegraphics[width=0.70\linewidth]{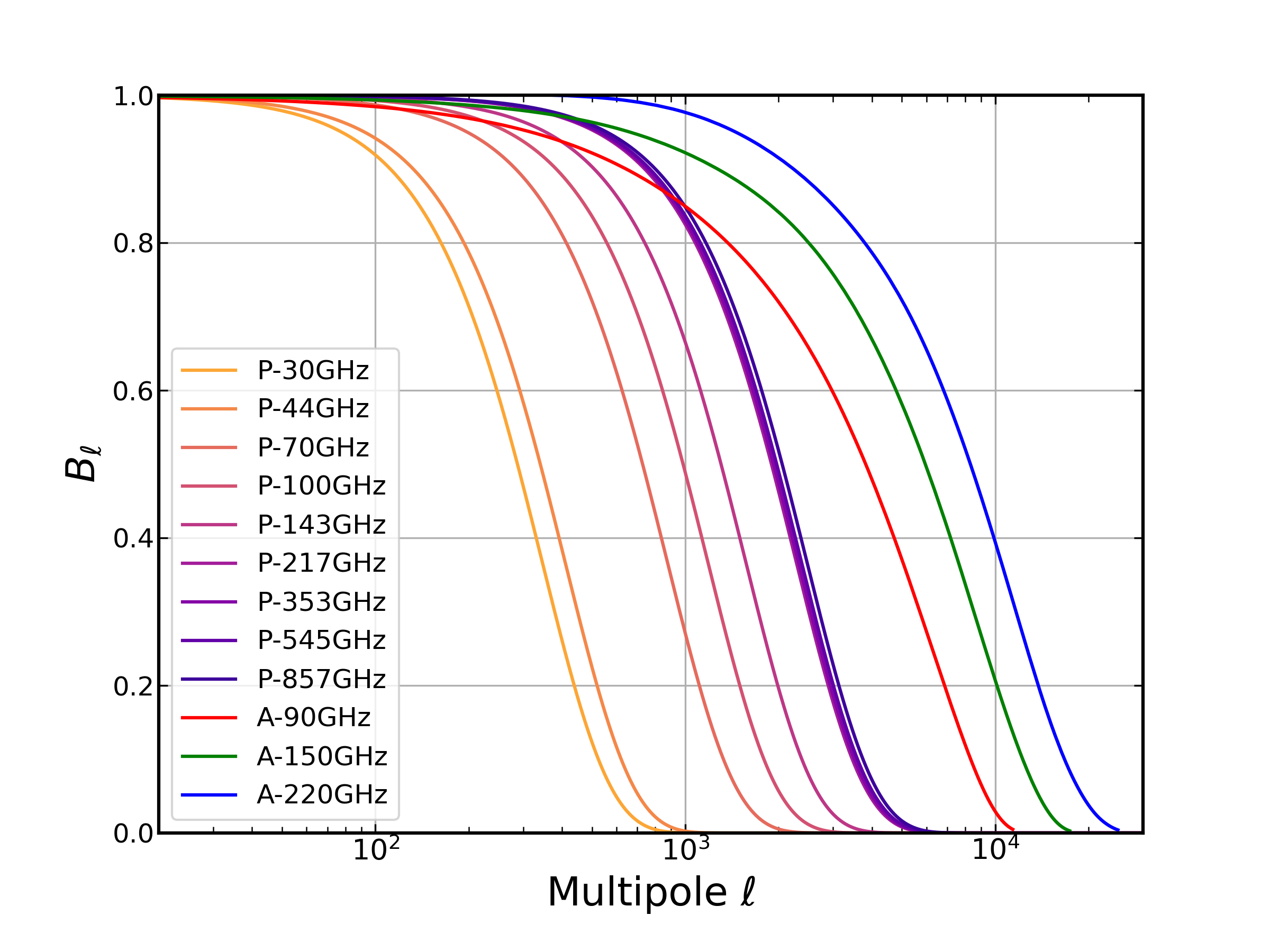}
    \caption{The beams used in this work for \textit{Planck} and ACT frequency maps. The \textit{Planck} beams are Gaussian with FWHM given in Sec. \ref{sec:planckmaps}. The curves P-Frequency denote the \textit{Planck} beams for nominal bands, and curves A-Frequency denote the ACT nominal bands.}
    \label{fig:allbeams}
\end{figure}

We model the \textit{Planck} beams as Gaussian, following \citet{mccarthy_component-separated_2023}. The beam FWHMs and \texttt{healpix} resolutions for the nominal bands are listed in Tab.~\ref{tab:plk_beam_nside_spec}, and the corresponding harmonic beam profiles are shown in Fig.~\ref{fig:allbeams}. The single-frequency maps are provided in \texttt{healpix} format, with $N_{\rm side}=1024$ for the LFI channels and $N_{\rm side}=2048$ for the HFI channels. The corresponding pixel window functions are computed separately for LFI and HFI using \href{https://github.com/healpy/healpy}{\texttt{healpy}} \citep{healpy_paper1,healpy_paper2}.



\begin{table}[ht]
\centering
\begin{tabular}{llcc}
\hline
Nominal band & Instrument & FWHM($'$) & healpix $N_{\rm side}$ \\
\hline
30GHz&LFI&$32.29$&$1024$\\
44GHz&LFI&$27.94$&$1024$\\
70GHz&LFI&$13.08$&$1024$\\
\hline
100GHz&HFI&$9.66$&$2048$\\
143GHz&HFI&$7.22$&$2048$\\
217GHz&HFI&$4.90$&$2048$\\
353GHz&HFI&$4.92$&$2048$\\
545GHz&HFI&$4.67$&$2048$\\
857GHz&HFI&$4.22$&$2048$\\
\hline
\end{tabular}
\caption{\textit{Planck} nominal frequency bands, beam FWHMs, and \texttt{healpix} resolutions used in this work..}
\label{tab:plk_beam_nside_spec}
\end{table}

\subsection{The unWISE catalog and the Low-z and Mid-z samples}

The unWISE catalog \citep{unWISE_catalog} is derived from unblurred coadds of five years of \textit{WISE} imaging \citep{unWISE_survey} of the entire sky in the $W1$ (3.3 $\mu$m) and $W2$ (4.8 $\mu$m) bands. The catalog
contains the positions and fluxes of $\sim2\times 10^9$ infrared objects.
unWISE is an ideal dataset for defining
large galaxy samples across a broad redshift range ($0 < z < 2$) and the entire sky, and has been extensively used for cross-correlation studies. In order to specifically study the tSZ effect around lower-mass halos, we use a brighter sample of galaxies, with different color cuts in $W1$ and $W2$, from the Blue, Green and Red samples of \cite{unWISE_catalog,krolewski_unwise_2020}.

\begin{figure}[ht]
    \centering

    \subfloat[
        Definition of the unWISE Low-z and unWISE Mid-z samples in the
        WISE bands $W1$ and $W2$; each sample lies above and to the left
        of the corresponding black lines. The heatmap shows the mean
        photometric redshift from COSMOS \protect\cite{Laigle2015} as a
        function of $W1$ and $W2$ colors. The higher-redshift Blue,
        Green, and Red samples defined in
        \protect\cite{unWISE_catalog} are also shown.
    ]{%
        \label{fig:unWISEsample0}%
        \begin{minipage}[t]{0.49\textwidth}
            \centering
            \includegraphics[
                width=0.95\linewidth
            ]{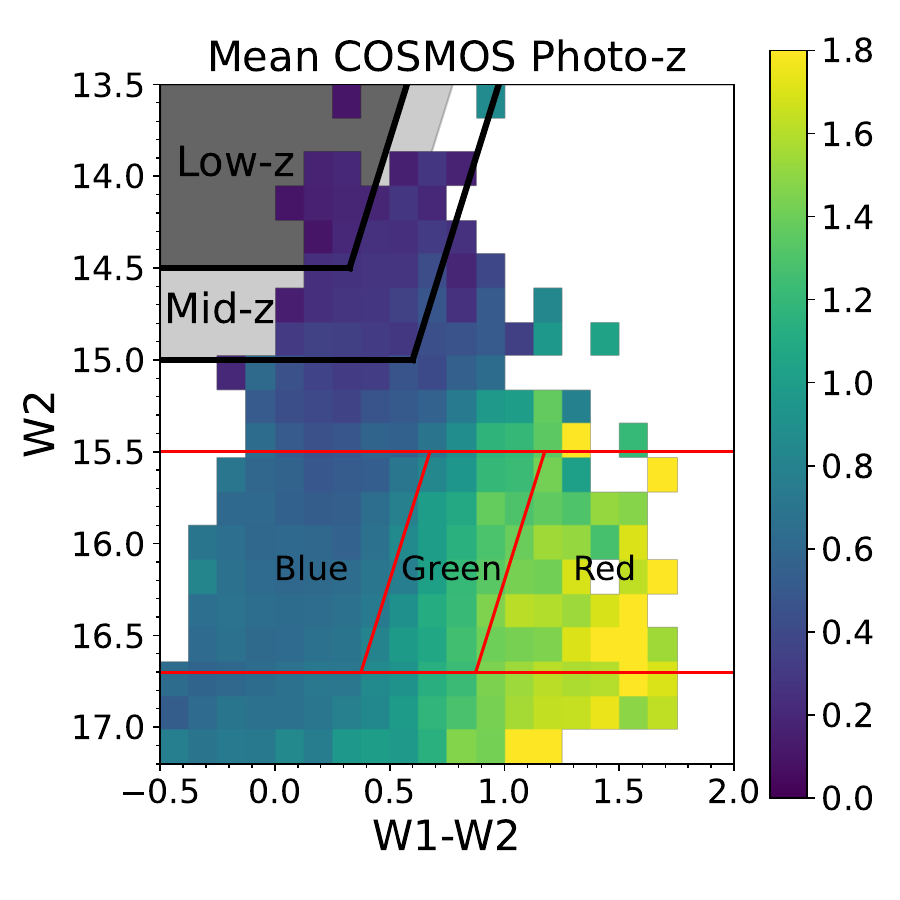}
        \end{minipage}%
    }
    \hfill
    \subfloat[
        The redshift distribution of the unWISE Low-z and Mid-z samples
        from cross-matches to DESI and SDSS.
    ]{%
        \label{fig:unWISEsample1}%
        \begin{minipage}[t]{0.49\textwidth}
            \centering
            \includegraphics[
                width=0.95\linewidth
            ]{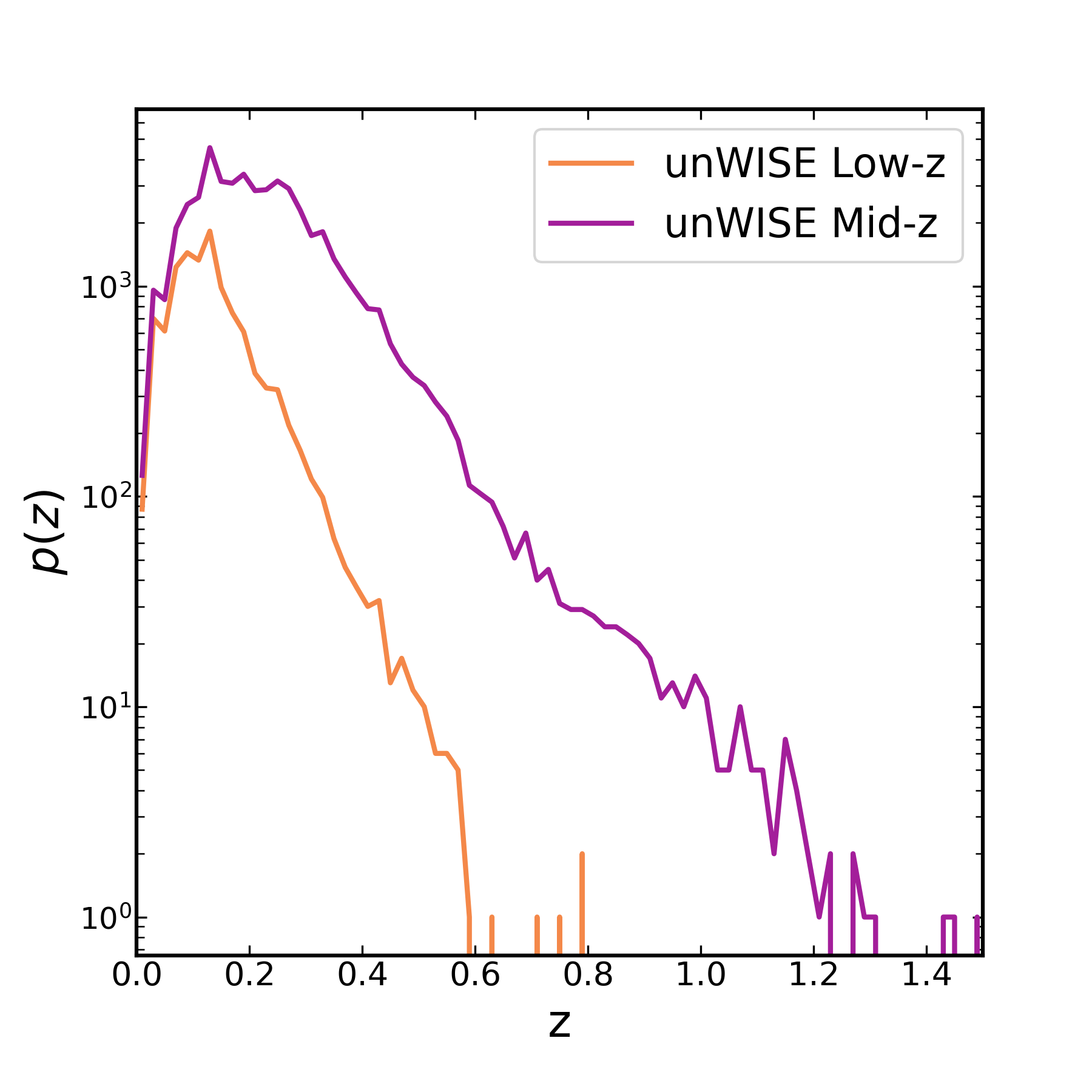}
        \end{minipage}%
    }
\end{figure}

In Fig.~\ref{fig:unWISEsample0}, we compare the unWISE Low-z and unWISE Mid-z color cuts to the Blue, Green, and Red samples. In this figure, redshifts are measured by cross-matching to the COSMOS multi-band photometric redshifts \citep{Laigle2015}, which are highly complete across the entire relevant $W1-W2$ color space, and are more accurate than typical photometric redshifts due to the large number ($\sim$30) of bands used. The Low-z color cuts were chosen by varying a constant $W2$ faint cut and a $W1-W2$ cut drawn approximately along contours of constant redshift (parallel to the $W1-W2$ cut for Blue, Green and Red), and picking samples that had the lowest possible mean redshift and mean stellar mass (using the best-fit stellar mass from the COSMOS photometry) while retaining a reasonable number density, $> 50$ deg$^{-2}$. Mid-z was chosen in a similar way, but seeking a higher number density, $\sim$200 deg$^{-2}$.
The resulting mean stellar masses, from the multi-band fits to COSMOS photometry, are $\log{M_\star/M_\odot} = 10.77$ for Low-z and $\log{M_\star/M_\odot} = 10.90$ for Mid-z.

We define Low-z by the following color cuts:
\begin{align}
   W1-W2 &< (17-W2)/4 - 0.3,  \\ \nonumber 
    W2 &< 14.5.
\end{align}
and likewise, we define Mid-z as:
\begin{align}
   W1-W2 &< (17-W2)/4 + 0.8,  \\ \nonumber 
    W2 &< 15.
\end{align}
such that the two samples are nested and the higher-density Mid-z sample includes Low-z.
As in \citet{unWISE_catalog}, we use Gaia to remove stars.
We find that these brighter, lower-redshift selections are more prone to stellar contamination than the Blue, Green and Red samples (as their colors are closer to the colors of typical Galactic stars). We therefore expand the pointlike Gaia selection from \citet{unWISE_catalog} to also remove any object brighter than a specified threshold in Gaia $G$ magnitude; for Low-z, this selection is:
\begin{equation}
\mathrm{pointlike}(G, A) = \begin{cases}
\text{True} & \text{if $G < 18.5$} \\
\log_{10} A < 0.5 & \text{if $18.5 < G < 19.25$} \\
\log_{10} A < 0.5 + \frac{5}{16} (G-19.25) & \text{otherwise} \, ,
\end{cases}
\end{equation}
and for Mid-z this selection is
\begin{equation}
\mathrm{pointlike}(G, A) = \begin{cases}
\text{True} & \text{if $G < 18.7$} \\
\log_{10} A < 0.5 & \text{if $18.7 < G < 19.25$} \\
\log_{10} A < 0.5 + \frac{5}{16} (G-19.25) & \text{otherwise} \, ,
\end{cases}
\end{equation}
and we then remove any object that is classified as point-like from Gaia.

When investigating areas of the sky with anomalously high densities of unWISE Low-z and Mid-z galaxies, we found that some of these overdensities were linked to nearby very large galaxies shredded into several sources. To remove these shredded galaxies, we use single-linkage agglomerative clustering with a linking length of 5'' to identify groups of galaxies. We find that typically all groups with three or more galaxies are shreds, and prune them down to a single member.

We measure the galaxy redshift distribution by cross-matching to DESI DR1 \citep{DESIDR1} and SDSS DR16 \citep{SDSSDR16} catalogs (Fig.~\ref{fig:unWISEsample1}). Most galaxies are in the DESI BGS\_ANY sample, with a smattering from DESI LRG and from SDSS; essentially every unWISE Low-z and Mid-z galaxy within the DESI footprint contains a match to DESI or SDSS. From these spectroscopic cross-matches, we find 1.1\% of the unWISE Low-z sample and 0.95\% of the unWISE Mid-z sample are stars.


\begin{figure}[htbp]
    \centering

    \subfloat[
        The unWISE Low-z sample galaxy overdensity map.
    ]{%
        \includegraphics[
            width=0.45\textwidth,
            trim=1.5cm 0.8cm 1.5cm 1cm,
            clip
        ]{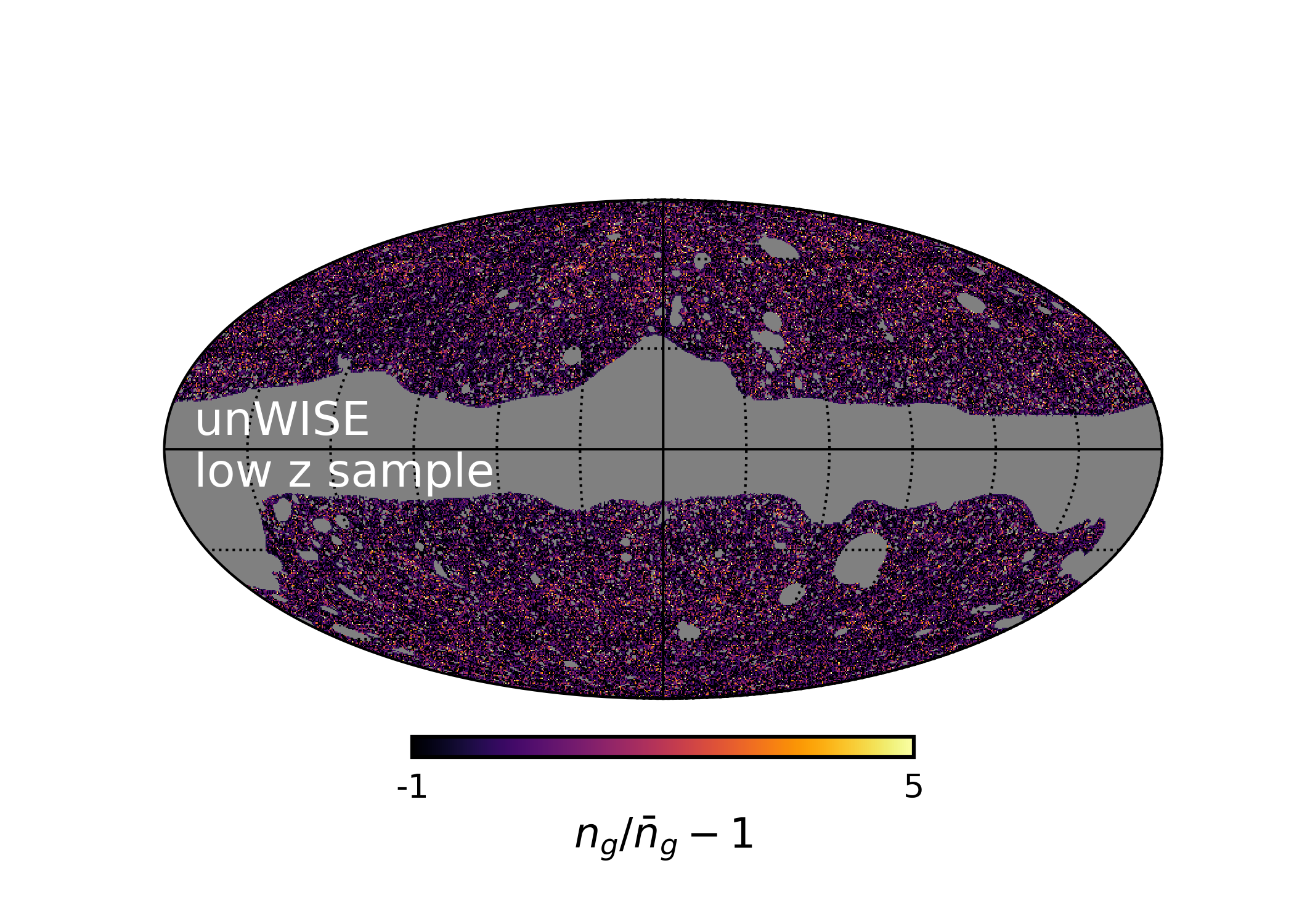}%
        \label{fig:unwise_lowz}%
    }%
    \hfill
    \subfloat[
        The unWISE Mid-z sample galaxy overdensity map.
    ]{%
        \includegraphics[
            width=0.45\textwidth,
            trim=1.5cm 0.8cm 1.5cm 1cm,
            clip
        ]{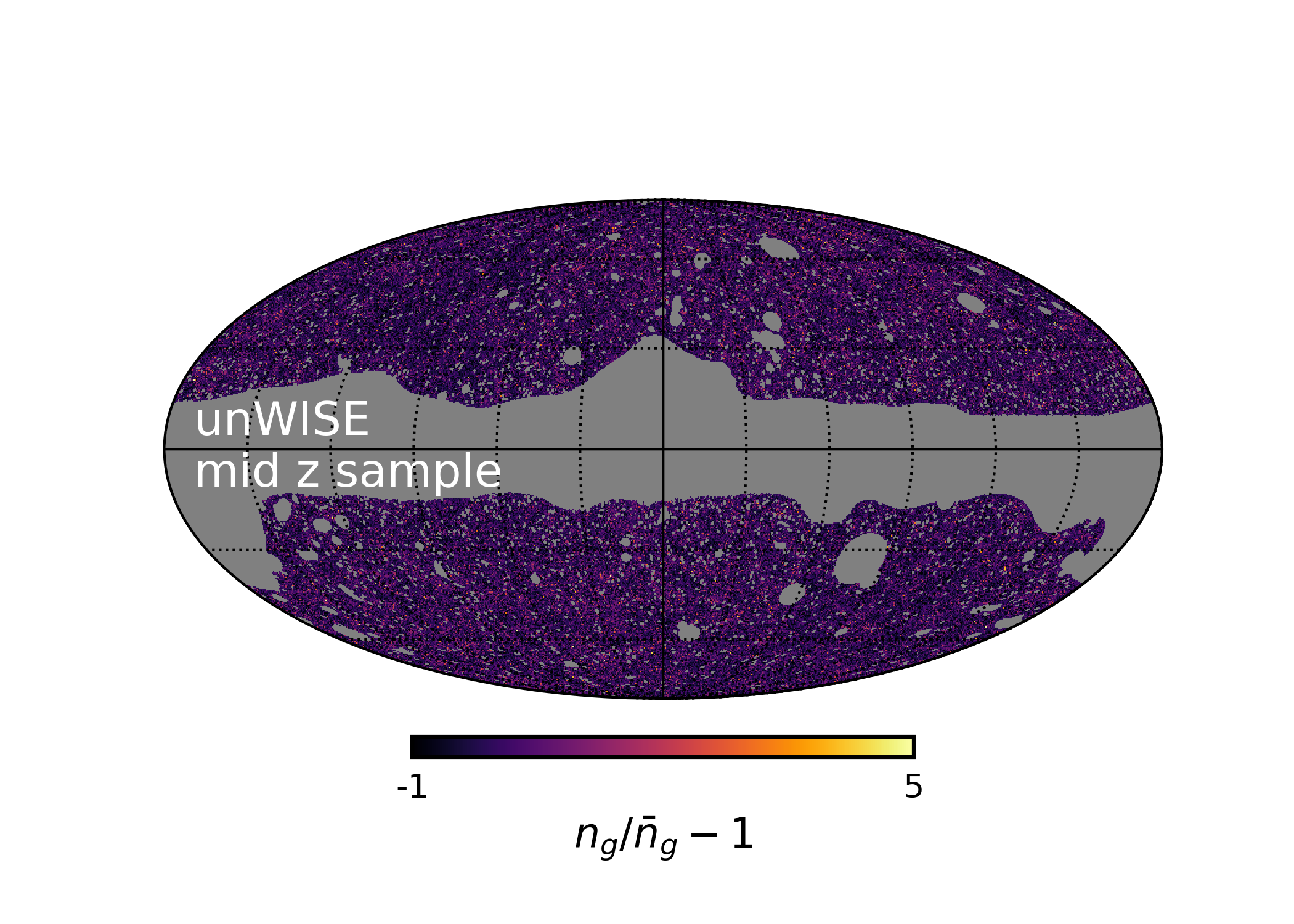}%
        \label{fig:unwise_midz}%
    }

    \caption{
        Galaxy overdensity
        $\delta_g=\frac{n_g}{\bar{n}_g}-1$
        of the two samples from the unWISE catalog.
        A Galactic and source mask is applied to both maps.
    }
    \label{fig:unWISE_dataset}
\end{figure}

\section{Measuring angular power spectra}\label{sec:correlation_func}

The power spectra studied in this work are estimated using the \href{https://github.com/LSSTDESC/NaMaster}{\texttt{NaMaster}} \citep{Namastercodepaper} pseudo-$C_\ell$ framework. 
\texttt{NaMaster} implements the MASTER algorithm \citep{Hivon02} to obtain the window-deconvolved binned angular power spectrum.
In order to accurately capture the full non-diagonal structure of the covariance matrix with a non-trivial mask, we use the \texttt{NaMaster} analytic window-convolved Gaussian covariance module \citep{GarciaGarcia19} using smooth fits to the measured auto power spectra as the input templates.
The covariance is shown in Fig.~\ref{fig:cl_gf_cov_planck_lowz} and~\ref{fig:cl_gf_cov_actplanck_lowz}.

\subsection{Galaxy-temperature cross-correlations}\label{sec:cl_gf}

\begin{figure}
    \centering

    \subfloat[
        The angular cross-power spectrum $C_\ell^{gT_f}$ in thermodynamic units of K.
    ]{%
        \includegraphics[
            width=0.4275\linewidth
        ]{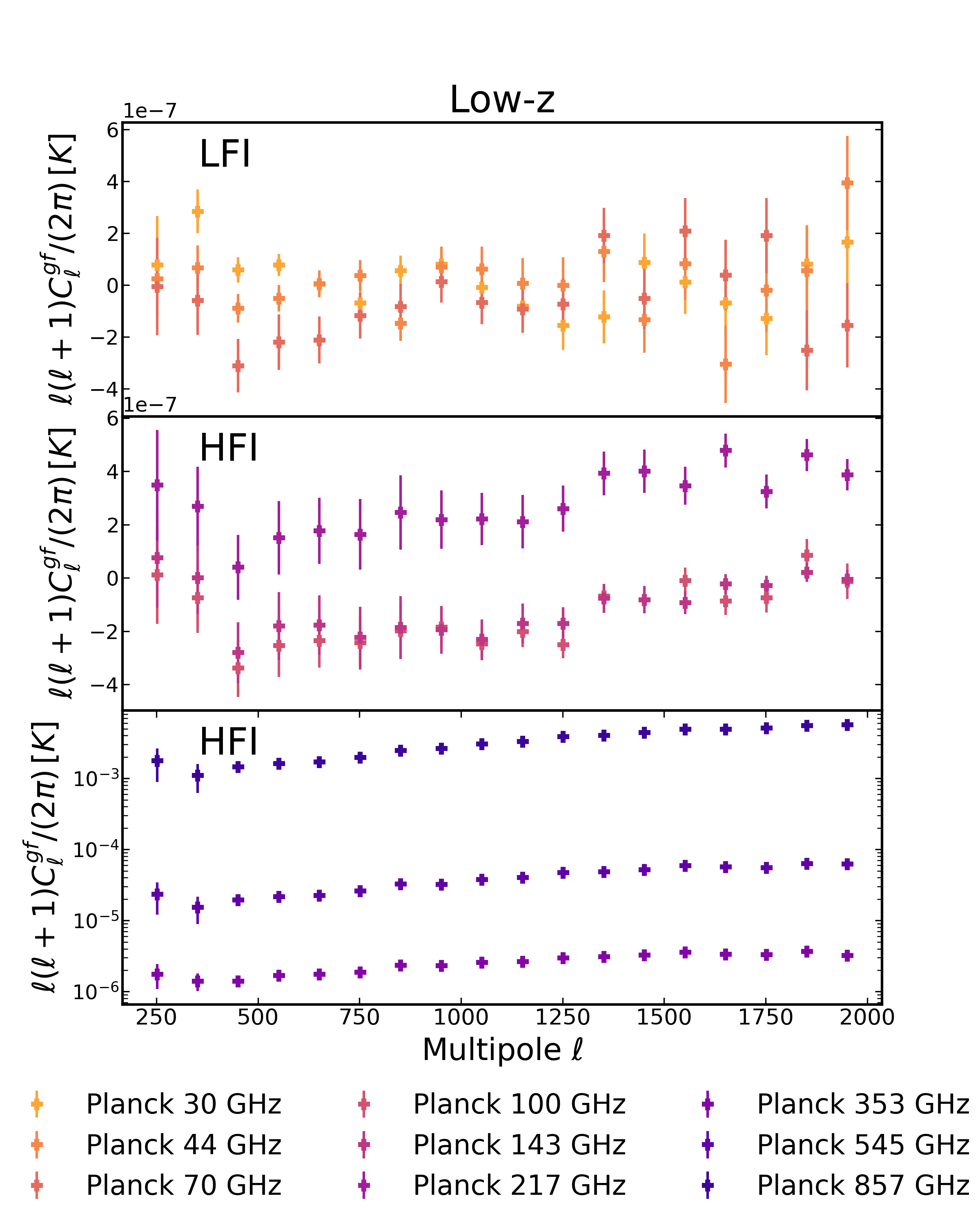}%
        \label{fig:cl_gf_cl_planck_lowz}%
    }
    \hfill
    \subfloat[
        The correlation matrix $\mathcal{C}_{\ell_1\ell_2}^{f_1f_2}$.
        Each block represents $\ell$ from 200 to 2000 for the frequency pair.
    ]{%
        \includegraphics[
            width=0.494\linewidth,
            trim=1.5cm 1cm 1cm 0cm,
            clip
        ]{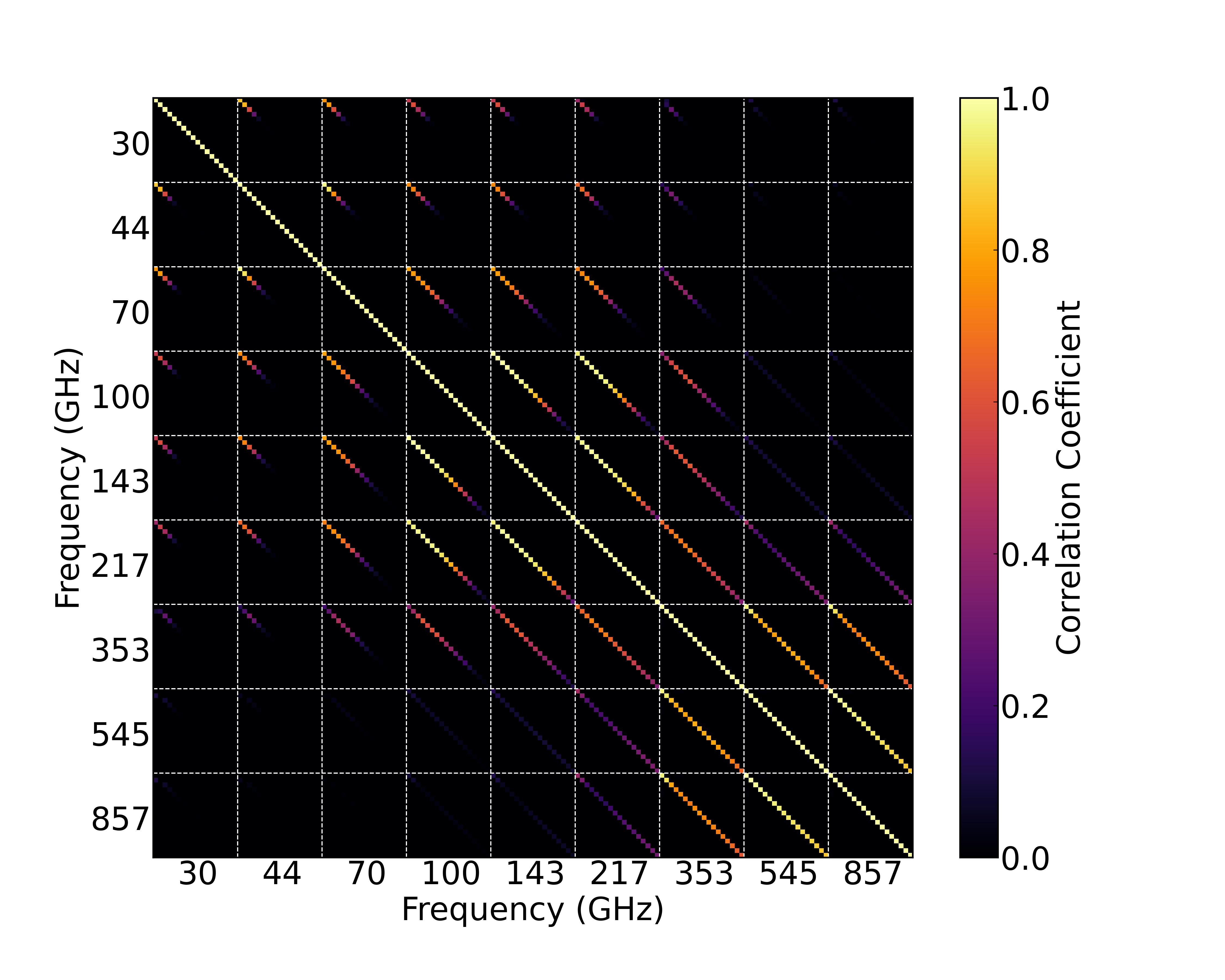}%
        \label{fig:cl_gf_cov_planck_lowz}%
    }

    \caption{
        The \texttt{NaMaster} result for the pseudo-$C_\ell$ of the
        unWISE Low-z galaxy overdensity cross-correlated with nine
        \textit{Planck} single-frequency temperature maps:
        30, 44, 70, 100, 143, 217, 353, 545, and 857\,GHz.
    }
    \label{fig:cl_gf_planck_lowz}
\end{figure}

By combining the CMB temperature and galaxy datasets described in Sec.~\ref{sec:dataset}, we compute the galaxy-temperature cross-correlations using \texttt{NaMaster} for each nominal band of \textit{Planck} and ACT+Planck. We show the calculated $C_\ell^{gT_f}$ between \textit{Planck} and unWISE Low-z sample in Fig.~\ref{fig:cl_gf_cl_planck_lowz}, and the correlation matrix of the $C_\ell^{gT_f}$s is shown in \ref{fig:cl_gf_cov_planck_lowz} for diagnostics. 

The harmonic beam and pixel window function of Planck-only maps is very close to zero at $\ell>2000$. To avoid singular matrices in the component separation, we adopted an $200<\ell<2000$ cut to each $C_\ell$ and $\mathbf{Cov}$ blocks, we bin the multipoles linearly using bin size $\Delta \ell =100$. 

\begin{figure}
    \centering

    \subfloat[
        The angular cross-power spectrum $C_\ell^{gT_f}$ in thermodynamic units of K.
    ]{%
        \label{fig:cl_gf_cl_actplanck_lowz}%
        \includegraphics[
            width=0.4275\linewidth
        ]{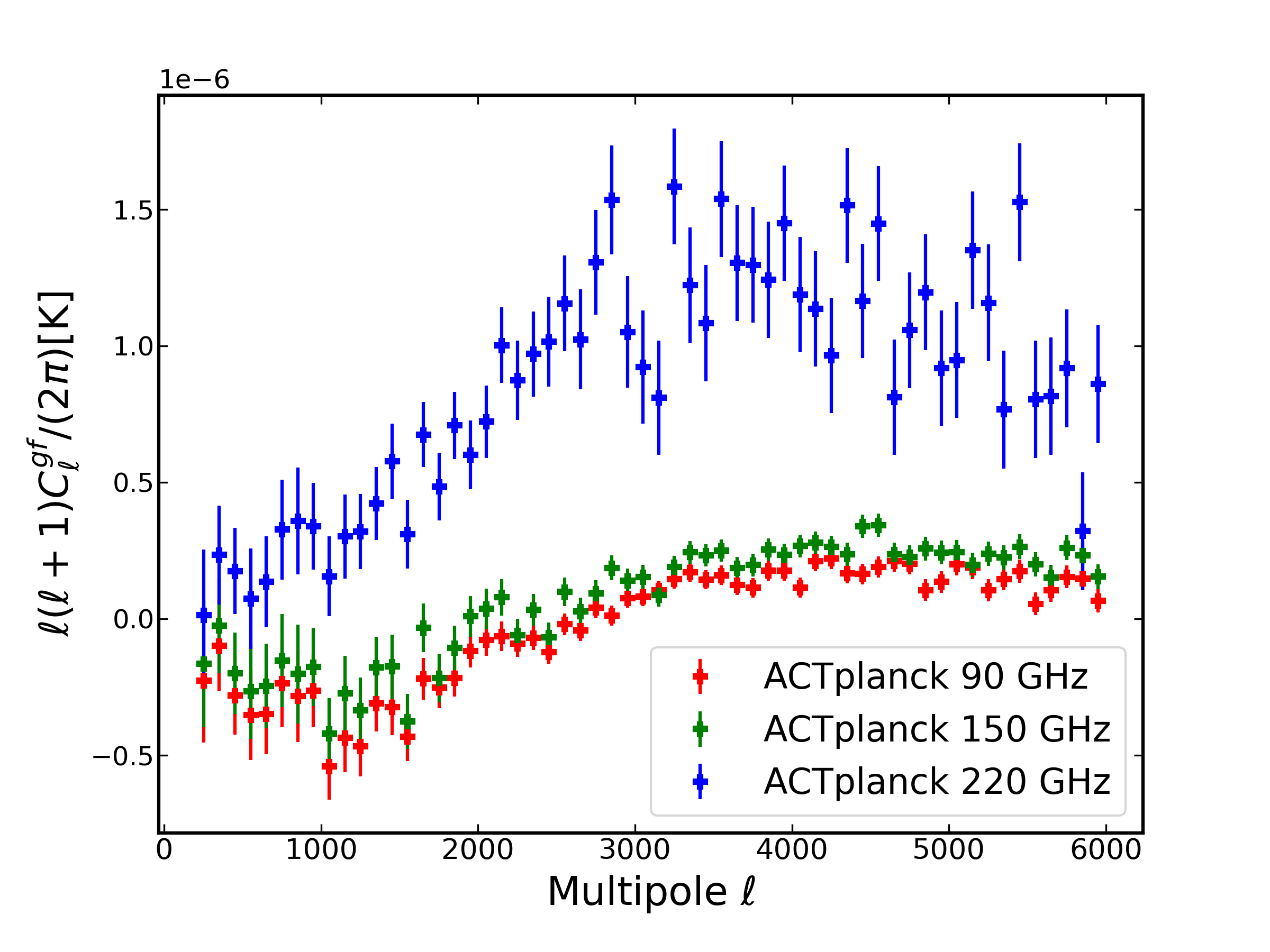}%
    }
    \hfill
    \subfloat[
        The correlation matrix $\mathcal{C}_{\ell_1\ell_2}^{f_1f_2}$.
        Each block represents $\ell$ from 200 to 6000 for the frequency pair.
    ]{%
        \label{fig:cl_gf_cov_actplanck_lowz}%
        \includegraphics[
            width=0.494\linewidth,
            trim=1.5cm 1cm 1cm 0cm,
            clip
        ]{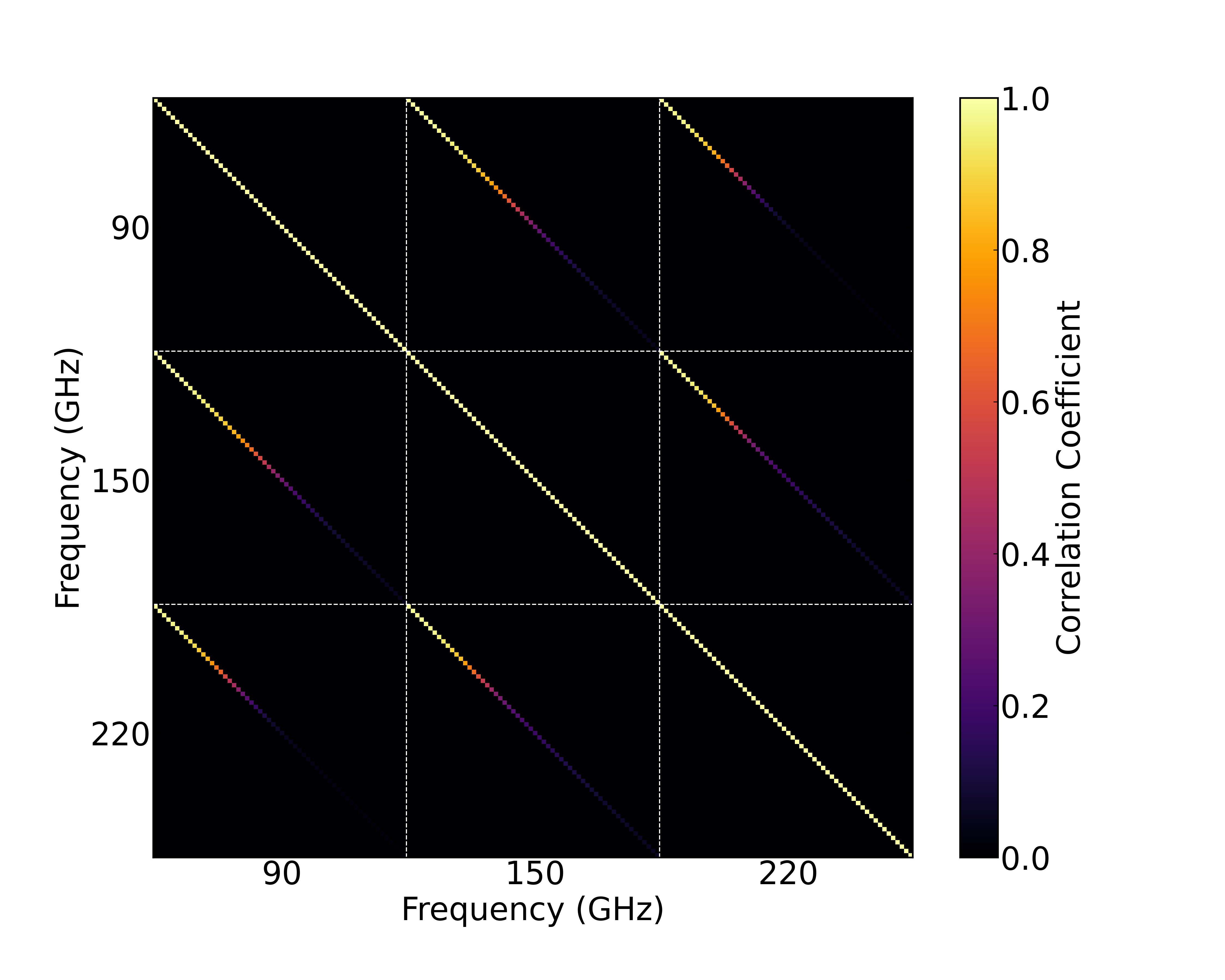}%
    }

    \caption{
        The \texttt{NaMaster} result for the pseudo-$C_\ell$ of the
        unWISE Low-z galaxy overdensity cross-correlated with three
        ACT+\textit{Planck} single-frequency temperature maps:
        90\,GHz, 150\,GHz, and 220\,GHz.
    }
    \label{fig:actplanck_lowz}
\end{figure}

We measure the cross-correlation between unWISE galaxy samples and three ACT frequencies using the ACT+\textit{Planck} coadd maps. We utilize the high angular resolution in ACT map using map resolution $N_{\rm side}=2048$ and calculate angular power spectrum for $200<\ell<6000$ with $\Delta \ell =100$ linear binning. We show the result for Low-z galaxy sample in Fig.~\ref{fig:cl_gf_cl_actplanck_lowz}, and we also show the correlation matrix of three frequencies for diagnosis in Fig.~\ref{fig:cl_gf_cov_actplanck_lowz}.

\subsubsection{SED model fitting}

Using 9 \textit{Planck} frequencies, there are enough measured degrees of freedom to give constraint on the choice and functional form of the responses. The best-fit quadratic $\tilde{\chi}^2(\beta_0,T_0,\beta_r)$ given by Equation~\ref{eq:marginalized_chisq} can be now be further used to constrain the SED parameters $\beta_0,T_0,\beta_r$. Here we present the best-fit model parameters using all 9 frequencies of \textit{Planck} in Tab.~\ref{tab:plk_model_fitting_results}. The uncertainties for best-fit parameters are computed using the Fisher matrix at the best fit.

\begin{table}[htbp]
\centering
\small
\setlength{\tabcolsep}{0pt}
\begin{tabular}{@{}L{0.08\textwidth}L{0.28\textwidth}C{0.14\textwidth}C{0.14\textwidth}C{0.13\textwidth}C{0.06\textwidth}C{0.07\textwidth}@{}}
\hline
\textbf{Sample} & \textbf{Recipe} & $\beta_0\pm \sigma_{\beta_0}$ & $T_0\pm \sigma_{T_0}[\rm K]$ & $\beta_r\pm\sigma_{\beta_{\rm r}}$  &$\chi^2_r$& \textbf{BIC} \\
\hline
Low-z & tSZ+Radio+CIB-amp+CIB-$\delta \beta$ & 1.29 $\pm$ 0.44 & 23.6 $\pm$ 5.7 & -2.18 $\pm$ 0.24 & 0.94  & 463.1 \\
\grayrowstart{Low-z} & tSZ+Radio+CIB-amp & 1.79 $\pm$ 0.29 & 22.0 $\pm$ 5.1 & -2.18 $\pm$ 0.26 & 1.00 & 395.2\\
Low-z & tSZ+CIB-amp+CIB-$\delta\beta$       & 0.86 $\pm$ 0.40 & 54 $\pm$ 21 & --               & 1.89 & 485.6 \\
Mid-z & tSZ+Radio+CIB-amp+CIB-$\delta\beta$ & 1.44 $\pm$ 0.69 & 26.1 $\pm$ 3.8 & -2.71 $\pm$ 0.28 & 0.928 & 462.3 \\
\grayrowstart{Mid-z} & tSZ+Radio+CIB-amp & 1.41 $\pm$ 0.13 & 28.0 $\pm$ 4.2 & -2.62 $\pm$ 0.26 & 1.11 & 407.0\\
Mid-z & tSZ+CIB-amp+CIB-$\delta\beta$       & 0.82 $\pm$ 0.24 & 42 $\pm$ 11 & --               & 2.28  & 527.0 \\
\hline
\end{tabular}
\caption{Best-fit CIB and radio spectral properties for Low-z and Mid-z samples for cross-correlation with \textit{Planck} frequency maps. The uncertainties are estimated from the Fisher matrix at the best fit. The highlighted rows show the best-fit minimal models from the samples, based on both reduced $\chi^2$ and Bayesian Information Criterion (BIC).}
\label{tab:plk_model_fitting_results}
\end{table}


Comparing the \ac{BIC} and reduced-$\chi^2$ result in Tab.~\ref{tab:plk_model_fitting_results}, we find that the recipe tSZ + Radio + CIB-amplitude is the most Bayesian-supported for both galaxy samples. Comparing the two three-component recipes, the recipe that includes a radio component is supported with $\Delta \rm BIC=90$ in Low-z and $\Delta \rm BIC=120$ in Mid-z, which translates to $9.5\sigma$ and $11.0\sigma$ respectively.

The four component recipe tSZ + Radio + CIB-amplitude + CIB-$\delta\beta$ yields a better fit, but is disfavored by \ac{BIC}, indicating an overfitting. This shows that the CIB-$\delta\beta$ component is not a necessary component for these galaxy samples and the tSZ + radio + CIB-amplitude recipe is optimal for application in ACT DR6 maps with low-redshift cross-correlation. We emphasize that this conclusion may be different for different galaxy samples, i.e.\ when cross-correlating with higher redshift galaxies, CIB-$\delta\beta$ may become more important than radio as it is more correlated with high-redshift galaxies.

The best-fit values of CIB dust temperature is $22.0\pm 5.1K$ for Low-z and $28.0\pm4.2 K$ for Mid-z, and are consistent with previous studies correlating unWISE blue, green and red galaxies with \textit{Planck} CIB by \citet{unwise_planckcib}.

The best-fit value $\beta_r$ is $-2.18\pm 0.24$ for Low-z and $-2.62\pm0.26$ for Mid-z. In contrast, the Rayleigh-Jeans spectral index is $\beta_{\rm ff}\approx-2.14$ for free-free emission and $\beta_{\rm syn}\approx-3.11$ for synchrotron emission in Milky-Way mass galaxies~\citep{Planckdiffusemap2015}. The best-fit effective spectral index indicates a preference for free-free emission in Low-z galaxies and a preference for synchrotron radiation in Mid-z galaxies.

\begin{figure}[h!]
    \centering

    \subfloat[
        The unWISE Low-z posterior.
    ]{%
        \includegraphics[
            width=0.46\textwidth
        ]{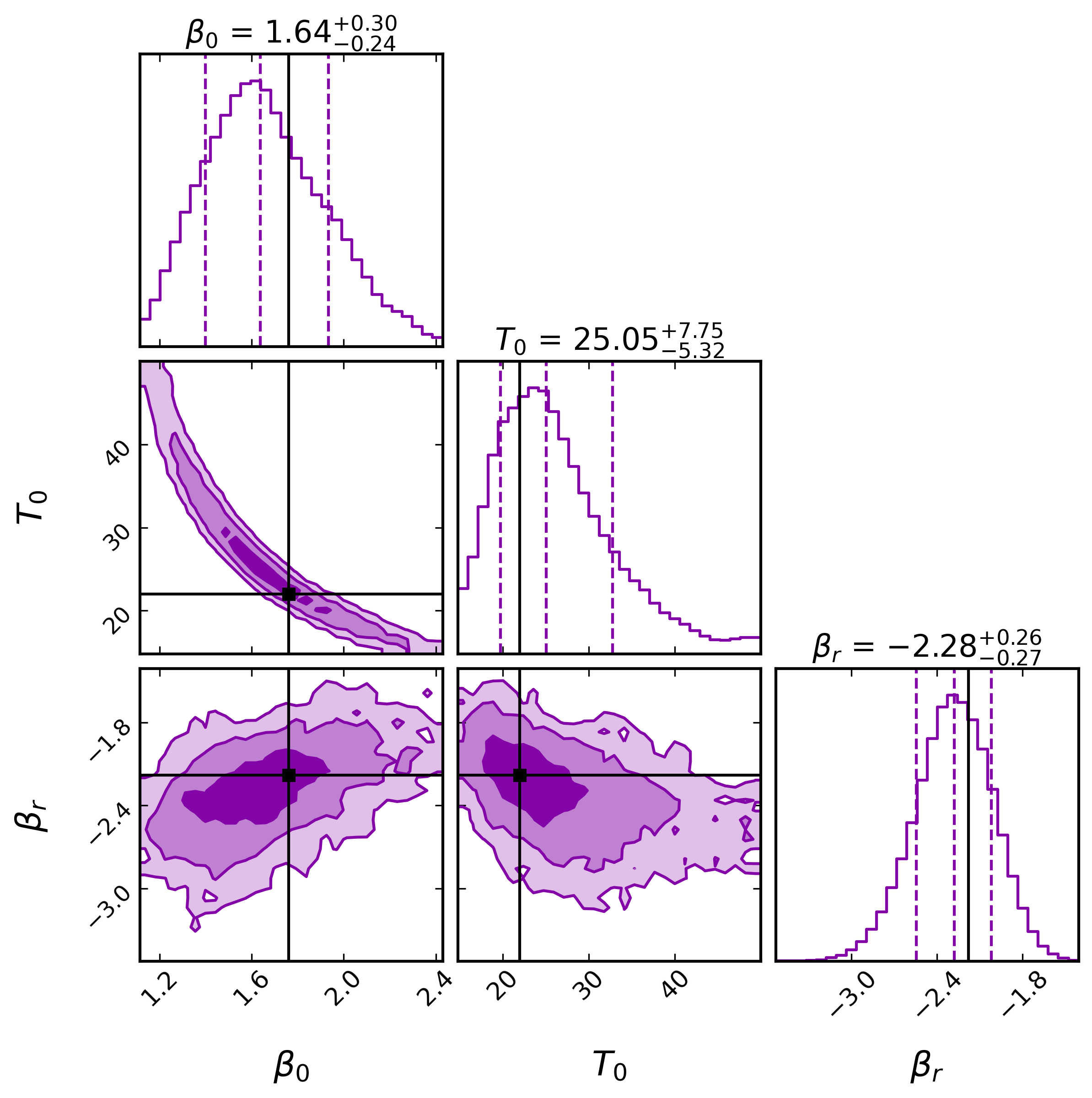}%
        \label{fig:lowztrbcleanpost}%
    }
    \hfill
    \subfloat[
        The unWISE Mid-z posterior.
    ]{%
        \includegraphics[
            width=0.46\textwidth
        ]{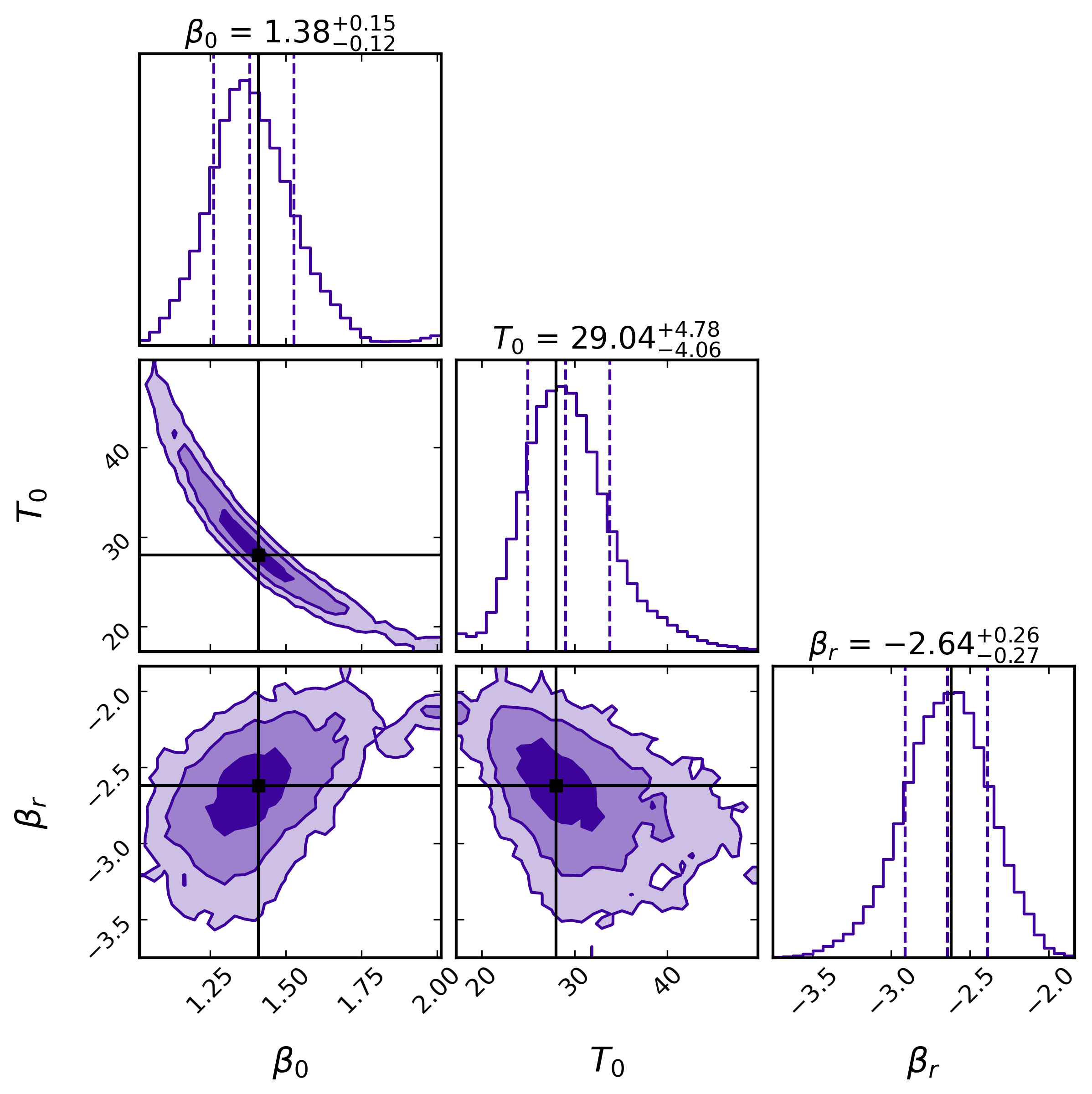}%
        \label{fig:midztrbcleanpost}%
    }

    \caption{
        Posterior MCMC samples obtained by minimizing the cleaning
        statistic $\tilde{\chi}^2$ for the unWISE samples
        cross-correlated with nine \textit{Planck} frequency maps.
        The statistic $\tilde{\chi}^2$ is derived from the residuals
        of the component separation in $C_\ell^{gT_f}$ space.
    }
    \label{fig:mcmc_clean}
\end{figure}

We run a \ac{MCMC} posterior estimation for the best fitting recipe tSZ+Radio+CIB-amplitude, as shown in Fig.~\ref{fig:mcmc_clean}. The \textit{Planck} sample cross-correlation is less capable of constraining CIB-parameters in the $T_0-\beta_0$ plane compared to the radio spectral index $\beta_r$.



\subsection{Component separation in harmonic space}\label{sec:component_separation_res}

Using the best-fit parameters for each recipe, we examine the separated components from Equation~\ref{eq:estimatorsol}. In addition to the \ac{tSZ} cleaned result ($C_\ell^{gy}$), we obtain the galaxy cross-correlation with other signal components: Radio ($C_\ell^{gr}$), CIB-amplitude ($C_\ell^{gB}$), and CIB-$\delta\beta$ ($C_\ell^{g\beta}$).

\begin{figure}[htbp]
    \centering

    \subfloat[
        The \textit{Planck} $\times$ unWISE Low-z components.
    ]{%
        \includegraphics[
            width=0.48\textwidth
        ]{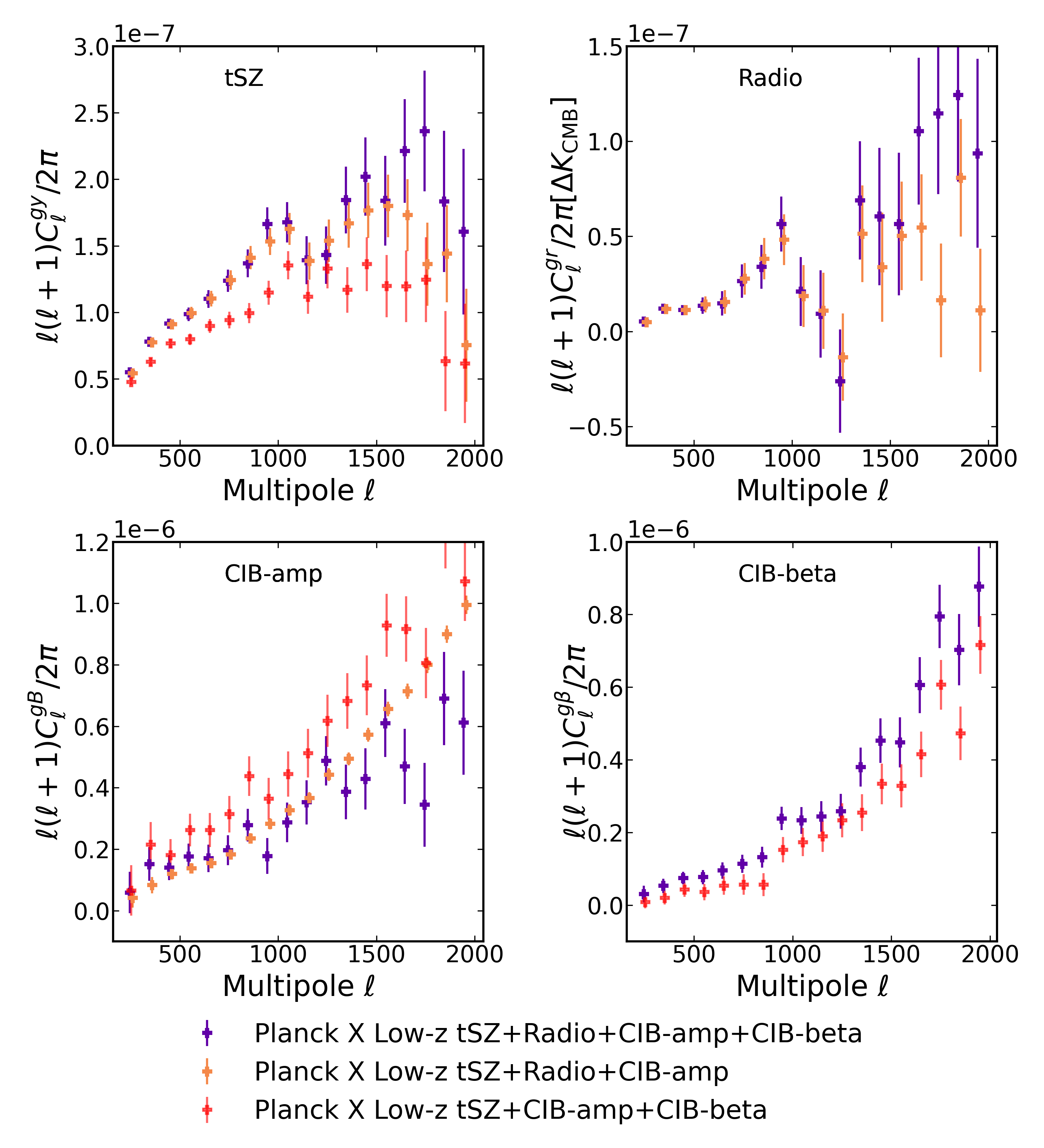}%
        \label{fig:plancklowzcomp}%
    }
    \hfill
    \subfloat[
        The \textit{Planck} $\times$ unWISE Mid-z components.
    ]{%
        \includegraphics[
            width=0.48\textwidth
        ]{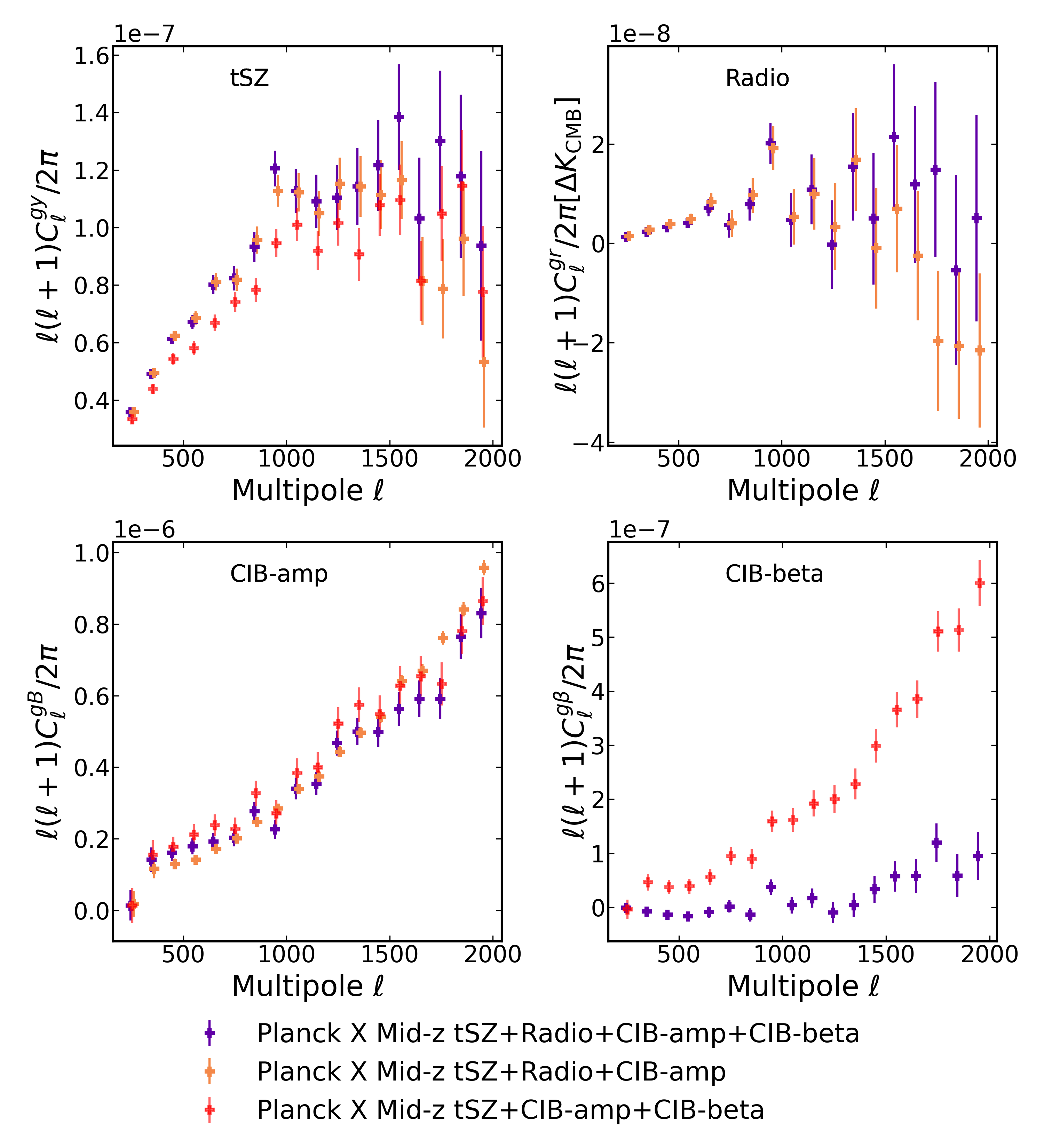}%
        \label{fig:planckmidzcomp}%
    }

    \caption{
        Component separation of the \textit{Planck}-only $\times$
        unWISE cross-correlations for different recipes, using the
        best-fit SED parameters in
        Table~\protect\ref{tab:plk_model_fitting_results}.
        The four panels show the galaxy--tSZ cross-spectrum
        $C_\ell^{gy}$, galaxy--radio amplitude $C_\ell^{gr}$,
        galaxy--CIB amplitude $C_\ell^{gB}$, and galaxy--CIB spectral
        index $C_\ell^{g\beta}$, respectively.
    }
    \label{fig:planckcomp}
\end{figure}

First we compare the results for \textit{Planck}-only component separation using different recipes in Fig.~\ref{fig:planckcomp}. In both Mid-z and Low-z, the tSZ + Radio + CIB-amplitude + CIB-$\delta\beta$ 4-component recipe agrees with tSZ + Radio + CIB-amplitude 3-component recipe within the uncertainty range, and deviates from the tSZ + CIB-amplitude + CIB-$\delta\beta$ results significantly. This is consistent with the superior goodness of fit from the first two recipes, and emphasizes the importance of the radio template.

The increased uncertainty at $\ell>1100$ corresponds to the limit of the \textit{Planck} beam, which for 857GHz nominal band is $B_\ell\sim0.4$ at $\ell=1100$. The overall uncertainty in 4-component recipe is higher than 3-component recipes due to increased fitting degrees of freedom. The combined evidence from BIC, goodness of fit and the level of uncertainty, again supports the tSZ + Radio + CIB-amplitude recipe to be the optimal signal combination for both galaxy samples. 


\begin{figure}[htbp]
    \centering

    \subfloat[
        The ACT+\textit{Planck} $\times$ unWISE Low-z
        cross-correlation components.
    ]{%
        \includegraphics[
            width=0.48\textwidth
        ]{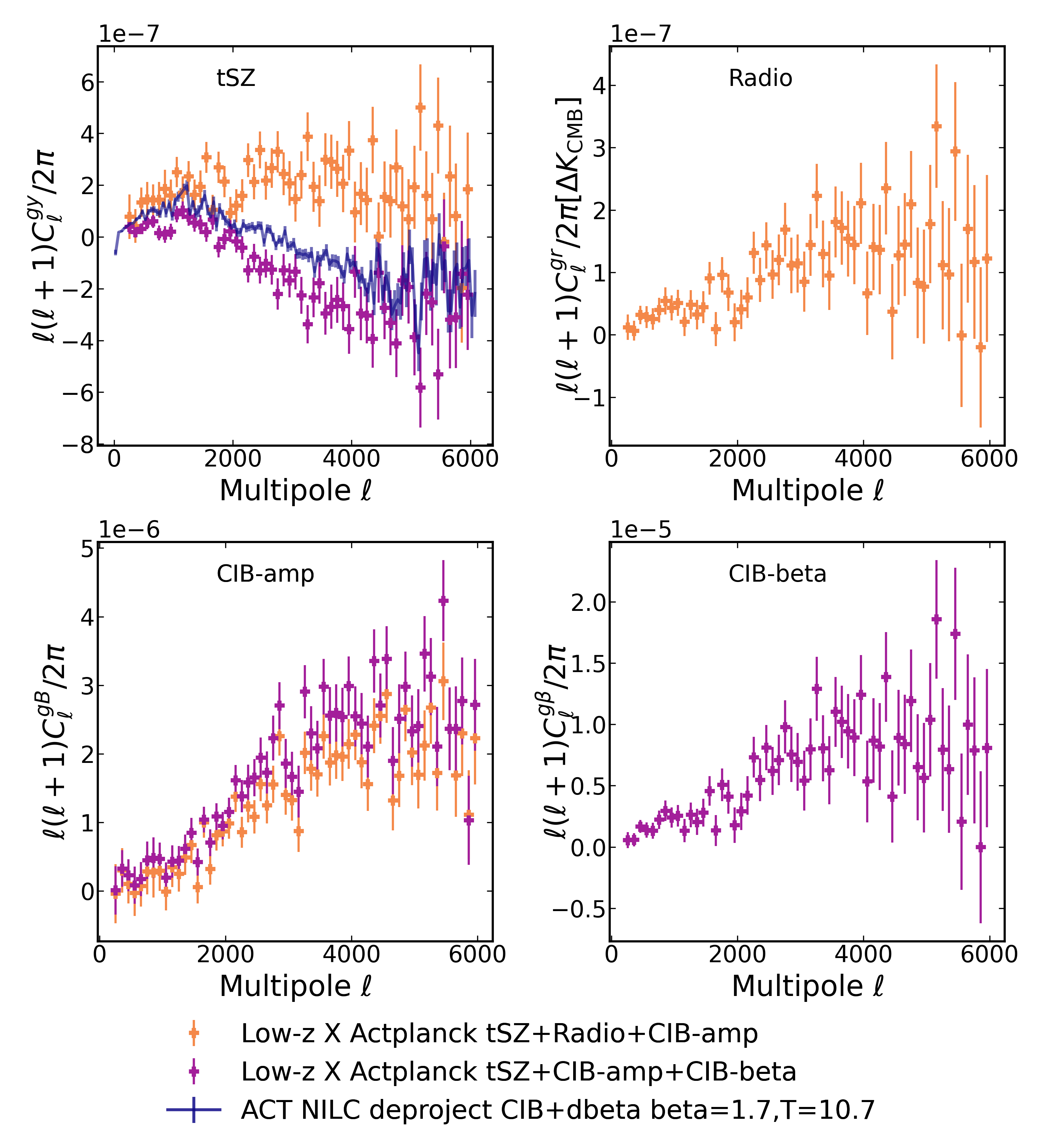}%
        \label{fig:actplancklowzcomp}%
    }
    \hfill
    \subfloat[
        The ACT+\textit{Planck} $\times$ unWISE Mid-z
        cross-correlation components.
    ]{%
        \includegraphics[
            width=0.48\textwidth
        ]{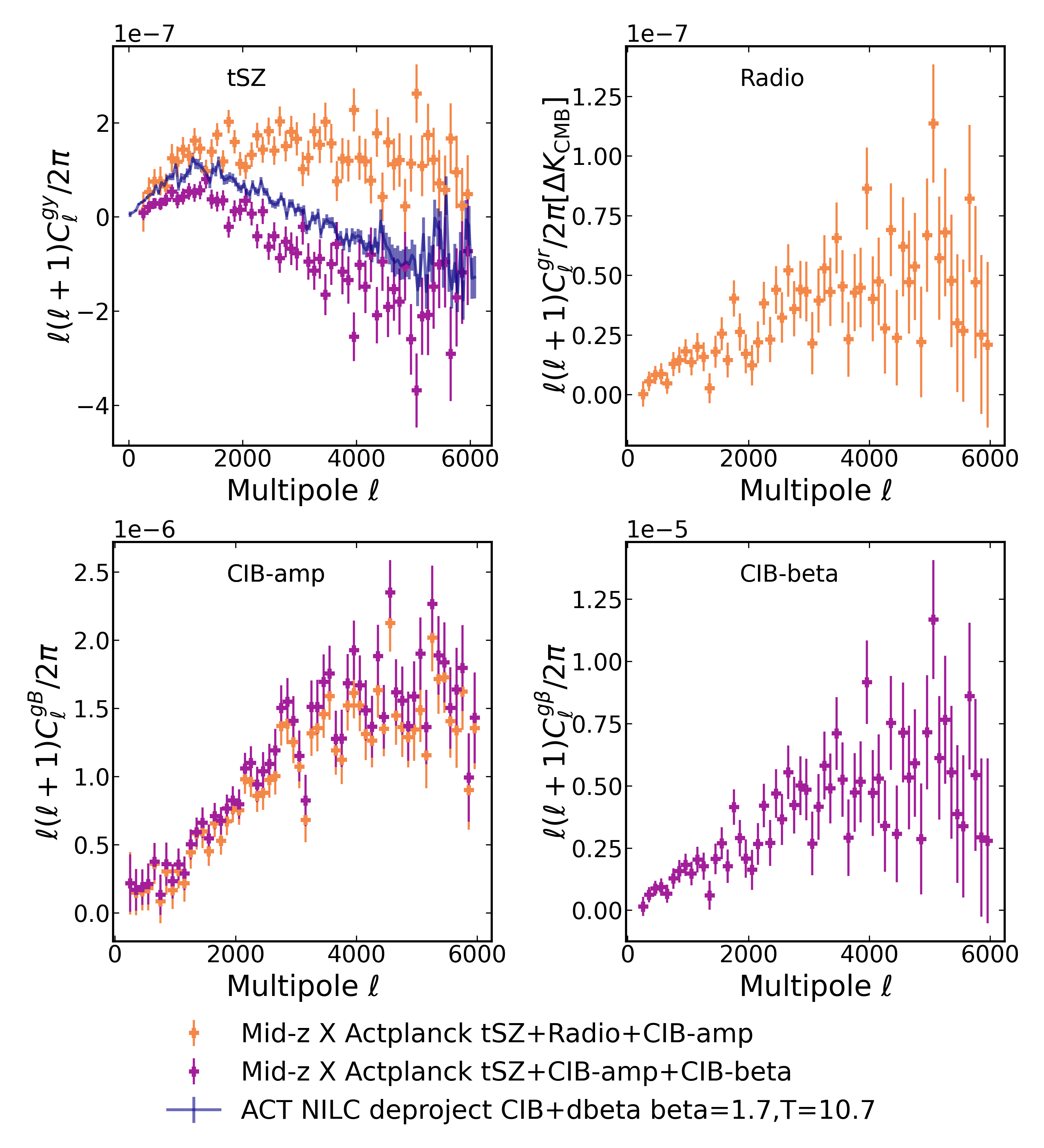}%
        \label{fig:actplanckmidzcomp}%
    }

    \caption{
        Component separation of the ACT+\textit{Planck} $\times$
        unWISE cross-correlations for different recipes, using the
        best-fit SED parameters in
        Table~\protect\ref{tab:plk_model_fitting_results}.
        The four panels show the galaxy--tSZ cross-spectrum
        $C_\ell^{gy}$, galaxy--radio amplitude $C_\ell^{gr}$,
        galaxy--CIB amplitude $C_\ell^{gB}$, and galaxy--CIB spectral
        index $C_\ell^{g\beta}$, respectively.
        The ACT DR6 NILC $\times$ unWISE cross-correlations are also
        shown for reference.
    }
    \label{fig:actplanckcomp}
\end{figure}

Then we adopt the best-fit parameters in Tab.~\ref{tab:plk_model_fitting_results} for 3-component recipes to the ACT+\textit{Planck} cross-correlations. Taking advantage of the significantly smaller ACT beam, the component separation is able to reach $\ell=6000$. We compare the different separation recipes in Fig. \ref{fig:actplanckcomp}. We also show the cross-correlation between galaxy samples and the CIB+CIB-$\delta\beta$ deprojected NILC y-map by ACT collaboration \citep{Coultontszcomponent} for comparison.\footnote{The ACT DR6 $y$ map is from \href{https://portal.nersc.gov/project/act/dr6_nilc/ymaps_20230220/ilc_actplanck_ymap.fits}{https://portal.nersc.gov/project/act/dr6\_nilc/ymaps\_20230220/ilc\_actplanck\_ymap.fits}, with the cross-correlation and covariance calculated using \texttt{NaMaster}. The $1.6' $ NILC beam is deconvolved for the $C_\ell^{gy}$ in Fig~\ref{fig:actplanckcomp}.}

The overall uncertainty of ACT+\textit{Planck} component-separated results are significantly larger than the \textit{Planck}-only results. There are two factors that contribute to this.
\begin{enumerate}
    \item In the process of separating three components from three frequencies, the noise is significantly amplified, as there are effectively as many fit degrees of freedom as there are data points.
    \item The ACT+\textit{Planck} footprint ($f_{\rm sky}\approx0.32$) is less than \textit{Planck} ($f_{\rm sky}\approx0.58$).
\end{enumerate}
 
The ACT+\textit{Planck} tSZ + CIB-amplitude + CIB-$\delta\beta$ shows a zero-crossing at $\sim \ell =2000$ for both Low-z and Mid-z, and there is a broadly similar negativity trend observed in NILC galaxy-$y$ cross-correlations. A major difference between the method in this work and the CIB deprojection from ACT DR6 is the exact values for the CIB SED, where $\beta_0=1.7,T_0=10.7$ is taken as fiducial value by \citet{Coultontszcomponent}.

The combination of the considerably better fit of tSZ+Radio+CIB-amp over the models with CIB-$\delta \beta$, and the fact that the $gy$ cross-correlation stays positive everywhere when using tSZ+Radio+CIB-amp (Fig.~\ref{fig:actplanckcomp}), leads us to strongly prefer the fits using the tSZ+Radio+CIB-amp model. We interpret the negative $gy$ cross-correlation  at $\ell > 2000$ when using the NILC map or the CIB-$\delta \beta$ models as an artifact arising from unmitigated radio foregrounds.

\begin{figure}[h]
    \centering
    \includegraphics[width=0.8\linewidth]{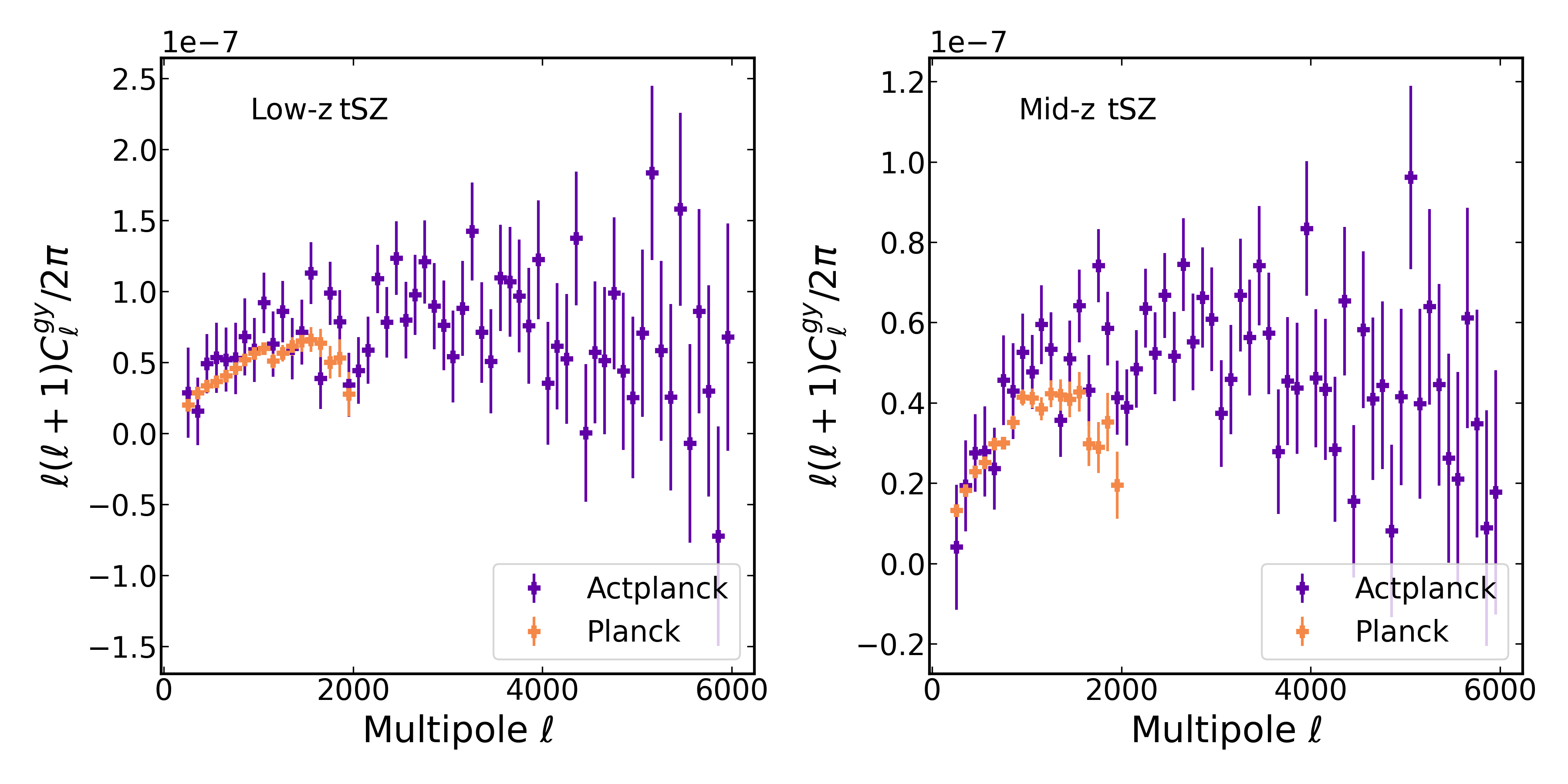}
    \caption{Comparison between the \textit{Planck} and ACT+\textit{Planck} component separation tSZ result with tSZ+Radio+CIB-amplitude recipe. In both the left and right panel, orange errorbars shows the component-separated \textit{Planck}-only galaxy-$y$ cross power spectra, and purple errorbars show the component-separated ACT+\textit{Planck} cross power spectra.}
    \label{fig:planck_vs_actplanck_separation}
\end{figure}

We also make a comparison between the component separation results from \textit{Planck} and ACT+\textit{Planck} in Fig. \ref{fig:planck_vs_actplanck_separation}.  The ACT+\textit{Planck} maps consist of a coadd of the ACT and \textit{Planck} observations, with \textit{Planck} dominating on large scales and the map signal dominated by ACT results at $\ell\gg 200$ \citep{Naess_ACTDR5}. In the results we see consistency between ACT+\textit{Planck} and \textit{Planck} across the $200<\ell<2000$ range for Low-z sample and $200<\ell<1500$ for Mid-z sample. This supports the stability of the harmonic space component separation in finding instrument-independent \ac{tSZ}. There is an enhancement of ACT+\textit{Planck} $\times$ Mid-z result in the range $1500<\ell<2000$. We attribute this artifact to the difference in beam modeling across instruments, where \textit{Planck} beam is starting to dominate in $\ell>1500$.

%
\section{Halo Model Analysis}\label{sec:halo_model_section}

We model the galaxy clustering $C_\ell^{gg}$ and galaxy-tSZ cross-correlation $C_\ell^{gy}$ using the halo model calculation in the package \href{https://github.com/LSSTDESC/CCL}{\texttt{pyccl}} \citep{pycclpaper}. In the halo model, the 2-point correlation function of two cosmological fields $u,v$ is broken into intra-halo (1-halo) and inter-halo (2-halo) terms, where the 3D power spectrum:
\begin{align}
    P_{uv}(k,a)&=P_{uv}^{1h}(k,a)+P_{uv}^{2h}(k,a)\nonumber\\
    &=\int dM\ n(M,a)\langle u(k,a|M)v(k,a|M)\rangle\nonumber\\
    &+\int dM_1 \ n(M_1,a)b(M_1,a)\langle u(k,a|M_1)\rangle\int dM_2 \ n(M_2,a)b(M_2,a)\langle v(k,a|M_2)\rangle P_{\rm lin}(k,a),
    \label{eq:halomodel_powerspec}
\end{align}
where $\langle u(k,a|M)\rangle,\langle v(k,a|M)\rangle$ are the field $u,v$ halo profile evaluated at redshift with scale-factor $a$, and $\langle u(k,a|M)v(k,a|M)\rangle$ is the 2-point moment of the two halo profiles, and $P_{\rm lin}(k,a)$ is the linear matter power spectrum. 

In the analyses of this work, we use the halo mass function $n(M,a)$ parametrization of \citet{tinker08hmf}, and the halo bias $b(M,a)$ of \citet{tinker10bias}. We adopt the $M_{500c}$ mass definition and define halo mass as:
\begin{equation}
    M_{500c}=\frac{4\pi}{3}r_{\rm vir}^3\rho_c(z)\times500,
\end{equation}
with critical density $\rho_c(z)=\frac{3H(z)^2}{8\pi G}$.

\subsection{Angular correlations}
The observed angular autocorrelation or cross-correlation $C_\ell^{uv}$ is projected from the 3D power spectrum $P_{uv}$:
\begin{equation}
    C_\ell^{uv}=\int \frac{d\chi}{\chi^2}W_u(\chi)W_v(\chi)P_{uv}\left(k=\frac{\ell+1/2}{\chi},z(\chi)\right)
\end{equation}

where $W_u,W_v$ are the radial kernels of the fields as functions of radial comoving distance $\chi$, and we adopt the Limber approximation~\citep{Limberapprox,extendedlimber} with reasonable accuracy due to slow-varying radial kernels and wide redshift distributions with width $\delta z\sim 0.2$.

The radial kernels of the galaxy clustering are calculated using the redshift distribution of galaxies previously shown in Fig.~\ref{fig:unWISEsample1}:
\begin{equation}
    W_g(\chi)=p(z)H(z),
\end{equation}
where $p(z)$ is the normalized redshift distribution, and $H(z)$ is the expansion rate in units of $\rm Mpc^{-1}$.

The \ac{tSZ} is a radially projected field, with the Compton-$y$ parameter \citep{szbirkshaw},
\begin{align}
    y &= \frac{\sigma_T}{m_ec^2}\int P_e(l)dl=\frac{\sigma_T}{m_ec^2}\int P_e(l)\frac{1}{1+z}d\chi\nonumber\\
    W_y(\chi)&=\frac{\sigma_T}{m_ec^2}\frac{1}{1+z}\label{eq:tszradialkernel}
\end{align}

\subsection{Halo model calibration with HOD}\label{sec:HOD_model}

To calibrate the galaxy sample for halo distribution information, we adopt the \ac{HOD} model introduced by \citet{Zhenghalomodel2007} and \citet{Andohod2017} to parametrize the profile $\langle \delta g(k,a|M)\rangle$. The population of galaxies in the \ac{HOD} is separated into $\langle N_c(M)\rangle$ central galaxies and $\langle N_s(M)\rangle$ satellite galaxies:
\begin{align}
    \langle N(M)\rangle&=\langle N_c(M)\rangle\left(f_c +\langle N_s(M)\rangle\right),\nonumber\\
    \langle N_c(M)\rangle&=\frac{1}{2}\left[1+\mathrm{erf}\left(\frac{\log M - \log M_{\rm min}}{\sigma_{\log M}}\right)\right],\label{eq:hod_nc}\\
    \langle N_s(M)\rangle&=\Theta(M-M_0)\left(\frac{M-M_0}{M_1}\right)^{\alpha}.\label{eq:hod_ns}
\end{align}
We simplify the model by fixing $\sigma_{\log M}=0.4,\alpha=1,M_0=10^7M_\odot$ (motivated by the best-fit values for BOSS CMASS \citep{Alam17} and DESI LRGs \citep{Yuan23}), and allow changes for $\log M_{\rm min}$ and $\log M_{1}$, which respectively control the abundance of central and satellite galaxies for given halo mass. We find that the data cannot constrain the model well with all parameters free due to parameter degeneracies. The parameter $f_c$ denotes the observed cental galaxy fraction. The number density distribution is modeled as central galaxies placed at center of halos and satellite galaxies distributed following the \ac{NFW} \cite{NFW} profile:
\begin{align}
    \langle N(r,a|M)\rangle&=\langle N_c(a|M)\rangle\left(f_c +u_s(r,a|M)\langle N_s(a|M)\rangle\right),\\
    u_s(r,a|M)&\propto \frac{1}{\frac{r}{r_s}\left(1+\frac{r}{r_s}\right)^2},
\end{align}
where $r_s$ is the \ac{NFW} scale radius. We adopt concentration-mass relation by~\citet{ishiyama21_cM} to for the calculation of the \ac{NFW} scale radius.

For galaxy autocorrelation, there is a non-trivial second moment of $\langle \delta g(k,a|M)^2\rangle$ in the 1-halo term, and the expression for the Fourier space second moment is:
\begin{equation}
    \langle N^2(k,a|M)\rangle=\langle N_c(a|M)\rangle\left[2f_c\langle N_s(a|M)\rangle u_s(k,a|M)+\left(\langle N_s(a|M)u_s(k,a|M)\rangle\right)^2\right].
\end{equation}

For the Low-z and Mid-z unWISE catalogs used in this work, the redshift distribution is relatively broad as shown in Fig.~\ref{fig:unWISEsample1}, and cannot be simplified as tomographic bins of similar redshift. Thus we model the parameters $\log_{10}M_{\min},\log_{10}M_{1}$ as redshift dependent, with linear relation to the scale factor $a$ \citep{Nicola20} 
\begin{align}
    \log_{10}M_{\min}(a) &=\log_{10}M_{\min,0}+(a-1.0)\frac{d\log_{10}M_{\min}}{da}\\
    \mu(a) &=\mu_0+(a-1.0)\frac{d\mu}{da},
\end{align}
where we define the parameter $\mu=\log_{10}{(M_{1}/M_{\min})}(z)$ as the mass difference between central and satellite thresholds, and set the derivatives as extra free parameters.

We constrain the parameter redshift evolution by abundance-matching the DESI-calibrated redshift distribution $dN/dz$. For each given set of HOD parameters $\mu_0,\frac{d\mu}{da}$, we solve for $\log_{10}M_{\min}(z)$ that the HOD number density matches the observed galaxy number density $n_g(z)$, and then fit this relation with \texttt{pyccl}'s linear-in-scale-factor parametrization. The resulting HOD is used to compute $P_{gg}$ and project to $C_\ell^{gg}$ to compare with measured galaxy clustering. This approach incorporates the redshift-dependent color-luminosity selection of unWISE samples while reducing the degeneracy between abundance and clustering constraints.\footnote{%
    We fix the central-galaxy fraction to $f_c=1$, corresponding to the assumption that all central galaxies satisfying the sample selection are observed. This choice is motivated by the strong degeneracy between $f_c$ and $\mu=\log_{10}(M_1/M_{\min})$, as both parameters regulate the relative contributions of central and satellite galaxies. Allowing $f_c$ to vary with a physically motivated uniform prior, $0<f_c<1$, yields a prior-bound best-fit value of $f_c=1$ for the Low-z galaxy sample. For Mid-z galaxy sample, $f_c\neq1$ yields a $\Delta \chi^2=0.3$ improvement of best-fit and is not supported by BIC. We therefore fix $f_c=1$ in the fiducial analysis.
}

\begin{figure}[h]
    \centering

    \subfloat[
        The unWISE Low-z $C_\ell^{gg}$ measurement compared with the
        best-fit HOD model.
    ]{%
        \includegraphics[
            width=0.46\textwidth
        ]{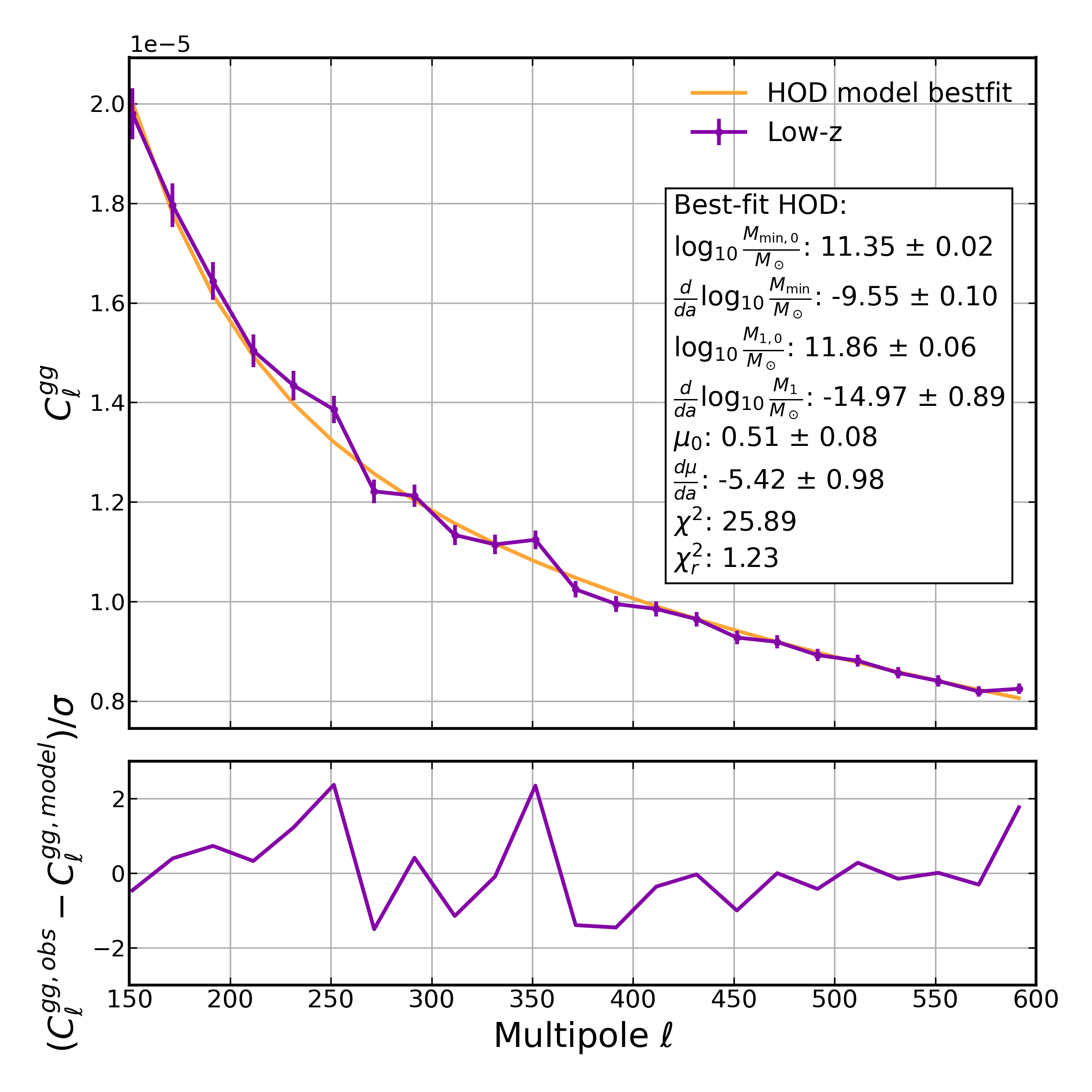}%
        \label{fig:lowzgghodfit}%
    }
    \hfill
    \subfloat[
        The unWISE Mid-z $C_\ell^{gg}$ measurement compared with the
        best-fit HOD model.
    ]{%
        \includegraphics[
            width=0.46\textwidth
        ]{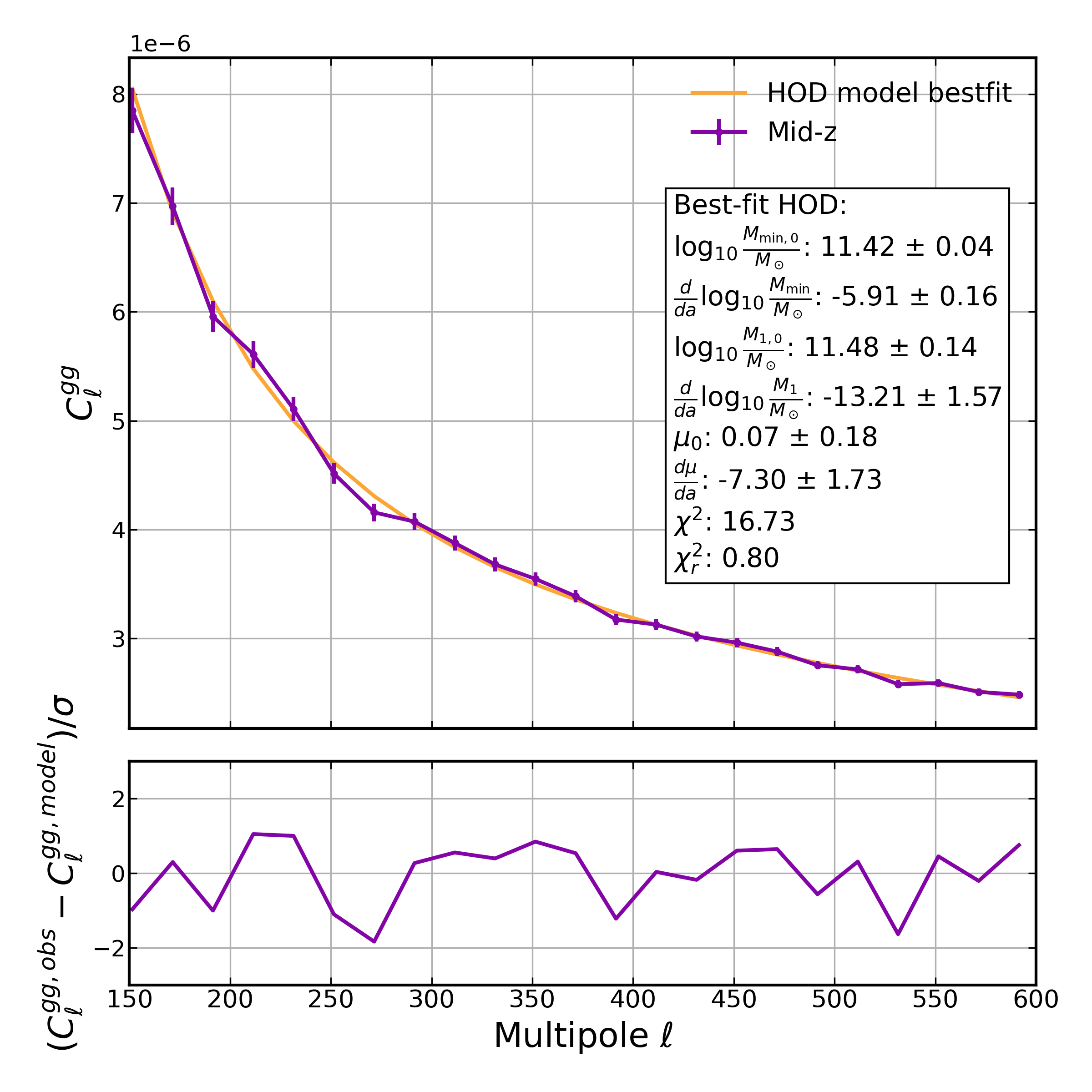}%
        \label{fig:midzgghodfit}%
    }

    \caption{
        HOD calibration fits for the unWISE galaxy samples used in this
        work. We fit the logarithmic mass difference between the central
        and satellite mass parameters,
        $\mu=\log_{10}(M_1/M_{\min})$, and parameterize its time evolution
        as
        $\mu(a)=\mu_0+(d\mu/da)(a-1)$.
        The central and satellite mass thresholds are determined by
        matching the HOD galaxy number density to the measured number
        density in each redshift bin. We then fit $\mu_0$ and $d\mu/da$
        to the measured galaxy clustering spectra $C_\ell^{gg}$.
        The uncertainties are estimated from the Fisher matrix.
    }
    \label{fig:gghod_fits}
\end{figure}

We fit the \ac{HOD} model by a residual $\chi^2$ defined as:
\begin{equation}
    \chi^2=\left[\hat{C}_\ell^{gg}-W_\ell C_{\ell,\rm hod}^{gg}\right]^{\rm T}\mathbf{Cov}^{-1}\left[\hat{C}_\ell^{gg}-W_\ell C_{\ell,\rm hod}^{gg}\right].
\end{equation}
The fitting results are shown in Fig.~\ref{fig:gghod_fits}.
The galaxy clustering auto power spectrum and covariance matrices are computed with \texttt{NaMaster} for both galaxy samples with a scale cut of $150<\ell<600$, and binned with $\Delta \ell = 20$. The large-scale $\ell_{\textrm{min}}$ is motivated both by the Limber approximation, and more importantly, by excess power on large scales due to Galactic foregrounds in the unWISE catalogs--this is consistent with previous unWISE analyses that used $\ell_{\textrm{min}} = 100$ \citep{krolewski_unwise_2020}.
We find additionally that using $\ell_{\textrm{min}} = 150$ improves the goodness-of-fit of the autocorrelation model, suggesting mild contamination in the auto-spectrum at $100 < \ell < 150$. The best-fits HOD parameters and the goodness of fit values for Low-z and Mid-z are listed in Tab.~\ref{tab:gghod_fits}.
\begin{table}[ht]
\centering
\begin{tabular}{lcccccc}
\hline
Sample & $\log_{10}(M_{\rm min,0}/M_\odot)$&$\frac{d}{da}\log_{10}(M_{\rm min}/M_\odot)$ & $\log_{10}(M_{1,0}/M_\odot)$ & $\frac{d}{da}\log_{10}(M_1/M_\odot)$ &$\chi^2$ & $\chi^2_r$ \\
\hline
Low-z  & $11.35\pm 0.02$& $-9.55\pm 0.10$ & $11.86\pm 0.06$ & $-14.97\pm 0.89$ & $25.89$  & $1.23$\\
Mid-z  & $11.42\pm 0.04$& $-5.91\pm 0.16$ & $11.48\pm 0.14$ & $-13.21\pm 1.57$ & $16.73$  & $0.80$\\
\hline
\end{tabular}
\caption{The best-fit parameter values for Low-z and Mid-z and the goodness of fit respectively.}
\label{tab:gghod_fits}
\end{table}

The underlying halo sample mass distribution is inferred from HOD parameters by:
\begin{equation}
    p(M_h)\propto \langle N_c(M) (1+N_s(M))\rangle,
\end{equation}
where $N_c$ and $N_s$ are defined by equations~\ref{eq:hod_nc} and~\ref{eq:hod_ns}. The inferred halo mass $1\sigma$ distribution is $\log_{10}M_h/M_\odot=12.97^{+0.74}_{-0.54}$ for Low-z and $\log_{10}M_h/M_\odot=12.96^{+0.64}_{-0.51}$. This indicates that Low-z and Mid-z are tracing a very similar population of halos at different mean redshift.

\subsection{Pressure profile fitting with GNFW}\label{sec:GNFW_model}

To model the electron pressure profile $\langle P_e(k,a|M)\rangle$, we adopt a generalized \ac{NFW} (GNFW) parametrization by \citet{ArnaudGNFW}. The physical space profile is modeled as:
\begin{equation}
    P_e(r)=1.65h_{70}^2\left(\frac{H(z)}{H_0}\right)^{8/3}\left[\frac{h_{70}M_{500c}}{3\times10^{14}M_\odot}\right]^{2/3+\alpha_P}\rm eV\cdot cm^{-3}\times\mathbbm{p}(x),
\end{equation}
where $x=r/R_{500}$ and:
\begin{equation}
    \mathbbm{p}(x) = P_0h_{70}^{-1}(c_{500}x)^{-\gamma}[1+(c_{500}x)^\alpha]^{(\gamma-\beta)/\alpha},\label{eq:gnfw_formfactor}
\end{equation}
is the pressure profile form factor. We reduce the fitted variables to two: the pressure profile normalization $P_0$ and GNFW outer-slope parameter $\beta$. We fix all other parameters using best-fit values from \citet{SPTSZ2023} measurement of $z<0.6$ galaxy clusters using joint measurement of SPT-SZ and Planck. We list the definitions and values of fixed parameters, as well as definition and fitting bounds of fitted parameters in Tab.~\ref{tab:pressureprofile_params}.

In galaxy-$y$ cross-correlation, the 1-halo average of galaxy $\times$ tSZ power spectrum is:
\begin{equation}
    \langle \delta_g(k,a|M)P_e(k,a|M)\rangle=(1+\rho_{gy})\langle \delta_g(k,a|M)\rangle \langle P_e(k,a|M)\rangle,
\end{equation}
where $\rho_{gy}$ is a phenomenological parameter quantifying the covariance between galaxy distribution and electron pressure profile distribution in an ensemble of halos, following the convention of \texttt{pyccl} \citep[i.e.][]{Koukoufillipas20,Yan21}. In our analysis of ACT + \textit{Planck} $\times$ unWISE, a scale cut $\ell_{\min}=200$ corresponds $k=\ell/\chi\sim0.4 \rm Mpc^{-1}$ for Low-z and $\sim 0.3 \rm Mpc^{-1}$ for Mid-z, and thus limits the 2-halo contribution. In a 1-halo dominated regime, the pressure profile normalization is completely degenerate with the $\rho_{gy}$ parameter. In this work, we absorb the $1+\rho_{gy}$ into $P_0$, corresponding to an effective 1-halo pressure normalization $P_{0,\rm eff}=(1+\rho_{gy})P_0$.

\begin{table}[ht]
\centering
\begin{tabular}{llcc}
\hline
Quantity & definition  & fit bounds/fixed values \\
\hline
$P_0$&Profile normalization, converted to electron pressure. & [0,40]\\
$c_{500}$&Concentration parameter.&1.07\\
$\alpha$&Generalized \ac{NFW} shape parameter.&1.05\\
$\beta$&Generalized \ac{NFW} shape parameter.&[0,20]\\
$\gamma$&Generalized \ac{NFW} shape parameter.&0.55\\
$\alpha_P$&Additional mass dependence parameter.&0.12\\
\hline
\end{tabular}
\caption{The parameters involved in the Generalized NFW electron pressure profile. The parameters $P_0$ and $\beta$ are marginalized while keeping the other parameters fixed. The fixed values are taken consistent with $z\leq 0.6$ sample of \citet{SPTSZ2023}. Note that the pressure normalization by \citet{SPTSZ2023} is based on gas thermal pressure, and the value $P_0$ shown here are converted into electron pressure profile using the \texttt{pyccl} normalization convention.}

\label{tab:pressureprofile_params}
\end{table}

We use the harmonic space component separated galaxy-tSZ cross-correlation results from Sec.~\ref{sec:method}. We choose the optimal signal recipe tSZ + Radio + CIB-amplitude with the best-fit parameters for the ACT+\textit{Planck} $\times$ unWISE sample.

\begin{figure}[htbp]
    \centering

    \subfloat[
        The ACT+\textit{Planck} $\times$ unWISE Low-z
        cross-correlation and best-fit GNFW model.
    ]{%
        \includegraphics[
            width=0.46\textwidth
        ]{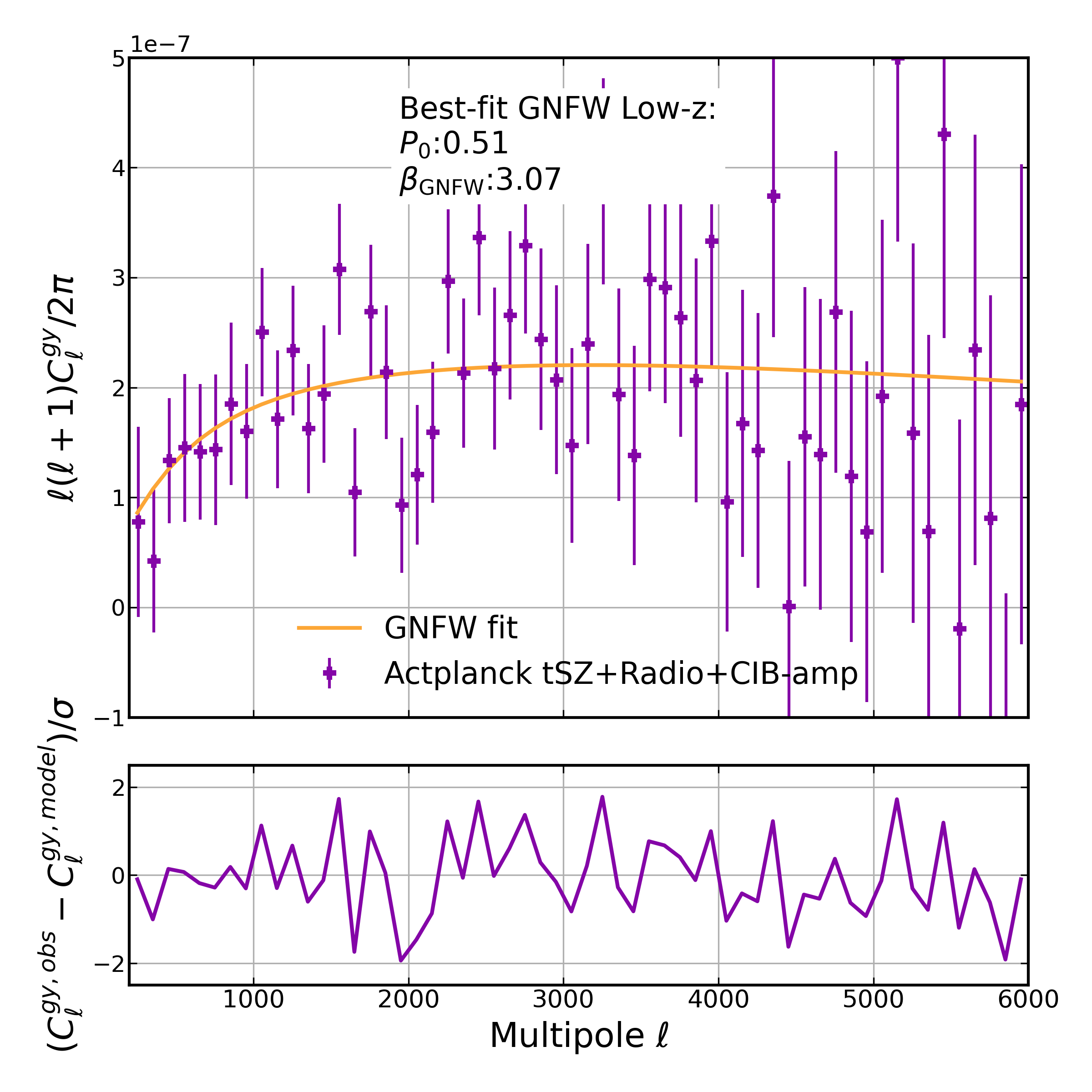}%
        \label{fig:lowzgyhodfits}%
    }
    \hfill
    \subfloat[
        The ACT+\textit{Planck} $\times$ unWISE Mid-z
        cross-correlation and best-fit GNFW model.
    ]{%
        \includegraphics[
            width=0.46\textwidth
        ]{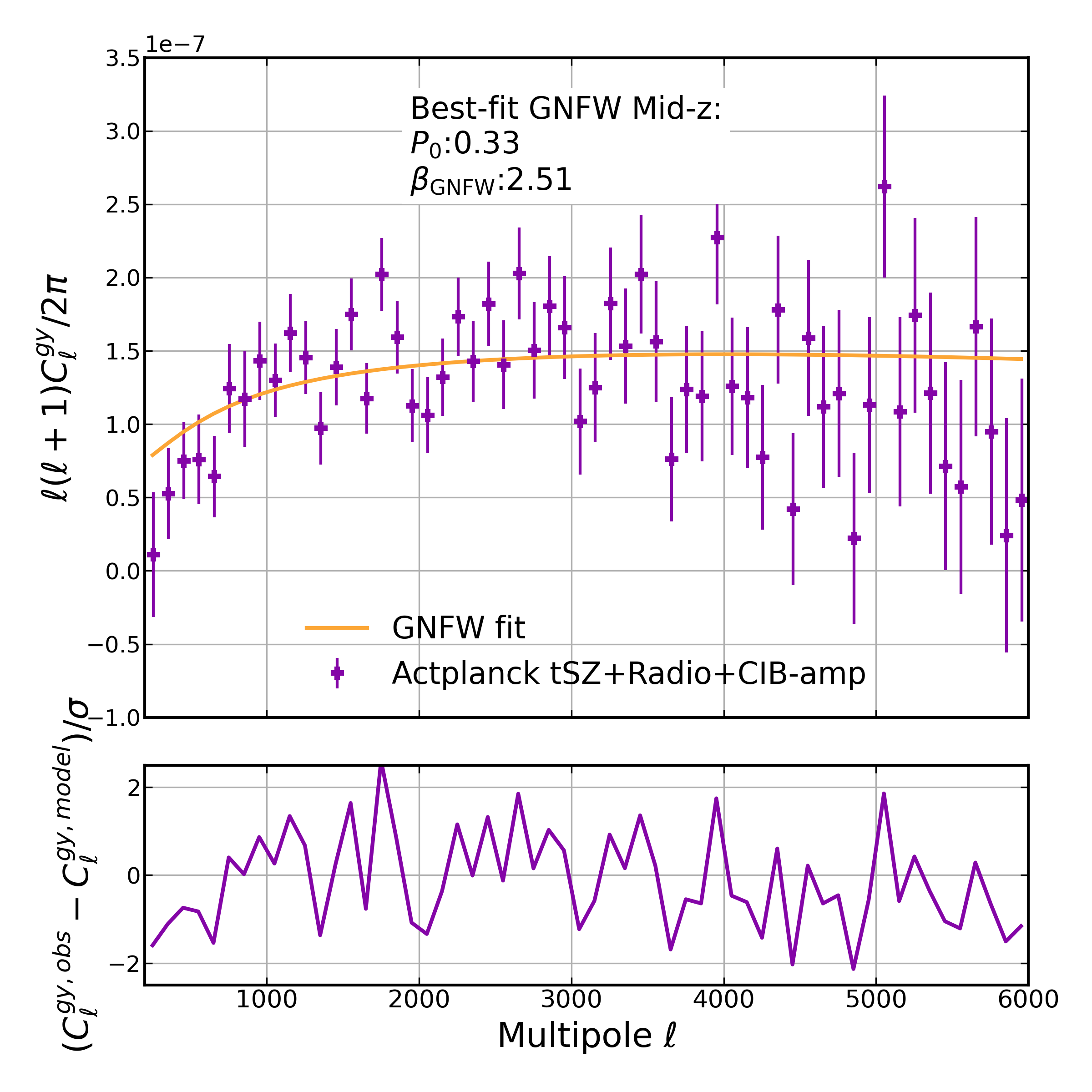}%
        \label{fig:midzgyhodfits}%
    }

    \caption{
        Component-separated ACT+\textit{Planck} $\times$ unWISE
        cross-correlations obtained using the best-fit SED parameters
        derived from \textit{Planck}, as listed in
        Table~\protect\ref{tab:plk_model_fitting_results}.
        A generalized Navarro--Frenk--White (GNFW) profile is fitted to
        each $C_\ell^{gy}$ measurement to constrain the pressure-profile
        normalization $P_0$ and the GNFW outer-slope parameter
        $\beta_{\rm GNFW}$. The remaining GNFW parameters are fixed to
        the values measured for the $z\leq 0.6$ sample by
        \protect\citet{SPTSZ2023}. The residuals are shown in the lower
        panels.
    }
    \label{fig:gyhodfits}
\end{figure}

\begin{figure}[htbp]
    \centering

    \subfloat[
        The ACT+\textit{Planck} $\times$ unWISE Low-z GNFW-fit
        parameter posterior distribution.
    ]{%
        \includegraphics[
            width=0.25\textwidth
        ]{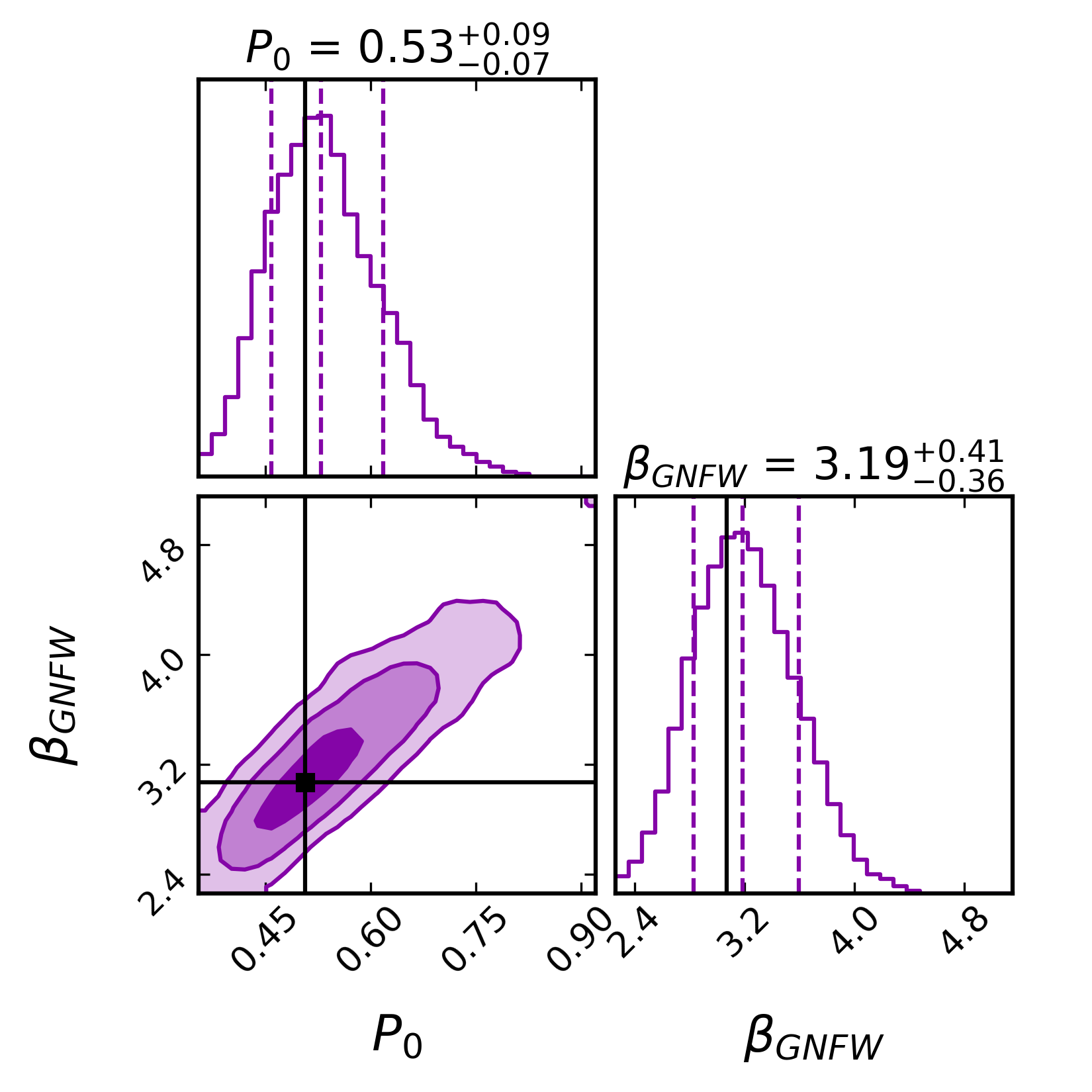}%
        \label{fig:lowzgnfwchain}%
    }
    \hfill
    \subfloat[
        The ACT+\textit{Planck} $\times$ unWISE Mid-z GNFW-fit
        parameter posterior distribution.
    ]{%
        \includegraphics[
            width=0.25\textwidth
        ]{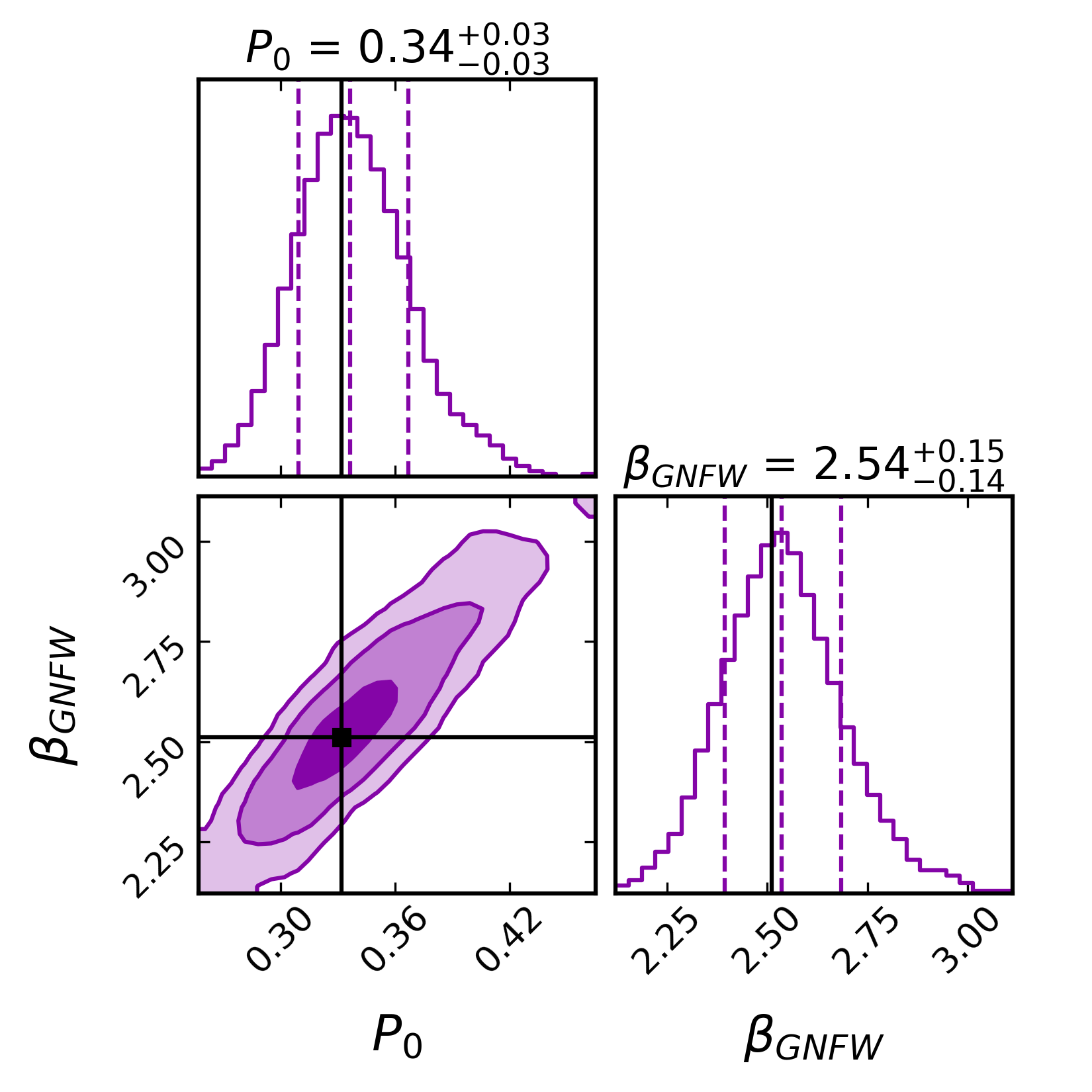}%
        \label{fig:midzgnfwchain}%
    }
    \hfill
    \subfloat[
        Posterior predictions for the pressure-profile form factor
        $\mathbbm{p}(x)$. The best-fit realizations are shown as solid
        lines. The result for the $z<0.6$ SPT sample from
        \protect\citet{SPTSZ2023} is shown for comparison.
    ]{%
        \includegraphics[
            width=0.45\textwidth
        ]{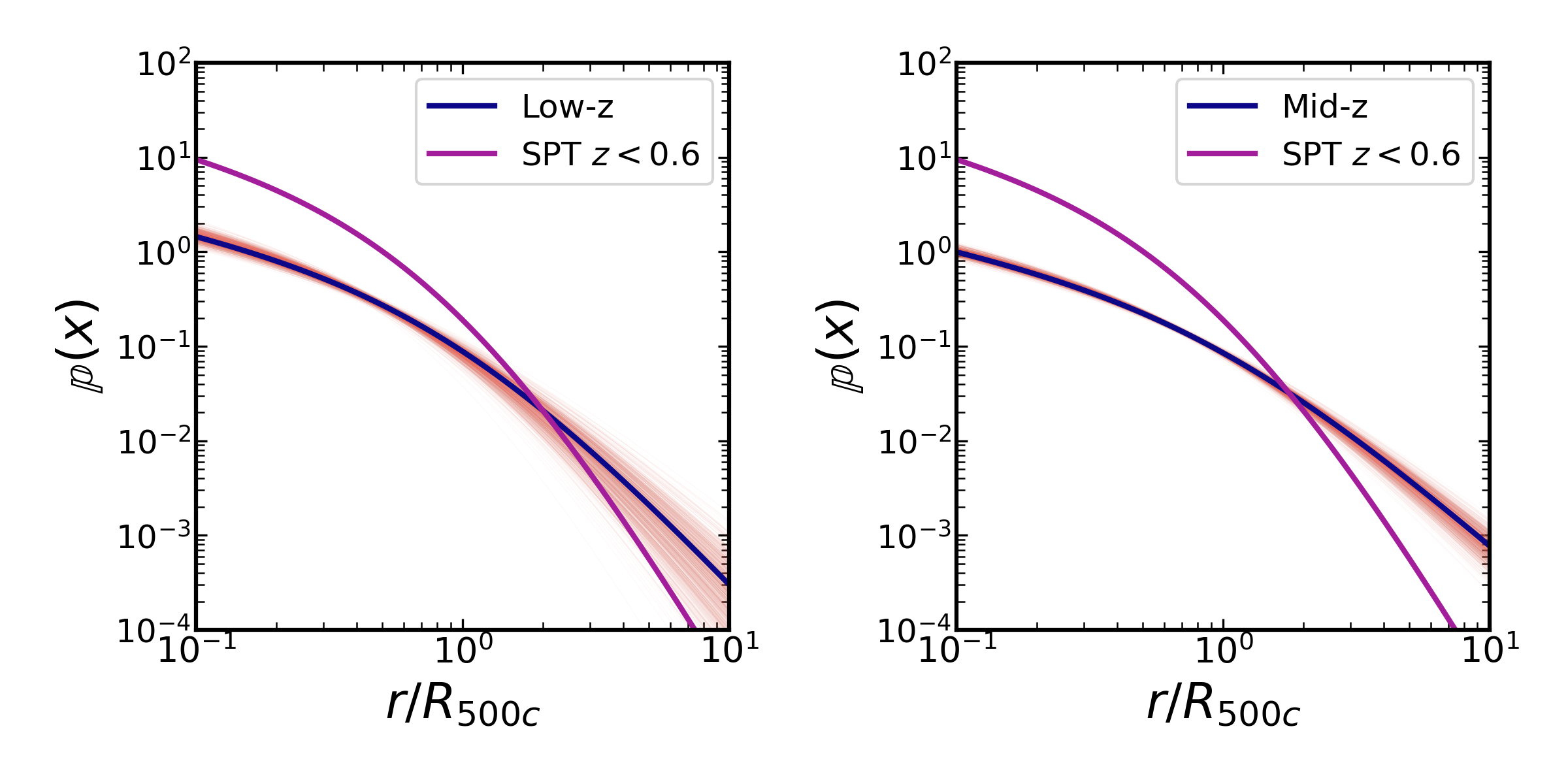}%
        \label{fig:lowzmidz_pereal}%
    }

    \caption{
        Posterior distributions of the generalized NFW electron-pressure
        profile inferred from the component-separated $C_\ell^{gy}$.
        The left two panels show the parameter posteriors sampled using
        \texttt{emcee}, while the right panel shows the posterior
        predictions for the real-space pressure-profile form factor
        defined in Equation~\protect\ref{eq:gnfw_formfactor}.
    }
    \label{fig:gyhod_posteriors}
\end{figure}

We adopt the best-fit \ac{HOD} parameters from Sec.~\ref{sec:HOD_model} measured from the autocorrelation of the Low-z and Mid-z samples. Parameter fits to $C_\ell^{gy}$ for both Low-z and Mid-z are shown in Fig.~\ref{fig:gyhodfits}. We use the MCMC package \texttt{emcee} \citep{emcee} to sample the posterior distribution of parameters $P_0,\beta$. We show the posterior distributions and posterior predicted samples of electron pressure profile form factor $\mathbbm{p}(x)$ of equation~\ref{eq:gnfw_formfactor} in Fig.~\ref{fig:gyhod_posteriors}. The posterior estimates are 
$P_0=0.53^{+0.09}_{-0.07}$ and $\beta_{\rm GNFW}=3.19^{+0.46}_{-0.36}$ for Low-$z$, 
and $P_0=0.34^{+0.03}_{-0.03}$ and $\beta_{\rm GNFW}=2.54^{+0.15}_{-0.14}$ for Mid-$z$. Relative to the $z<0.6$ subsample of~\citet{SPTSZ2023}, our inferred normalization $P_0$ is lower by $2.3\sigma$ for Low-$z$ and $2.5\sigma$ for Mid-$z$, while the outer-slope parameter $\beta_{\rm GNFW}$ is lower by $3.1\sigma$ and $4.0\sigma$ respectively, implying different gas properties of the unWISE selected samples from cluster samples of~\citet{SPTSZ2023}.
This leads to a lower central pressure and more extended pressure profile for the unWISE samples as compared to the clusters.
This is consistent with the emerging picture of stronger feedback affecting the gas distribution around $\sim10^{13}$ $M_\odot$ halos than in cluster-mass halos \citep{Bigwood24,Bigwood25,Hadzhiyska25b,Hadzhiyska25a,RiedGuachalla25,Qu26,Siegel26b,Siegel26a}.

Since we marginalize only a subset of the GNFW parameters, direct interpretation of the fitted $P_0$ and $\beta_{\rm GNFW}$ is subject to systematic uncertainty from the fixed pressure-profile parameters. Moreover, the one-halo galaxy $\times$ \ac{tSZ} amplitude constrains the effective combination $(1+\rho_{gy})P_0$, rather than the intrinsic pressure normalization $P_0$ alone. We therefore present the corresponding $\tilde Y_{500}-M_h$ relation in Appendix~\ref{sec:ym_relation} only as a model-dependent diagnostic. This comparison illustrates how the inferred SZ amplitude depends on the HOD-based halo mass assignment and on the assumed galaxy--pressure correlation, but it should not be interpreted as an independent constraint on the $Y_{\rm SZ}-M_h$ scaling relation.



\section{Conclusion}

Our result using galaxy-single frequency cross-correlations shows that the multi-frequency SED fitting to the \textit{Planck} cross-spectra favors a three-component recipe consisting of tSZ, radio emission, and a CIB amplitude term, using the broad frequency availability of \textit{Planck}. By BIC standards, the preference against a radio-less recipe is $9.5\sigma$ for Low-z galaxy sample and $11\sigma$ for Mid-z galaxy sample. 

To extract a physically meaningful galaxy-tSZ cross-correlation, the \textit{Planck} analysis sets the preferred foreground model for necessary foreground components before moving to higher-resolution data from ACT. 
In recipes that omit radio contamination, the recovered small-scale reconstructed $g \times y$ signal can become strongly suppressed or even spuriously negative. With the preferred SED recipe from \textit{Planck} cross-correlations, tSZ + Radio + CIB-amplitude, the spurious negative behavior is removed. By including the radio term, the reconstructed $g \times y$ spectrum remains physically sensible on small angular scales and can be described by a standard halo-model interpretation. We therefore conclude that radio contamination is not a subdominant correction in this analysis, but a necessary component of the cleaning procedure.\footnote{As an alternative approach, \citet{LiustackACTtsz} mask the central region of each galaxy approximately within $R_{\rm vir}$. This yields consistent 90 and 150 GHz profiles in Compton-$y$ units and a 220 GHz signal consistent with zero, showing that masking is effective at removing CIB and radio contamination, which tend to be more concentrated around the galaxy than the tSZ.}

The comparison with NILC-based analyses further clarifies this point. As shown in Appendix~\ref{appendix:NILC}, the harmonic-space component-separation result obtained with the tSZ + CIB + CIB-$\delta\beta$ recipe is consistent with the CIB-deprojected NILC cross-correlation from \cite{Coultontszcomponent}. This agreement suggests that the NILC-based result effectively tracks a cleaning scheme in which dust-related contamination is treated, but radio emission is not explicitly accounted for. Combined with the result that \textit{Planck} model-selection favors tSZ + radio + CIB-amplitude, and that this recipe removes the negative small-scale feature in ACT+Planck, this altogether implies that NILC-based $g \times y$ analyses in ACT DR6 should be interpreted with caution when used for small-scale cosmological or halo-model inference. In particular, agreement with CIB-deprojected NILC alone should not be taken as evidence that the recovered signal is free from radio-induced bias, especially for low mass and low redshift halo populations.



Importantly, the components treated here as contaminants (CIB, Radio) to the galaxy-tSZ signal also contain potentially valuable information about the astrophysics of galaxies and their surrounding environments, including star-formation activity, non-thermal emission, and feedback processes. In the present work, we focus primarily on the analysis of the galaxy-tSZ signal. However, with a more robust model of the halo profiles, these same cross-correlations for CIB and Radio will be more physically informative. An important direction for future work will be to investigate how these components can be used together with tSZ to constrain feedback and baryonic processes in low-mass halos. In addition, the current modeling of Radio signal as an effective spectral index could be further improved to extract physical information about the Radio contributions.

An important limitation of our analysis is the limited frequency in ACT. The limitation to three nominal bands restricts the maximum component to three, and cannot differentiate between different CIB and Radio spectral parameters. The forced application of \textit{Planck} best-fit spectral parameters to ACT + \textit{Planck} analysis in Sec.~\ref{sec:component_separation_res} is based on the assumption that the radio source-removal and masking level around the unWISE galaxy samples are consistent across the two surveys. This analysis will benefit greatly from the low noise and broad wavelength coverage of future CMB observatories such as Simons Observatory~\citep{SimonsObservatory_Scigoals}, which observes in 6 nominal bands from 23GHz to 280GHz.

\section*{Acknowledgements}

We thank Simone Ferraro for suggesting that we explicitly fit foreground templates to the Planck single-frequency maps as an alternative to the publicly available Compton-$y$ maps.
We thank Fiona McCarthy for the critical advice on the CMB single frequency maps.
We also thank Raphael Bertrand-Delgado, David S\'anchez Cid, and Felipe Andrade-Oliveira for discussions on the halo modeling. 

GZ acknoweldges Perimeter Institute and organizers of Cosmic Ecosystem conference for hosting him and stimulating discussions. GZ acknowledges Will Coulton and Henry Liu for fruitful discussions during the conference.
NA would like to acknowledge Natacha Altamirano and the late Chiamaka Okoli, whose collaboration was the original inspiration for this work. 
AK was supported as a CITA National Fellow by the Natural Sciences and Engineering Research Council of Canada (NSERC), funding reference \#DIS-2022-568580.

This research is greatly supported by computational resources from ETH Z\"urich Department of Physics, including the Euler cluster and the Department of Physics ISG cluster.
This research is partially supported by the Natural Sciences and Engineering Research Council of Canada and Perimeter Institute for Theoretical Physics. Research at Perimeter Institute is supported in part by the Government of Canada through the Department of Innovation, Science and Economic Development Canada and by the Province of Ontario through the Ministry of Colleges and Universities. Parts of this paper are based on observations obtained with \href{https://www.esa.int/Planck}{\textit{Planck}}, an ESA science mission with instruments and contributions directly funded by ESA Member States, NASA and Canada. 

\bibliographystyle{aa.bst}
\bibliography{main}

\appendix



\section{ACT passband selection influence on response}\label{sec:ACTpassbandcomp}

\begin{figure}[htbp]
  \centering

  \subfloat[
    The ACT passbands used in this work (solid lines), compared with
    those of other ACT instruments.
  ]{%
    \includegraphics[
      width=0.48\textwidth,
      trim=1cm 0.6cm 1cm 0.5cm,
      clip
    ]{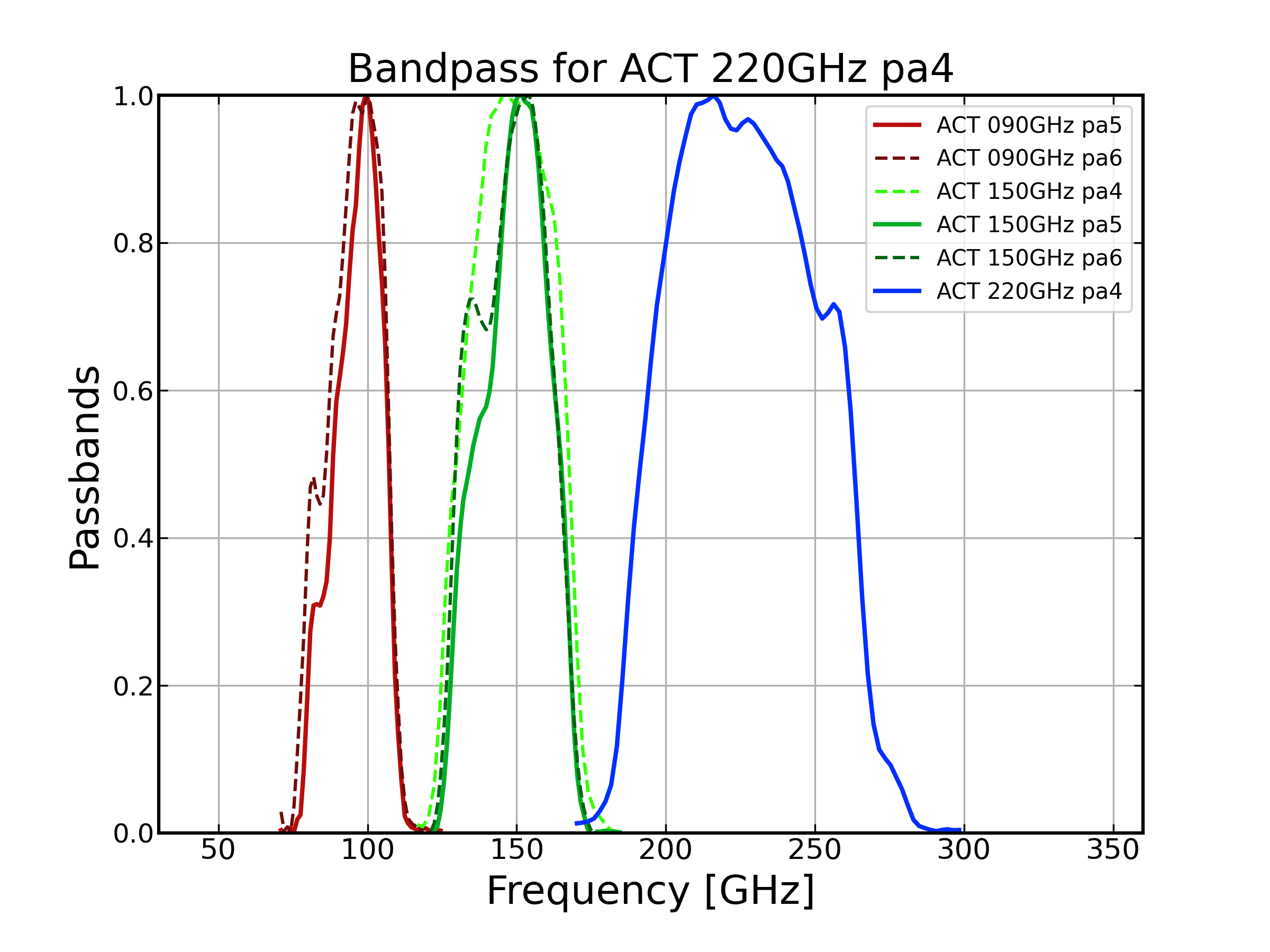}%
    \label{fig:actpasscompare}%
  }%
  \hfill
  \subfloat[
    The ACT band-averaged responses of the tSZ, free--free,
    CIB-amplitude, and CIB-$\delta\beta$ components evaluated using
    different ACT passbands, compared with the values adopted in this
    work.
  ]{%
    \includegraphics[
      width=0.48\textwidth
    ]{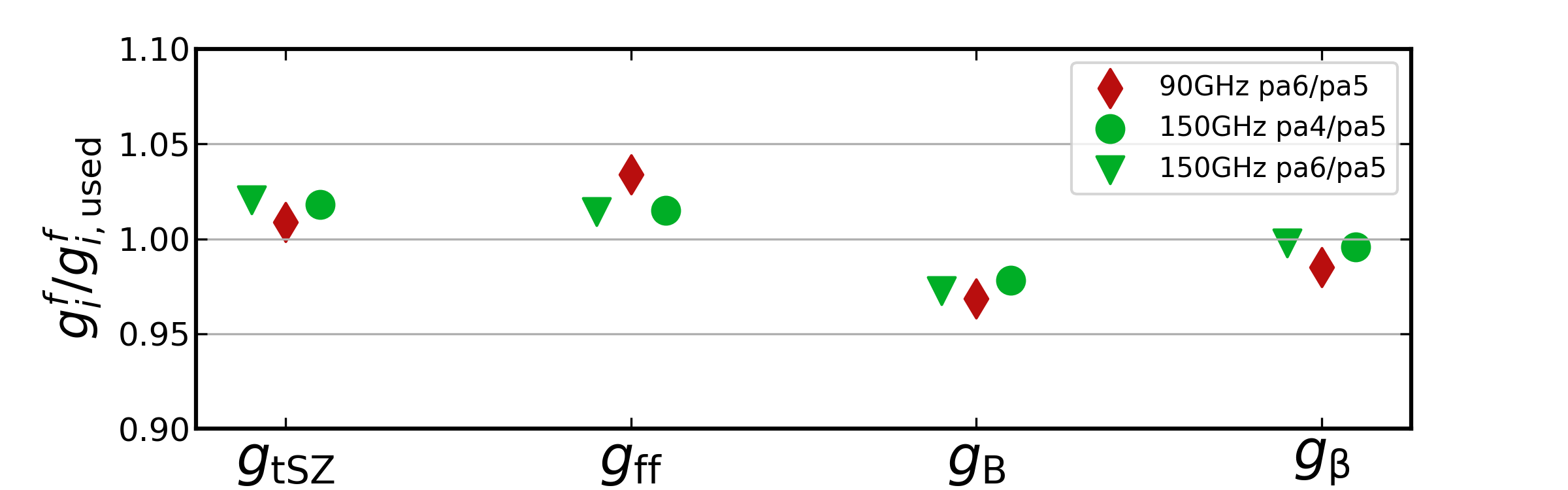}%
    \label{fig:actresponsecompare}%
  }

  \caption{
    Comparison of passband-induced response systematics.
  }
  \label{fig:actbandpassedappendix}
\end{figure}

The ACT DR6 released single frequency maps used in this work are coadded with multiple ACT instruments. For 90GHz nominal band, the ACT instruments involved are pa6 and pa5. For 150GHz nominal band, the ACT instruments involved are pa4, pa5 and pa6. In this work we adopt the passbands: pa5 for 90GHz and 150GHz, pa4 for 220GHz, based on the array with the highest sensitivity for a given band. The passbands of all the ACT instruments are shown in Fig.~\ref{fig:actbandpassedappendix}(a).

Since the signal response is a band averaged result of the response function, the slight difference of passbands induces systematics in the analysis. We examine the passbands-averaged response functions using equation \ref{eq:bandaverage}.


\begin{table}[h!]
    \centering
    \begin{tabular}{cccccc}
         \hline
         Nominal band & instrument   & $g_{\rm tSZ}$ & $g_{\rm ff}$&$g_{\rm B}$&$g_{\rm \beta}$ \\
         \hline
         90GHz & pa5 & -1.55 & 7.72 & 0.0750 &-0.0622\\
         90GHz & pa6 & -1.56 & 7.98 & 0.0727 &-0.0613\\
         150GHz& pa4 & -1.01 & 4.13 & 0.200  &-0.0798\\
         150GHz& pa5 & -0.98 & 4.07 & 0.204  &-0.0801\\
         150GHz& pa6 & -1.00 & 4.13 & 0.198 & -0.0800\\
         \hline
    \end{tabular}
    \caption{Band averaged response values on different ACT passbands.}
    \label{tab:actresponsespassbands}
\end{table}

\section{comparison with NILC deprojected results}\label{appendix:NILC}

\begin{figure}
    \centering
    \includegraphics[width=0.5\linewidth]{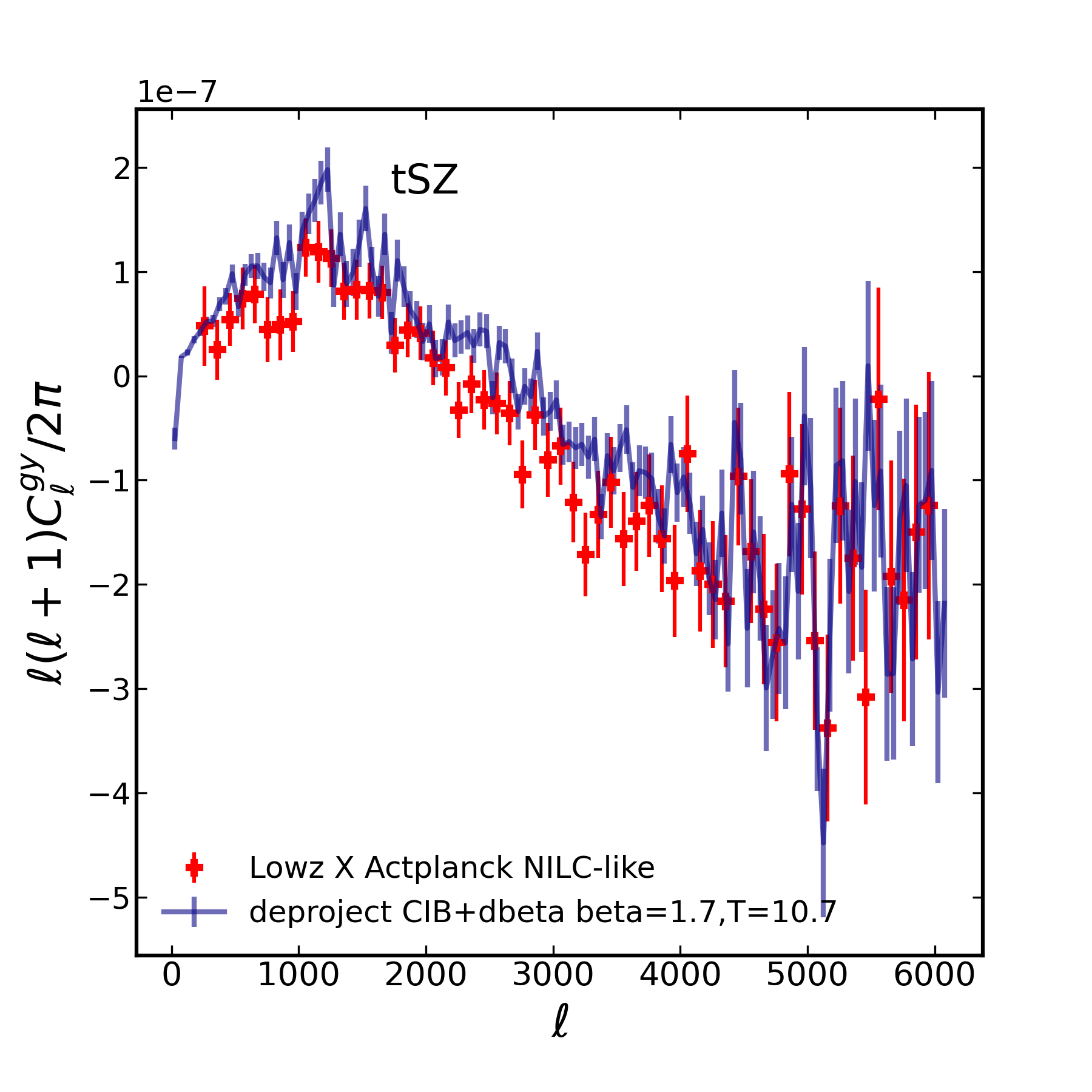}
    \caption{Comparison between ACT DR6 NILC deprojected cross-correlation, described in details by \citet{Coultontszcomponent}. The CIB SED parameters are the same between our component separation and ACT NILC.}
    \label{fig:actplanck_clean_vs_deproj}
\end{figure}

Here we present a comparison between our component separation ansatz with the ACT DR6 NILC map with deprojection. The NILC map used in this cross-correlation is provided by ACT DR6 \citep{Coultontszcomponent}, following the methodology of \cite{mccarthy_component}. We compare the deprojection map of CIB+CIB-d$\beta$ using spectral parameters $\beta_0=1.7,T_0=10.7$ to the harmonic-space component separation in this work using the same SED parameters. In the result shown in Fig.~\ref{fig:actplanck_clean_vs_deproj}, we show that our result is consistent with the galaxy-$y$ cross-correlation compared to NILC deprojected reconstruction up to multipole $\ell \sim 6000$. The uncertainty of our component separation is larger due to the added degrees of freedom in the CIB-amplitude and CIB-$\delta\beta$ (which we allow to vary in each bandpower rather than having a single spatial profile). Combining with Fig.~\ref{fig:actplanckcomp}, this is another strong argument that the negativity in the galaxy-$y$ cross-correlation using NILC deprojected maps is due to radio contamination.

\section{Y-M relation}\label{sec:ym_relation}

The $\tilde Y_{500}-M_h$ relation shown in this appendix should be interpreted as a model-dependent diagnostic because of two sources of ambiguity. First, the halo mass distribution inferred from the HOD fit is constrained only through galaxy clustering. In the standard HOD form used in Sec.~\ref{sec:halo_model_section}, the central occupation does not contain a strong high-mass cutoff. Therefore, without additional information such as weak-lensing masses, group catalogs, or cluster abundance constraints, the high-mass tail of the galaxy--halo connection is not tightly determined. This makes the effective halo mass entering a $Y_{\rm SZ}-M_h$ comparison sensitive to the adopted weighting scheme.

Second, the pressure-profile amplitude inferred from the galaxy--\ac{tSZ} cross-correlation is degenerate with the galaxy--pressure profile correlation parameter. At fixed pressure-profile shape, the one-halo contribution to $C_\ell^{gy}$ constrains the combination
\begin{equation}
P_{0,\rm eff}\equiv (1+\rho_{gy})P_0 ,
\end{equation}
rather than $P_0$ alone. Therefore, an inferred $Y_{\rm SZ}-M_h$ normalization based on $P_0$ is biased as $\rho_{gy}$ is absorbed into the pressure amplitude. With the \texttt{pyccl} GNFW convention $\alpha_P=0.12$, the mass dependence of the one-halo pressure contribution follows the same \textit{Planck} scaling as the input pressure model, and varying $P_0$ only changes the amplitude. Thus, the only directly constrained quantity relevant for this appendix is the effective amplitude $(1+\rho_{gy})P_0$.

For comparison with previous measurements, here we compute the integrated SZ signal corresponding to the best-fit pressure profile, with the assumption that $\rho_{gy}=0$. We adopt the same \ac{tSZ} flux normalization as in \citet{Lim2018ym} and spherically integrate the pressure profile to $R_{500}$:
\begin{align}
\tilde Y_{500}(M)
&=
Y_{500}(M,z)
\left[\frac{H(z)}{H_0}\right]^{-2/3}
\left(\frac{d_{\rm A}(z)}{500{\rm Mpc}}\right)^2,\\
d_A(z)^2Y_{500}(M,z)
&=
\frac{\sigma_T}{m_ec^2}
\int_0^{R_{500}}4\pi r^2\ dr\ P_e(r|M,z).
\end{align}
The halo mass distribution associated with each galaxy sample is estimated from the best-fit HOD model,
\begin{align}
p(M_h)
&\propto
\int dz,\frac{dN}{dz},
\left[N_c(M,a)+N_s(M,a)\right]n_h(M,a),
\label{eq}
\end{align}
where $n_h(M,a)$ is the halo mass function~\citep{tinker08hmf}, and $dN/dz$ is the DESI-calibrated redshift distribution shown in Fig.~\ref{fig:unWISEsample1}. This gives
$
\log_{10}(M_h/M_\odot)=12.98^{+0.74}_{-0.54}$
for Low-z, and
$
\log_{10}(M_h/M_\odot)=12.96^{+0.64}_{-0.51}
$
for Mid-z. These values indicate that the two samples occupy broadly similar HOD-weighted halo populations.

Because the SZ signal scales with halo mass approximately with $Y_{\rm SZ}\propto M_h^{5/3}$ under self-similar evolution, the same HOD distribution gives a much higher effective mass when weighted by the expected SZ contribution. After applying an $M_h^{5/3}$ weight and re-normalizing the distribution, we obtain
$
\log_{10}(M_h/M_\odot)=14.59^{+0.39}_{-0.52}
$
for Low-z, and
$
\log_{10}(M_h/M_\odot)=14.51^{+0.45}_{-0.57}
$
for Mid-z. Removing the satellite contribution gives similar values,
$
\log_{10}(M_h/M_\odot)=14.44^{+0.39}_{-0.55}
$
for Low-z, and
$
\log_{10}(M_h/M_\odot)=14.15^{+0.42}_{-0.56}
$
for Mid-z. This shows that the SZ-weighted effective mass is driven mainly by the high-mass tail allowed by the central HOD occupation, combined with the steep $M_h^{5/3}$ weighting, rather than by satellite galaxies alone.

\begin{figure}[htbp]
    \centering

    \subfloat[
        The $\tilde{Y}_{500}$--$M_h$ relations for the Mid-z and Low-z
        samples (blue and purple), compared with external measurements
        from \textit{Planck} LBGs
        \protect\citep{PlanckLBG2013} and
        \textit{Planck} $\times$ galaxy groups
        \protect\citep{Lim2018ym} (orange and red).
        The halo-mass distributions inferred from the HOD models are
        indicated by the shaded dashed curves.
    ]{%
        \includegraphics[
            width=0.4655\textwidth
        ]{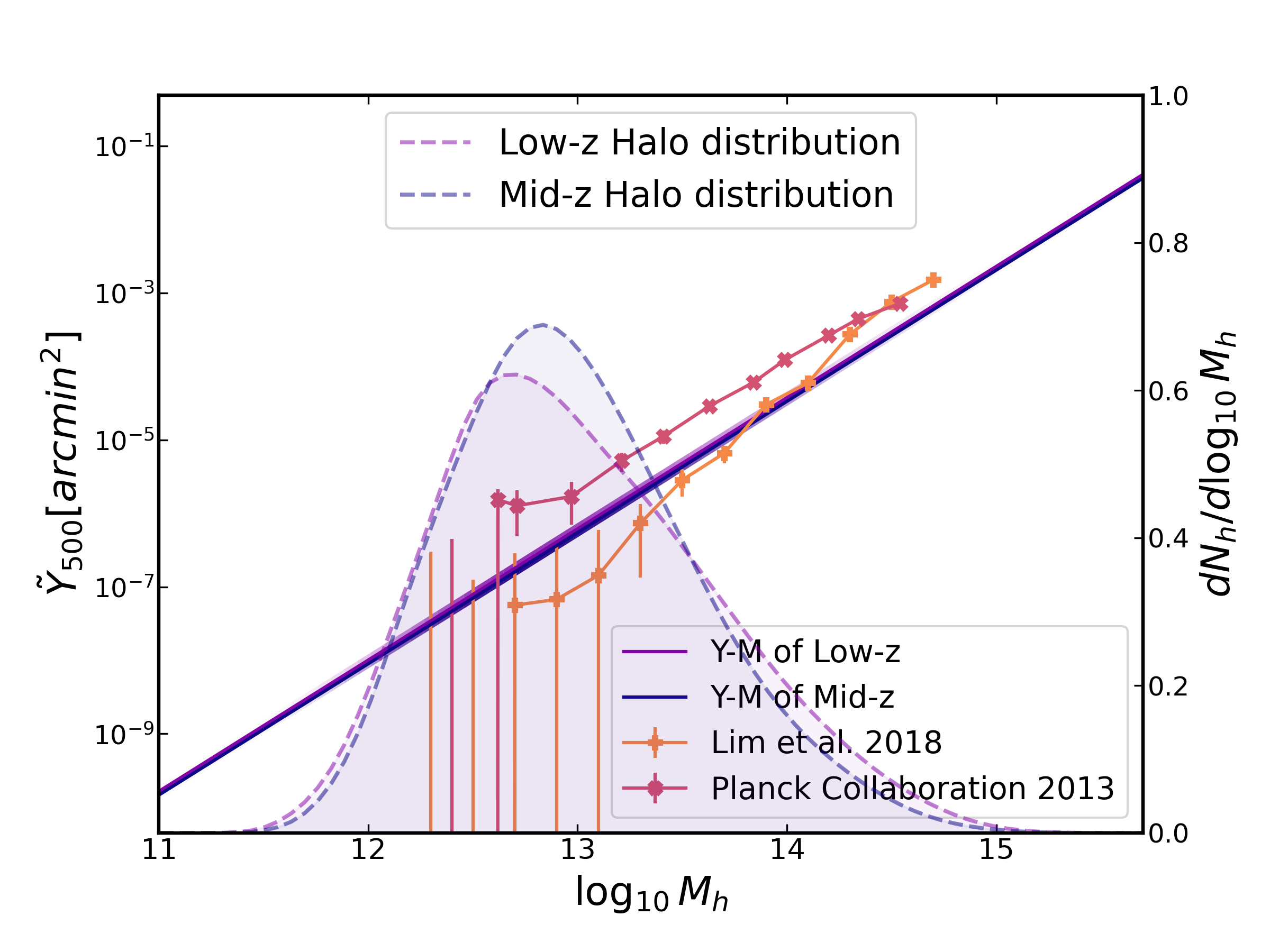}%
        \label{fig:ymrelation_mhalo}%
    }
    \hfill
    \subfloat[
        The $\tilde{Y}_{500}$--$M_h$ relations for the Mid-z and Low-z
        samples. The halo-mass distributions are weighted by the
        self-similar scaling $M_h^{5/3}$ to approximate their
        contributions to the SZ signal. The weighted distributions are
        normalized to unit area.
    ]{%
        \includegraphics[
            width=0.4655\textwidth
        ]{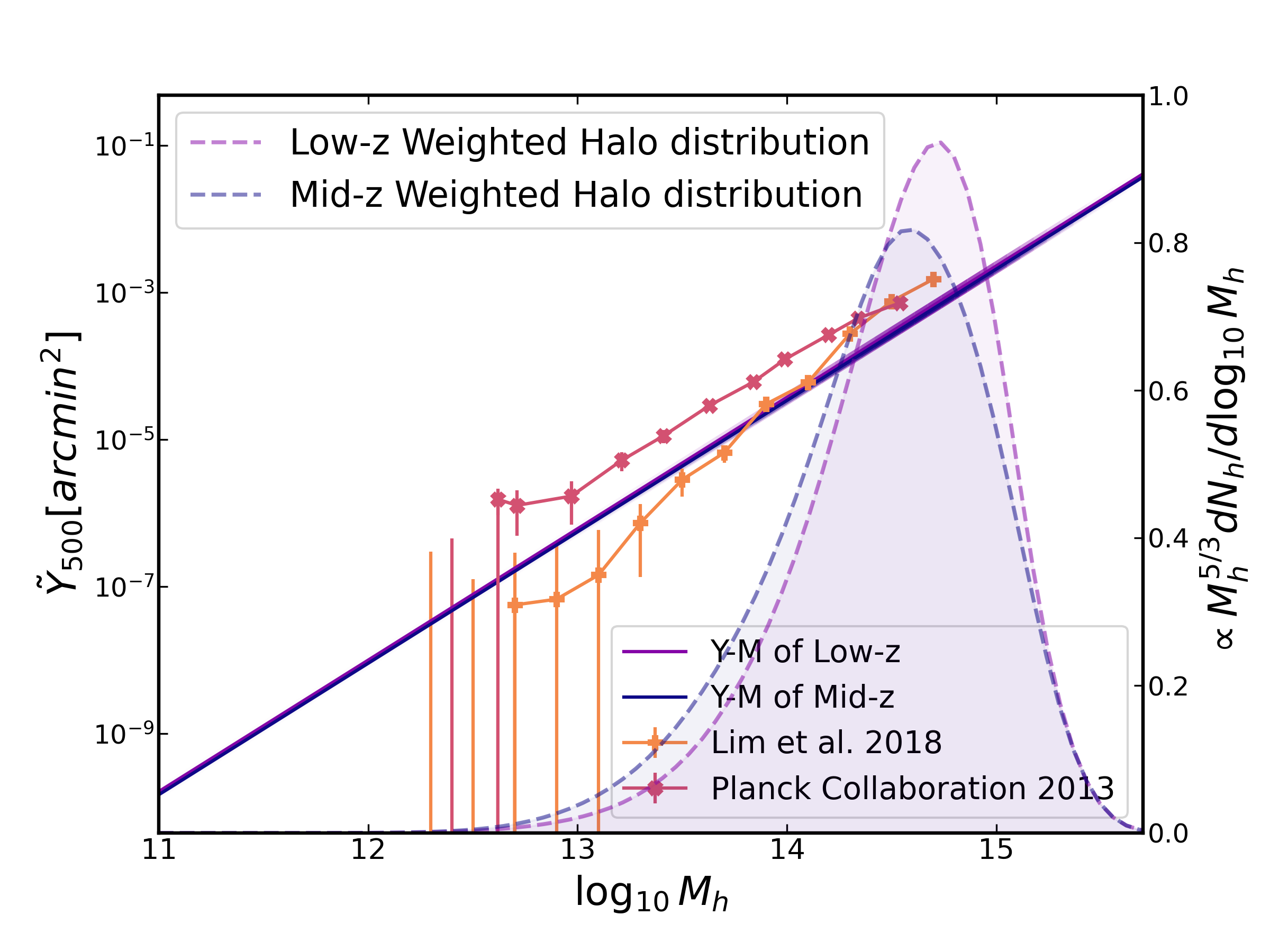}%
        \label{fig:ymrelation_mhalo53}%
    }

    \caption{
        Model-dependent comparison of the
        $\tilde{Y}_{500}$--$M_h$ relations for the Low-z and Mid-z
        samples. The left panel uses the unweighted HOD halo-mass
        distribution, while the right panel uses an
        $M_h^{5/3}$-weighted distribution to illustrate the halo masses
        that dominate the expected SZ flux. The shaded regions around
        the Low-z and Mid-z relations show the $1\sigma$ uncertainties
        from the GNFW fit. Because the one-halo cross-correlation
        constrains the effective amplitude $(1+\rho_{gy})P_0$, and
        because the HOD model does not impose a strong high-mass cutoff,
        this figure should be interpreted as a sensitivity test rather
        than as an independent calibration of the
        $Y_{\rm SZ}$--$M_h$ scaling relation.
    }
    \label{fig:ymrelation}
\end{figure}

We therefore show two versions of the $\tilde Y_{500}-M_h$ comparison in Fig.~\ref{fig:ymrelation}: one using the unweighted HOD halo mass distribution, and one using the $M_h^{5/3}$-weighted distribution. The difference between the two panels illustrates the model dependence of assigning an effective halo mass to a stacked \ac{tSZ} measurement. In the unweighted case, the inferred amplitude lies within $\sim1\sigma$ of the \textit{Planck} LBG measurement~\citep{PlanckLBG2013} and the \textit{Planck} $\times$ galaxy-group measurement~\citep{Lim2018ym} near $M_h\sim10^{13}M_\odot$. In the $M_h^{5/3}$-weighted case, the same measured \ac{tSZ} amplitude is compared at a much larger effective halo mass, leading to an apparent lower normalization relative to the literature relations. We do not interpret this difference as direct evidence for a tension in the $Y_{\rm SZ}-M_h$ relation, because both the effective mass and the pressure amplitude depend on model assumptions.

In summary, Fig.~\ref{fig:ymrelation} should be read as a consistency and sensitivity check. The current analysis robustly constrains the galaxy $\times$ \ac{tSZ} cross-correlation amplitude, but it does not separately determine the high-mass tail of the HOD distribution, the intrinsic pressure normalization $P_0$, and the correlation factor $\rho_{gy}$. A calibrated $Y_{\rm SZ}-M_h$ interpretation would require external mass information or an independent prior on $\rho_{gy}$.

\end{document}